\documentclass[prd,aps,floatfix,twocolumn,nofootinbib,10pt,longbibliography,superscriptaddress]{revtex4-2}

\usepackage[utf8]{inputenc}

\usepackage[dvipsnames]{xcolor}
\usepackage{mathrsfs}
\usepackage{graphicx}
\usepackage{ragged2e}
\usepackage{subcaption}
\DeclareCaptionJustification{justified}{\justifying}
\usepackage[pdftex,breaklinks,colorlinks,
linkcolor=Blue,
citecolor=teal,
anchorcolor=red,
urlcolor=cyan]{hyperref}
\usepackage{microtype}
\usepackage{nicefrac}
\usepackage{amsmath}
\usepackage{amssymb}
\usepackage{textcomp}
\usepackage{tensor}
\usepackage{bm}
\usepackage{natbib}

\newcommand{\beq}{\begin{equation}}
\newcommand{\eeq}{\end{equation}}
\newcommand{\<}{\left\langle}
\renewcommand{\>}{\right\rangle}

\renewcommand{\a}{\alpha}
\renewcommand{\b}{\beta}
\renewcommand{\c}{\gamma}
\renewcommand{\d}{\delta}

\newcommand{\g}{{g}}
\renewcommand{\k}{{k}}
\newcommand{\n}{{n}}
\newcommand{\e}{e}
\newcommand{\eps}{\epsilon}
\renewcommand{\H}{{\cal H}}
\newcommand{\G}{{\cal G}}

\begin{document}
	
	\title{Deformed Schwarzschild horizons in second-order perturbation theory: mass, geometry, and teleology}
	\author{Riccardo Bonetto}
	\affiliation{Bernoulli Institute for Mathematics, Computer Science and Artificial Intelligence, University of Groningen, Nijenborgh 9, 9747AG, Groningen, The Netherlands}
	\author{Adam Pound}
	\affiliation{School of Mathematical Sciences and STAG Research Centre, University of Southampton, Southampton, SO17 1BJ, United Kingdom}
	\author{Zeyd Sam}
	\affiliation{School of Mathematical Sciences and STAG Research Centre, University of Southampton, Southampton, SO17 1BJ, United Kingdom}
	\affiliation{Institute for Physics and Astronomy, University of Potsdam, D-14476 Potsdam, Germany}
	\date{\today}

\begin{abstract}

	In recent years, gravitational-wave astronomy has motivated increasingly accurate perturbative studies of gravitational dynamics in compact binaries. This in turn has enabled more detailed analyses of the dynamical black holes in these systems. For example, Pound et al. [Phys. Rev. Lett. 124, 021101 (2020)] recently computed the surface area of a Schwarzschild black hole's apparent horizon, perturbed by an orbiting body, to second order in the binary's mass ratio. In this paper, we take that as the starting point for a comprehensive study of a perturbed Schwarzschild black hole's apparent and event horizon at second perturbative order, deriving generic formulas for the first- and second-order corrections to the horizons' radial profiles, surface areas, Hawking masses, and intrinsic curvatures. We find that the two horizons are remarkably similar, and that any teleological behavior of the event horizon is suppressed in several ways. Critically, we establish that at all orders, the perturbed event horizon in a small-mass-ratio binary is effectively localized in time. Even more pointedly, the event horizon is identical to the apparent horizon at linear order regardless of the source of perturbation, implying that the seemingly teleological ``tidal lead", previously observed in linearly perturbed event horizons, is not genuinely teleological in origin. The two horizons do generically differ at second order, but their Hawking masses remain identical, implying that the event horizon obeys the same energy-flux balance law as the apparent horizon. At least in the case of a binary system, the difference between their surface areas remains extremely small even in the late stages of inspiral. In the course of our analysis, we also numerically illustrate puzzling behaviour in the black hole's motion around the binary's center of mass.
\end{abstract}
	
	\maketitle
	

\tableofcontents

\section{Introduction}

\subsection{Black holes in modern experimental physics}

Over the past few decades, black holes have gone from hypothetical objects of theoretical interest to ubiquitous elements of observational astrophysics.  
LIGO and Virgo now regularly detect binary black hole mergers~\cite{Abbott:2020niy}, and the Event Horizon Telescope has provided the first radio image of a black hole~\cite{EHT}. Future technological advances will enable far more precise observations, both with next-generation gravitational-wave detectors~\cite{Berti:2016lat,Maggiore:2019uih} and very-long-baseline radio interferometers~\cite{Psaltis:2020lvx, Volkel:2020xlc,Gralla:2020srx}. These will allow us to more stringently test whether the dark  objects we observe are genuine black holes or some other exotic compact objects~\cite{Cardoso:2019rvt}, and assuming they are black holes, whether they are accurately described by general relativity (GR).

In the near term, the most exacting measurements of black hole geometries will be made possible with the launch of the space-based gravitational-wave detector LISA in the early 2030s~\cite{Gair:2012nm,Barausse:2020rsu}. LISA will be sensitive to the merger of supermassive black holes, and the post-merger ringdown spectra from such systems will encode precise information about the nature of the final, merged object. Even more accurate measurements will be possible with LISA observations of extreme-mass-ratio inspirals (EMRIs), in which stellar-mass objects slowly spiral into massive black holes~\cite{Amaro-Seoane:2014ela,Amaro-Seoane:2020zbo}. The companion in an EMRI acts as a  probe of the massive black hole's geometry, performing $\sim 10^4$ or more intricate orbits while in the LISA band, most or all of them within 10 Schwarzschild radii of the black hole. The emitted, long-lived waveforms have a rich harmonic structure carrying detailed information about the black hole (or exotic compact object) spacetime. 

Measurements of this kind, with precise characterizations of deviations from GR's black hole spacetimes, are possible because of the remarkable simplicity of isolated black holes in GR. In the post-merger ringdown phase, and through all phases of an EMRI, the spacetime can be approximated as that of an isolated, stationary black hole subject to small perturbations---either quasinormal modes after a merger or the perturbations generated by the companion in an EMRI. In GR, an isolated black hole is uniquely described by the Kerr-Newman metric, which is fully specified by its mass, spin, and charge. In an astrophysical scenario, a black hole is unlikely to carry charge, as any nonzero charge will be quickly neutralized. Therefore an astrophysical black hole in GR should be uniquely described as a Kerr black hole, which has a unique set of quasinormal modes and a unique multipole structure that are fully determined by the black hole's mass and spin. By measuring the ringdown spectrum or the black hole's multipole structure, we can detect small deviations from the Kerr geometry. 

These prospects have motivated increasingly accurate theoretical studies of dynamically perturbed black holes in GR, moving beyond traditional linearized black hole perturbation theory onto {\em second-order} perturbation theory. There are ongoing efforts to calculate second-order effects in the ringdown of Kerr black holes~\cite{Green_2020,Loutrel:2020wbw,Ripley:2020xby} (building on earlier work in Ref.~\cite{Campanelli_1999}). And there is now a large body of work on EMRI models using gravitational self-force theory. In this method, the metric is expanded in powers of the binary's mass ratio $m/M$, where $M$ is the central black hole's mass and $m$ the small companion's. The perturbations effectively exert a self-force on the companion, accelerating it away from geodesic motion in the Kerr background. Surveys of self-force theory and EMRI modelling can be found in the recent reviews~\cite{Barack_2018,Pound-Wardell:2021}.

It is well known that EMRI models sufficiently accurate for LISA science must be carried to second perturbative order in $m/M$. Recently, Pound et al. reported the first calculation of a physical quantity at that order~\cite{Pound_2020}: the second-order contribution to the binding energy of quasicircular orbits around a Schwarzschild black hole. This calculation required a measurement of the Bondi mass of the binary, but also of the mass of the central black hole. In this paper, we take that calculation as the starting point for a more comprehensive study of perturbed Schwarzschild black holes at second order in perturbation theory. More specifically, we analyze the location and properties of a black hole's perturbed horizon(s).

\subsection{Black hole horizons}

In principle, the defining feature of a black hole is its event horizon. However, event horizons are intrinsically teleological surfaces: we can only identify their precise location if we know the entire future history of the universe. This has motivated the development of alternative ways of characterizing dynamically evolving black holes based on locally identifiable criteria~\cite{Hawking:1973,Hawking:1973uf,Hayward:1993mw,Ashtekar:2004cn,Booth:2005qc,Schnetter:2006yt,Krishnan:2007va,Gourgoulhon:2008pu,Hayward:2008ti}. The common characterization is via an apparent horizon or marginally outer trapped surface. Such a surface is defined in terms of the local-in-time, rather than global-in-time, behavior of null rays: given a slice of spacetime, an orientable, closed, spacelike 2-surface within the slice is a marginally outer trapped surface if future-directed null rays passing orthogonally outward through it have zero expansion. The apparent horizon is the outermost of these surfaces in the slice.

In GR, assuming appropriate energy conditions, the existence of an apparent horizon implies the presence of an event horizon, and the apparent horizon always either coincides with the event horizon or lies entirely inside the black hole. The converse is not true: the apparent horizon depends on one's foliation of spacetime, and one can find pathological foliations in which no apparent horizon is present on a slice even though the slice cuts through the event horizon~\cite{Wald:1991zz}. However, for reasonable choices of foliation~\cite{Ashtekar:2005ez}, the apparent horizon in a binary spacetime is found to be an excellent proxy for the event horizon except at moments very near merger~\cite{Cohen-Pfeiffer-Scheel:08}.

Event and apparent horizons have been extensively studied and compared in numerical relativity; see, for example, Refs.~\cite{Thornburg:03,Thornburg:2006zb,Schnetter:2006yt,Cohen-Pfeiffer-Scheel:08,Bohn-Kidder-Teukolsky:16,Pook-Kolb:2020zhm,Pook-Kolb:2020jlr}. There is also a long history of research on linear perturbations of horizons. 
In recent years, perturbed horizons have been studied in Refs.~\cite{Poisson:04,Vega_2011,Hughes1,Hughes2,Hughes3}, building on much  earlier work in Refs.~\cite{Hawking-Hartle:72,Hartle:1973zz,Hartle:1974gy,Teukolsky-Press:74,Price-Thorne:86}, for example. The series of papers~\cite{Hughes1,Hughes2,Hughes3} by O'Sullivan, Hughes, and Penna, in particular, have examined linearly perturbed black hole horizons specifically in EMRI scenarios at times significantly before the companion plunges into the black hole. A related line of work has examined linearly perturbed horizons in the specific case of a plunging companion~\cite{Hamerly-Chen:10,Emparan-Martinez:16,Hussain-Booth:17,Emparan-Martinez-Zilhao:17}.

Our treatment extends these analyses to second perturbative order, restricted to a Schwarzschild background and largely following Poisson and collaborators' treatment of the linear case in Refs.~\cite{Poisson:04,Vega_2011}. The extension to second order is necessary for the calculation of mechanical quantities in binaries, such as the binding energy in Ref.~\cite{Pound_2020}. But the extension is also interesting for one other major reason: in a binary inspiral, at times after merger or significantly before, it is precisely at second order that the apparent horizon (on a naturally chosen slicing) becomes distinguishable from the event horizon. It is also at this order that various definitions of the black hole's properties begin to differ. For example, there is a natural, unambiguous measure of the black hole's mass at linear order, but at second order one can define various nonequivalent measures of mass. In Ref.~\cite{Pound_2020}, Pound et al. specifically calculated the irreducible mass of the apparent horizon. Here, we show that the irreducible masses of the apparent and event horizon differ at second order, but that their Hawking mass (and Hawking-Hayward mass) remains identical.

\subsection{Outline and summary}

In Sec.~\ref{sec:perturbation_theory} we begin by reviewing basic methods in black hole perturbation theory. We emphasize two-timescale expansions, which play an important role in our characterization of the event horizon in small-mass-ratio binaries. In Sec.~\ref{sec:geometry}, we analyze the geometry of a generic 3-surface, foliated by spacelike 2-surfaces, near the background horizon; in later sections, this will be either the apparent or event horizon. We derive second-order perturbative formulas for the surface's intrinsic metric, surface area, Hawking mass, and scalar curvature.

In Secs.~\ref{sec:apparent horizon} and \ref{sec:event horizon}, we then obtain second-order formulas for the location of the apparent and event horizon in terms of the perturbations of the metric. We show that in a small-mass-ratio binary, the event horizon can be localized in time except in the final phase of inspiral, when the companion transitions into a plunging orbit. 

Section~\ref{sec:horizon locking} constructs simplified, gauge-invariant versions of the horizons' area, mass, and curvature. Using these measures, we show that in a dynamical region of spacetime, the apparent and event horizon first differ from one another at second perturbative order. We derive simple formulas relating their radial profiles, surface areas, masses, and curvatures. As alluded to above, we show that despite their other differences, the two horizons have identical Hawking mass through second order.

Next, in Sec.~\ref{sec:quasicircular_binaries}, we specialize to the case of quasicircular inspirals. In that context, we demonstrate numerically that the difference between the horizons' surface areas remains extremely small until near the innermost stable circular orbit, and we explain how the same is true for generic inspirals. We also examine the shape and location of the horizon at linear order, where we highlight two aspects. First, we emphasize that because the two horizons are identical at linear order, behavior that has previously been described as teleological must actually be entirely causal, and we suggest a nonteleological explanation for it. Second, we discuss a physically meaningful effect that has been omitted from previous depictions of the horizon at linear order: the motion of the black hole around the binary's center of mass. We show that well-motivated definitions of the black hole's ``position" do not have the expected Newtonian limit for large orbital radii, suggesting a more robust analysis is required. 

We conclude in Sec.~\ref{sec:conclusion} with a discussion of the implications of our results and their possible future developments. Among other applications, our results may be useful in sharpening the observed symmetry between quantities on the horizon and quantities at asymptotic infinity~\cite{Prasad:2020xgr} and in concretely calculating black hole memory effects in realistic scenarios~\cite{Hawking:2016msc,Rahman:2019bmk}.

Although motivated by EMRIs (and other small-mass-ratio systems such as intermediate-mass-ratio inspirals), most of our analysis applies to completely generic perturbations. Our only restriction is that in the intervals of time we consider, the number of null generators on the event horizon must remain fixed. This restriction is violated when additional generators join the horizon via caustics, as occurs when a body plunges into the black hole~\cite{Hamerly-Chen:10,Emparan-Martinez:16,Hussain-Booth:17,Emparan-Martinez-Zilhao:17}. In the case of a binary inspiral, our analysis therefore fails in an interval containing the final plunge. However, for reasons explained in Sec.~\ref{sec:event horizon}, our analysis remains valid until shortly before the transition to plunge.


\section{Second-order perturbations of Schwarzschild spacetime}\label{sec:perturbation_theory}

We begin in this section with an overview of the basic tools and conventions we use: second-order perturbation theory, the Schwarzschild metric, tensor spherical harmonics, and two-timescale expansions. Throughout the paper, we use geometric units with $G=c=1$.

\subsection{Perturbation theory through second order}

We assume the spacetime metric $g_{\alpha\beta}(\epsilon)$ depends on a small parameter $\epsilon$, and we expand the metric up to second order,
\begin{equation}\label{metric_expansion}
	g_{\alpha \beta} (\epsilon) = g^{(0)}_{\alpha \beta } + \epsilon  h^{(1)}_{\alpha \beta} + \epsilon^2 h^{(2)}_{\alpha \beta} + O(\epsilon^3) ,
\end{equation} 
where $g^{(0)}_{\alpha\beta}$ is the background metric and $ h^{(1)}_{\alpha \beta}$, $h^{(2)}_{\alpha \beta}$ are, respectively, the first- and second-order metric perturbations. In the context of a small-mass-ratio binary, $\eps$ will be the small mass ratio $m/M$, but in most of our analysis we work with generic perturbations due to an unspecified source. It will sometimes be convenient to use the total perturbation
\beq
h_{\alpha\beta} = \epsilon  h^{(1)}_{\alpha \beta} + \epsilon^2 h^{(2)}_{\alpha \beta} + O(\epsilon^3).
\eeq

We focus on a region around the black hole, where we assume the spacetime is vacuum, satisfying the vacuum Einstein equation
\begin{equation}\label{EFE_compact}
	R_{\alpha\beta}[g(\epsilon)]=0.
\end{equation}
Substituting the expansion \eqref{metric_expansion} into the Ricci tensor, one obtains
\begin{multline}
R_{\alpha\beta}[g(\epsilon)] = R_{\alpha\beta} +\eps \delta R_{\alpha\beta}[h^{(1)}]+\eps^2\delta R_{\alpha\beta}[h^{(2)}]\\
+\eps^2\delta^2 R_{\alpha\beta}[h^{(1)}]+O(\eps^3),
\end{multline}
where $\delta R_{\alpha\beta}$ is linear in its argument and $\delta^2 R_{\alpha\beta}$ is quadratic in its argument; this expansion is reviewed in Appendix~\ref{sec:curvature tensors}. Equation~\eqref{EFE_compact} then becomes
a sequence of equations, one at each order in $\eps$:
\begin{align}
    R_{\alpha\beta}[g^{(0)}] &=0,\\
	\delta R_{\alpha\beta}[h^{(1)}] &=0, \label{first_order_EFE} \\
	\delta R_{\alpha\beta}[h^{(2)}] &= -\delta^2R_{\alpha\beta}[h^{(1)}]. \label{second_order_EFE}
\end{align}
The zeroth-order equation states that the background metric must be a vacuum solution. The first-order equation is the standard linearized Einstein equation. In the second-order equation, the second-order perturbation $h^{(2)}_{\alpha\beta}$ is sourced by quadratic combinations of the first-order perturbation.

In this paper, we will not focus our attention on solving these equations; we refer to Refs.~\cite{Martel_2005,Brizuela:09,Pound-Wardell:2021} for descriptions of practical methods of obtaining solutions at first and second order in the case of a Schwarzschild background. Instead, taking the solution as a given, we analyze the effect the perturbations have on the black hole's horizons. For the most part, as stated above, we allow the perturbation to be completely generic. However, at various points, we specialize to an important class of perturbations that depend on two disparate time scales, and in the final section of the paper we numerically compute properties of the horizon in the specific scenario of a quasicircular inspiral. 

Most of our analysis will also leave the gauge of the metric perturbations unspecified. But one aspect of our calculations will make critical use of the gauge freedom within perturbation theory. In the perturbative context a gauge transformation corresponds to the infinitesimal coordinate transformation~\cite{Bruni-etal:96}
\begin{equation}
	x'^{\mu} = x^\mu - \epsilon \xi_{(1)}^{\mu} - \epsilon^2 \left(\xi_{(2)}^{\mu} - \frac{1}{2} \xi_{(1)}^{\nu} \partial_\nu \xi_{(1)}^{\mu}  \right) +O(\eps^3),
\end{equation}
under which the metric perturbations transform to
\begin{align}
	h'^{(1)}_{\alpha \beta} &= h^{(1)}_{\alpha \beta} + \mathcal{L}_{\xi_{(1)}} g^{(0)}_{\alpha \beta},\label{Delta h1}\\
	h'^{(2)}_{\alpha \beta} &= h^{(2)}_{\alpha \beta} + \mathcal{L}_{\xi_{(2)}} g^{(0)}_{\alpha \beta} + \frac{1}{2} \mathcal{L}^2_{\xi_{(1)}} g^{(0)}_{\alpha \beta} + \mathcal{L}_{\xi_{(1)}} h^{(1)}_{\alpha \beta}.\label{Delta h2}
\end{align}
Our conventions here follow Ref.~\cite{Pound_2015}.

\subsection{Schwarzschild background}

Throughout this paper, we take $g^{(0)}_{\alpha\beta}$ to be the Schwarzschild metric. We follow Refs.~\cite{Martel_2005,Vega_2011} by writing the spacetime manifold as the Cartesian product ${\cal M}={\cal M}^2\times S^2$, with $\mathcal{M}^2$ charted by, for example, $x^a=(t,r)$ or $x^a=(v,r)$, and $S^2$ charted by, for example, polar coordinates $\theta^A=(\theta, \phi)$. The background 4-metric is then divided into an induced metric on each submanifold,
\begin{equation}
	g^{(0)}_{\mu \nu}dx^\mu dx^\nu = g^{(0)}_{ab}dx^a dx^b + r^2 \Omega_{AB}d\theta^A d\theta^B.
\end{equation}
Here $r$ is the areal radius and $\Omega_{AB}$ is the metric on the unit sphere. 
On ${\cal M}^2$ we exclusively use ingoing Eddington-Finkelstein coordinates, $(v,r)$, such that 
\begin{equation}\label{schwarzschild_metric}
	g^{(0)}_{ab}dx^a dx^b = -f(r) dv^2 +  2 dv dr, 
\end{equation}
where $f(r) = 1-\frac{2M}{r}$. On $S^2$ we for the most part work covariantly, without specifying coordinates. 

Our conventions for covariant derivatives and for raising and lowering indices are somewhat nonstandard. $\nabla_\mu$ denotes the covariant derivative compatible with the exact metric, $g_{\alpha\beta}$, and we use $g_{\alpha\beta}$ and its inverse, $g^{\alpha\beta}$, to raise and lower Greek indices on nonperturbative quantities. ${}^{(0)}\nabla_\mu$ and a semicolon denote the covariant derivative compatible with $g^{(0)}_{\alpha\beta}$. For the most part, we avoid raising or lowering indices on perturbative quantities, but for the sake of brevity we occasionally use  $g^{(0)}_{\alpha\beta}$ and its inverse, $g_{(0)}^{\alpha\beta}$, for that purpose; in such instances, we explicitly warn the reader that we have done so. We also introduce $D_A$ as the covariant derivative compatible with the unit-sphere metric $\Omega_{AB}$. We use $\Omega_{AB}$ and its inverse, $\Omega^{AB}$, to raise and lower indices on quantities associated with $\Omega_{AB}$, such as $D_A$ and $\epsilon_{AB}$ (the Levi-Civita tensor associated with $\Omega_{AB}$). We do not use them to raise or lower capital Latin indices on other quantities.

We refer to Section II of Ref.~\cite{Martel_2005} for additional useful identities related to the 2+2 split of the Schwarzschild metric.

\subsection{Tensor spherical harmonics}

Given Schwarzschild spacetime's spherical symmetry, it is often convenient to expand quantities in spherical harmonics. With an appropriate choice of spherical basis functions, the Einstein equations separate into decoupled equations for each $lm$ mode. In this paper, we assume that the metric perturbation is obtained mode by mode in this way. Such a harmonic decomposition will also allow us to easily solve the differential equations governing the location of the horizon and to evaluate various integrals over the horizon surface.

For the harmonics we adopt the conventions of Martel and Poisson~\cite{Martel_2005}. We start with the scalar spherical harmonics, $Y^{lm}(\theta^A)$, which satisfy the eigenvalue equation
\beq\label{eigenvalue}
D^2Y^{lm} = -\lambda_1^2Y^{lm},
\eeq
where $D^2:=D^AD_A$ and $\lambda_s$ is defined for any integer $s$ in Eq.~\eqref{lambda_s}. From the scalar harmonics we define the vector harmonics
\begin{align}
	 Y^{lm}_A &:=D_A Y^{lm} ,\label{YA}\\
	 X^{lm}_A &:=-\epsilon\indices{_A^B}D_BY^{lm}.\label{XA}
\end{align}
$Y^{lm}_A$, $X^{lm}_A$ are respectively referred to as even-parity and odd-parity vector harmonics. Finally we define the even- and odd-parity tensor harmonics
\begin{align}
	Y^{lm}_{AB} &:=D_A D_B Y^{lm} + \frac{1}{2} \lambda_1^2\, \Omega_{AB}Y^{lm} , \\
	X^{lm}_{AB} &:=\frac{1}{2} \left(D_A X^{lm}_B+ D_B X^{lm}_A\right) ,\label{XAB}
\end{align}
which are both symmetric and trace-free with respect to $\Omega^{AB}$.

These harmonics satisfy the orthogonality relations 
\begin{align}
	\int  \bar Y^{lm} Y^{l'm'} d\Omega &= \delta^{ll'} \delta^{mm'} , \label{orthogonality-first} \\
	\int  \bar Y^{lm}_A Y_{l'm'}^A  d\Omega &=  \lambda_1^2 \delta^{ll'} \delta^{mm'} ,  \\
	\int  \bar X^{lm}_A X_{l'm'}^A d\Omega &=  \lambda_1^2 \delta^{ll'} \delta^{mm'} ,   \\
	\int \bar X^{lm}_A Y_{l'm'}^A d\Omega &= 0 , \\
	\int \bar Y^{lm}_{AB} Y_{l'm'}^{AB} d\Omega &= \frac{1}{2} \lambda_2^2 \delta^{ll'} \delta^{mm'} , \\
	\int  \bar X^{lm}_{AB} X_{l'm'}^{AB} d\Omega &= \frac{1}{2} \lambda_2^2 \delta^{ll'} \delta^{mm'} , \\
	\int  \bar Y^{lm}_{AB} X_{l'm'}^{AB} d\Omega &= 0 \label{orthogonality-last}.
\end{align}
Here $d\Omega$ is the surface element on the unit sphere, and we use $\Omega^{AB}$ to raise indices. An overbar denotes complex conjugation; the harmonics of all ranks satisfy identities of the form
\beq\label{Yconj}
\bar Y^{lm} = (-1)^m Y^{l-m}.
\eeq

It will be useful to split tensors on $S^2$ into their trace-free and traceful parts,
\beq
T_{AB}=T_{\langle AB\rangle}+T_\circ \Omega_{AB},
\eeq
where $\Omega^{AB}T_{\langle AB\rangle}=0$ and $T_\circ :=\frac{1}{2}\Omega^{AB}T_{AB}$. Given this split, we expand the metric perturbations in spherical harmonics according to
\begin{align}
	h^{(n)}_{ab} &= \sum_{lm} h^{(nlm)}_{ab} Y^{lm}, \label{scalar_metric_harm} \\
	h^{(n)}_{\circ} &= \sum_{lm} h^{(nlm)}_{\circ} Y^{lm},\\
	h^{(n)}_{aA} &= \sum_{lm} \left(h^{(nlm)}_{a +} Y^{lm}_A + h^{(nlm)}_{a -} X^{lm}_A\right), \label{vector_metric_harm} \\
	h^{(n)}_{\langle AB\rangle } &= \sum_{lm}\left( h^{(nlm)}_{+} Y^{lm}_{AB} +h^{(nlm)}_{-} X^{lm}_{AB}\right) \label{metric_harm}. 
\end{align}
Each of the coefficients $h^{(nlm)}_{ab}$, $h^{(nlm)}_\circ$, $h^{(nlm)}_{a\pm}$, and $h^{(nlm)}_{\pm}$ is a function of $v$ and $r$.

Our second-order calculations will naturally involve products of functions on $S^2$, and we will need to decompose such products into harmonics. As the simplest example, consider $\int \psi(\theta^A)\chi(\theta^A)d\Omega$, which is (up to a factor of $1/\sqrt{4\pi}$) the scalar monopole mode of the product $\psi\,\chi$. Expanding each function in harmonics, as $\psi = \sum_{lm} \psi^{lm} Y^{lm}$ and $\chi=\sum_{lm} \chi^{lm} Y^{lm}$, and using Eqs.~\eqref{orthogonality-first} and \eqref{Yconj}, we obtain 
\begin{equation}\label{conjugate}
	\int \psi\, \chi\,  d\Omega = \sum_{lm}  \psi^{lm}\bar\chi^{lm} = \sum_{lm}  \bar\psi^{lm}\chi^{lm},
\end{equation}
where we have used the fact that $\bar\chi^{lm}=(-1)^m\chi^{l-m}$ for any real-valued function $\chi(\theta^A)$. The analogous  rule applies for integrals of the form $\int \psi_A \Omega^{AB}\chi_B d\Omega $ and $\int \psi_{AC}\Omega^{AB}\Omega^{CD} \chi_{BD} d\Omega$. In Appendix~\ref{sec:coupling} we describe our procedure for evaluating more general integrals.

\subsection{Two-timescale expansion}\label{two_time_section}

Most of our analysis utilizes regular perturbation theory, in which the coefficients $h^{(n)}_{\alpha\beta}$ in Eq.~\eqref{metric_expansion} are independent of the small parameter $\eps$. However, at various points we adopt an expansion that is better suited to a small-mass-ratio binary: a two-timescale expansion. 
We refer to textbooks on singular perturbation theory for introductions to the method (e.g.,~\cite{Kevorkian-Cole:96}). In the particular context of an inspiral into a black hole, the use of the method is inspired by the fact that during the inspiral, the system evolves on two distinct timescales: the short \emph{orbital timescale} $\sim 1/\Omega\sim M$ associated with the companion's orbital frequency $\Omega$; and the long \emph{radiation-reaction time} $t_{\rm rr}\sim \Omega/(d\Omega/dt)\sim M/\eps$, over which the orbital frequencies evolve due to gravitational-wave emission. A two-timescale expansion allows us to maintain accuracy on both timescales, while regular perturbation theory would break down well before a radiation-reaction time~\cite{Pound-Wardell:2021}.

Concretely, an orbiting, nonspinning body in the equatorial plane of Schwarzschild spacetime has two independent frequencies, $\Omega_r$ and $\Omega_\phi$, associated with radial and azimuthal motion. Specialized to a region around the horizon, the two-timescale ansatz for the metric is then 
\beq\label{metric_expansion_two-timescale}
g_{\alpha\beta} = g^{(0)}_{\alpha\beta} + \sum_{n\geq1}\sum_{\bm{k}}\eps^n h^{(n,\bm{k})}_{\alpha\beta}(\tilde v,r,\theta^A)e^{-i\varphi_{\bm{k}}},
\eeq
where $\varphi_{\bm{k}}:=k^i\varphi_i$, and the sum runs over all pairs of integers $\bm{k}=k^i=(k^r,k^\phi)$. In this expansion we have introduced the {\em slow time} variable $\tilde v:=\eps v$ and the orbital phases $\varphi_i=(\varphi_r,\varphi_\phi)$, which are given by
\beq\label{varphi_i}
\varphi_i(v,\eps) = \int^v_0 \Omega_i(\eps v') dv' + \varphi_i(0,\eps)
\eeq
or equivalently, $d\varphi_i/dv=\Omega_i$; the unspecified lower limit in Eq.~\eqref{varphi_i} represents an arbitrary choice of initial condition.
The metric perturbations in Eq.~\eqref{metric_expansion_two-timescale} hence have the character of a sum of slowly varying amplitudes multiplied by rapidly varying phase factors. We refer to Refs.~\cite{miller2020twotimescale,Pound-Wardell:2021} for more detailed descriptions of such two-timescale expansions of the metric, which are further developments of Hinderer and Flanagan's seminal work on the two-timescale expansion of inspiral orbits~\cite{Hinderer_2008}. Such an approximation should be uniformly accurate until a time shortly before the inspiraling body transitions to a plunging orbit~\cite{Ori_2000}; we describe that cutoff in Sec.~\ref{sec:event horizon}.

Treated as a function of the coordinates $x^\alpha$ on ${\cal M}$, the coefficients of $\eps^n$ in the expansion~\eqref{metric_expansion_two-timescale} depend on $\eps$, meaning ~\eqref{metric_expansion_two-timescale} is not a Taylor series around $\eps=0$; this is a defining characteristic of singular perturbation theory~\cite{Pound:10}. However, the dependence on $\eps$ comes in a circumscribed form that allows us to solve the field equations through any desired order on the radiation-reaction timescale. We can also view Eq.~\eqref{metric_expansion_two-timescale} as a regular Taylor expansion of a field on a higher-dimensional manifold $\tilde{\cal M}$ charted by $(\tilde v, \varphi_i, r, \theta^A)$. ${\cal M}$ is embedded into this larger manifold with a map $T_\eps : \mathcal{M} \to \tilde{\mathcal{M}}$ defined by $T_\eps(v,r,\theta^A)=[\eps v,\varphi_i(v,\eps),r,\theta^A]$. After performing the ordinary Taylor expansion of the function on $\tilde {\cal M}$, we then pull it back to its restriction on the physical spacetime manifold ${\cal M}$.

Quantities on the horizon inherit the metric's two-timescale form, which has several important consequences. First, $v$ derivatives involve terms that are suppressed by one order in $\eps$. To see this, let $\tilde \chi$ be a field on $\tilde{\mathcal{M}}$, and let $\chi$ be its restriction to ${\cal M}$, such that $\tilde \chi[\eps v,\varphi_i(v,\eps),r,\theta^A] = \chi(v,r,\theta^A)$. The $v$ derivative of $\chi$ then becomes
\begin{align}
	\frac{d \chi}{d v} &= \epsilon \frac{\partial \tilde \chi}{\partial \tilde v}  + \Omega_i \frac{\partial \tilde \chi}{\partial \varphi_i}.\label{two_derivative}
\end{align}
The first term, which characterizes the field's slow evolution, will be demoted to the next order, such that if $\chi$ appears on the horizon at first perturbative order, for example, then its $\tilde v$ derivative will contribute to second-order quantities on the horizon. For notational simplicity, we will not distinguish between $\tilde\chi$ and its pullback $\chi$. 

The second consequence of the two-timescale expansion is that it transforms differential equations in $v$ into algebraic ones. Suppose we have a differential equation governing a quantity's behavior on the horizon, of the form
\begin{equation}\label{diff_eq}
	\frac{d \chi(v,\eps)}{d v} = S(v,\eps). 
\end{equation}
If we expand in the Fourier series $\chi(v) = \sum_{\bm{k}}\chi_{\bm{k}}(\tilde v)e^{-i\varphi_{\bm{k}}}$ and $S(v)=\sum_{\bm{k}}S_{\bm{k}}(\tilde v)e^{-i\varphi_{\bm{k}}}$, then at leading order Eq.~\eqref{diff_eq} becomes 
\beq
-i\Omega_{\bm{k}} \chi_{\bm{k}} = S_{\bm{k}},
\eeq
where $\Omega_{\bm{k}}:=k^i\Omega_i$. This kind of transformation is one of the key utilities of the two-timescale expansion. For example, it puts the Einstein field equations~\eqref{first_order_EFE} into precisely the same form they would have in a standard frequency-domain treatment, while correctly capturing the system's slow evolution.

Transforming differential equations in this way implicitly localizes them in time: rather than having to integrate over $v$ from some initial condition, we algebraically determine the solution at a given value of slow time $\tilde v$. This is especially relevant for the event horizon, which is inherently a nonlocal-in-time surface that depends on the spacetime's distant future. The underlying reason for this localization in time is that integrals over large ranges of $v$ collapse to local-in-$\tilde v$ quantities when the integral contains multiple time scales. For example, consider the integral $\int_v^\infty\!\! F(\eps v')e^{-i\varphi_{\bm{k}}(v',\eps)}dv'$, which extends from the present time $v$ into the infinite future. If $F/\Omega_{\bm{k}}$ vanishes when $v\to\infty$, then we can repeatedly integrate by parts to obtain\footnote{For resonant modes that pass through $\Omega_{\bm{k}}=0$ for some value of $\tilde v$~\cite{vandeMeent:14}, this approximation breaks down. The integral should in that case be approximated using the stationary-phase approximation~\cite{Pound-Wardell:2021}. For simplicity we exclude resonances from our analysis.}
\begin{multline}\label{integral_approximation1}
\int_v^\infty\!\! F(\eps v')e^{-i\varphi_{\bm{k}}(v',\eps)}dv' \\ =e^{-i \varphi_{\bm{k}}(v,\eps)}\left[ \frac{F(\tilde v)}{i\Omega_{\bm{k}}(\tilde v)}+\frac{\eps}{i\Omega_{\bm{k}}}\frac{d}{d\tilde v}\frac{F(\tilde v)}{i\Omega_{\bm{k}}(\tilde v)} +o(\eps)\right].
\end{multline}
Depending on the behavior of $F$ and $\Omega_i$ in the far future, this approximation can be carried to arbitrary order in $\eps$, making nonlocal effects arbitrarily small. We will see in Sec.~\ref{sec:event horizon} how this type of approximation applies to the location of the event horizon.

Because it cleanly separates slow and fast evolution, a two-timescale expansion also allows us to unambiguously identify the time average of a quantity as the zero mode in its fast-time Fourier series:
\beq\label{average_definition}
\left\langle \chi \right\rangle := \frac{1}{(2\pi)^2}\oint \chi(\tilde v,\varphi_i) \,d^2\varphi = \chi_{\bm{0}}(\tilde v).
\eeq
This will enable us to characterize the black hole's average evolution, discarding fluctuations on the orbital timescale.

Finally, we note that the mode number $k^\phi$ is precisely the azimuthal mode number $m$. This is because, due to the background spacetime's axisymmetry, the small body's stress-energy can only depend on $\phi$ and $\varphi_\phi$ in the combination $(\phi-\varphi_\phi)$, and the metric perturbations inherit that dependence; see Sec. 7.1 of Ref.~\cite{Pound-Wardell:2021}. As a consequence, we can write the expansion~\eqref{metric_expansion_two-timescale} in the form 
\beq
h_{\alpha\beta}=\sum_{n\geq1}\sum_{m,k}\eps^n h^{(nmk)}_{\alpha\beta}(\tilde v,r,\theta^A)e^{-i(m\varphi_\phi + k\varphi_r)}.
\eeq
For the quasicircular orbits we consider in Sec.~\ref{sec:quasicircular_binaries}, this reduces to 
\beq\label{two-timescale h quasicircular}
h_{\alpha\beta}=\sum_{n\geq1}\sum_{m=-\infty}^\infty\eps^n h^{(nm)}_{\alpha\beta}(\tilde v,r,\theta^A)e^{-im\varphi_\phi}.
\eeq


\section{Geometry of a perturbed horizon}\label{sec:geometry}

Before considering the apparent horizon and event horizon in detail, we begin by describing the geometry of a generic 3-surface ${\cal H}$ close to the background horizon, which may be either of the two horizons. This serves to set our notation and to present formulas that will be common to both horizons. 

Over the course of the section, we introduce a convenient basis of vectors and the induced metric on the surface, and we then derive perturbative formulas for the horizon's surface area and intrinsic curvature. We conclude by showing the consistency between our formulas for the area and curvature, as dictated by the Gauss-Bonnet theorem. 

\subsection{Embedding and induced metric}

As coordinates on ${\cal H}$, we use the extrinsic coordinates $y^i=(v,\theta^A)$. ${\cal H}$ is then described by an embedding
\beq
x_{\H}^\alpha(y^i,\eps) = [v, r_{\H}(y^i,\eps), \theta^A]. \label{embedding_relation}
\eeq
We assume that the perturbed horizon's radial profile can be written as an expansion around the background horizon radius: 
\begin{equation}\label{radius_expansion}
	r_\H = 2M + \epsilon r^{(1)}(y^i) + \epsilon^2 r^{(2)}(y^i) + O(\epsilon^3).
\end{equation}
The perturbations $r^{(n)}$ will depend on whether ${\cal H}$ is the apparent horizon or the event horizon. In Secs.~\ref{sec:apparent horizon} and \ref{sec:event horizon}, we express $r^{(n)}$ in terms of the metric perturbations $h^{(n)}_{\alpha\beta}$ in each of the two cases.

The embedding~\eqref{embedding_relation} defines a basis of vectors fields tangent to ${\cal H}$,
\beq
\e^{\alpha}_{i}\partial_\alpha = \frac{\partial x_\H^\alpha}{\partial y^i}\partial_\alpha = \partial_i + \frac{\partial r_\H}{\partial y^i}\partial_r.
\eeq
In terms of these tangent vectors, the induced metric on ${\cal H}$ is
\begin{equation}\label{induced 3-metric}
	\gamma_{ij}=\e_i^\alpha\e_j^\beta\g_{\alpha\beta}.
\end{equation}
However, we will be more interested in the foliation of ${\cal H}$ into spacelike slices of constant $v$, ${\cal H}_v$, on which we use coordinates $\theta^A$. The basis of vectors tangent to ${\cal H}_v$ is
\beq
\e^{\alpha}_{A}\partial_\alpha = \frac{\partial x_\H^\alpha}{\partial \theta^A}\partial_\alpha = \partial_A + \frac{\partial r_\H}{\partial \theta^A}\partial_r,\label{basis vectors}
\eeq
and the induced metric on ${\cal H}_v$ is
\begin{equation}\label{intrinsic_metric_basic}
	\gamma_{AB}=\e_A^\alpha\e_B^\beta\g_{\alpha\beta}.
\end{equation}
If we substitute the expanded metric ~\eqref{metric_expansion}, we can write this as
\beq
\gamma_{AB} = \gamma^{(0)}_{AB} +\eps \gamma^{(1)}_{AB} + \eps^2 \gamma^{(2)}_{AB} + O(\eps^3),\label{gamma expansion on H}
\eeq
where $\gamma^{(0)}_{AB}:=\e^\alpha_A \e^\beta_B g_{\alpha\beta}(x^\mu_\H)$ and $\gamma^{(n)}_{AB}:=\e^\alpha_A \e^\beta_B h^{(n)}_{\alpha\beta}(x^\mu_\H)$ for $n>0$. The components of $\gamma_{AB}$ are functions of the coordinates $\theta^A$ on ${\cal H}_v$, and they inherit an additional parametric dependence on $v$.

We do not provide more explicit expressions for the coefficients $\gamma^{(n)}_{AB}$  because we ultimately perform an additional expansion of them. This second expansion is called for because Eq.~\eqref{gamma expansion on H} is written in terms of tensors at points on ${\cal H}_v$. It will generally be more useful to expand all such tensors around their values on the background horizon. By substituting the expansion~\eqref{radius_expansion}, we can expand any tensor's components $T^\alpha{}_\beta$ on ${\cal H}_v$ as
\begin{align}
T^\alpha{}_\beta(v,r_\H,\theta^A) &=  T^\alpha{}_\beta(v,2M,\theta^A) + \eps r^{(1)}\partial_r T^\alpha{}_\beta \nonumber\\
&\quad + \eps^2 r^{(2)}\partial_r T^\alpha{}_\beta + \frac{1}{2}\eps^2\big(r^{(1)}\big)^2\partial^2_r T^\alpha{}_\beta \nonumber\\
&\qquad\qquad\qquad\qquad\qquad +O(\eps^3),\label{expansions around H0}
\end{align}
where derivatives of $T^\alpha{}_\beta$ on the right are evaluated at $(v,2M,\theta^A)$. Geometrically, this represents an expansion of the pullback $\varphi^* T^\alpha{}_\beta$, where  $\varphi:(v,2M,\theta^A)\mapsto (v,r_\H,\theta^A)$ maps points on the background horizon to points on the perturbed horizon. The pullback can be expressed in terms of Lie derivatives as $\varphi^*T^\alpha{}_\beta= e^{\delta r_\H{\cal L}_\Xi}T^\alpha{}_\beta$ with $\delta r_\H := r_\H-2M$ and $\Xi^\alpha\partial_\alpha=\partial_r$.

Performing such an expansion for the induced metric, we find
\beq\label{intrinsic_metric}
\gamma_{AB} = 4 M^2 \Omega_{AB} +\eps \breve\gamma^{(1)}_{AB} + \eps^2 \breve\gamma^{(2)}_{AB} + O(\eps^3),
\eeq
where
\begin{align}
	\breve\gamma^{(1)}_{AB} &= h^{(1)}_{AB} + 4M r^{(1)} \Omega_{AB},\\ 
	\breve\gamma^{(2)}_{AB} &= h^{(2)}_{AB} + 4M r^{(2)} \Omega_{AB} + 2 h^{(1)}_{r ( A} D_{B )} r^{(1)} \nonumber \\
	&\quad + r^{(1)} \partial_r h^{(1)}_{AB}  + (r^{(1)})^2 \Omega_{AB}.
\end{align}
All quantities on the right are evaluated at $(v,2M,\theta^A)$. The inverse metric is then
\begin{align}
\gamma^{AB} &= \frac{1}{4M^2}\Omega^{AB} - \frac{\eps}{(4M^2)^2}\Omega^{AC}\Omega^{BD} \breve\gamma^{(1)}_{CD} \nonumber\\
&\quad - \frac{\eps^2}{(4M^2)^2}\Omega^{AC}\Omega^{BD} \breve\gamma^{(2)}_{CD} \nonumber\\
&\quad + \frac{\eps^2}{(4M^2)^3}\Omega^{AC}\Omega^{BE}\Omega^{DF} \breve\gamma^{(1)}_{CD}\breve\gamma^{(1)}_{EF}+O(\eps^3).\label{gamma inverse}
\end{align}
Similarly, the basis vectors~\eqref{basis vectors} on ${\cal H}_v$ become
\beq
\e^\alpha_A = \breve e^{(0)\alpha}_A +\eps\, \breve e^{(1)\alpha}_A + \eps^2\, \breve e^{(2)\alpha}_A + O(\eps^3),\label{e expansion around H0}
\eeq
with
\begin{align}
	\breve\e^{(0)\alpha}_A\partial_\alpha &=\partial_A , \label{tangent_1} \\
	\breve\e^{(1)\alpha}_A\partial_\alpha &=  D_A r^{(1)}\partial_r , \label{tangent_2} \\
	\breve\e^{(2)\alpha}_A\partial_\alpha &=D_A r^{(2)}\partial_r. \label{tangent_3}
\end{align}

Later calculations will require the expansion of these quantities in spherical harmonics. We write the radial perturbations as
\beq\label{r harmonic expansion}
r^{(n)}(v,\theta^A) = \sum_{lm}r^{(n)}_{lm}(v)Y^{lm}(\theta^A),
\eeq
and we write $\breve\gamma^{(n)}_{AB}=\breve\gamma^{(n)}_{\langle AB\rangle} + \breve\gamma^{(n)}_\circ\Omega_{AB}$ as
\begin{subequations}\label{gamma harmonic expansions}
\begin{align}
\breve{\gamma}^{(n)}_{\circ} &=\sum_{lm} \breve{\gamma}^{(nlm)}_\circ(v) Y^{lm}(\theta^A),\\
\breve{\gamma}^{(n)}_{\langle AB\rangle} &=\sum_{lm}\left[ \breve{\gamma}^{(nlm)}_+(v) Y^{lm}_{AB}+\breve{\gamma}^{(nlm)}_-(v) X^{lm}_{AB}\right].
\end{align}
\end{subequations}

The coefficients $\breve{\gamma}_\circ^{(nlm)}$ and $\breve{\gamma}_\pm^{(nlm)}$ are straightforwardly expressed in terms of the coefficients in the harmonic expansions of $h^{(n)}_{\alpha\beta}$ and $r^{(n)}$. At first order,
\begin{subequations}\label{gamma1 harmonic expansions}
\begin{align}
    \breve{\gamma}^{(1lm)}_\circ &= h^{(1lm)}_\circ + 4M r^{(1)}_{lm},\\
    \breve{\gamma}^{(1lm)}_\pm &= h^{(1lm)}_\pm.
\end{align}
\end{subequations}
At second order, we must decompose products of harmonics into pure harmonics, as described in Appendix~\ref{sec:coupling}. The result is
\begin{subequations}\label{gamma2 harmonic expansions}
\begin{align}
    \breve{\gamma}^{(2lm)}_\circ &= h^{(2lm)}_\circ + 4M r^{(2)}_{lm} \nonumber\\
    &\quad + \sum_{\substack{l'm'\\l''m''}}\Big[{}^{(1)}\Gamma^{lm}_{l'm'l''m''}(v) C^{lm0}_{l'm'1l''m''-1}\nonumber\\
    &\qquad\qquad +{}^{(0)}\Gamma^{lm}_{l'm'l''m''}(v) C^{lm0}_{l'm'0l''m''0}\Big],\\
    \breve{\gamma}^{(2lm)}_\pm &= h^{(2lm)}_\pm +\sum_{\substack{l'm'\\l''m''}}\Big[{}^{(2)}\Gamma^{lm\pm}_{l'm'l''m''}(v)C^{lm2}_{l'm'2l''m''0}\nonumber\\
    &\qquad\qquad\quad +{}^{(1)}\Gamma^{lm\pm}_{l'm'l''m''}(v)C^{lm2}_{l'm'1l''m''1}\Big],
\end{align}
\end{subequations}
where the $C$ symbols are given in Eq.~\eqref{coupling}, and their $v$-dependent coefficients are given by
\begin{subequations}\label{gamma2 coeffs}
\begin{align}
{}^{(0)}\Gamma^{lm}_{l'm'l''m''} &= r^{(1)}_{l'm'}\partial_r h''_{\circ} + r^{(1)}_{l'm'}r^{(1)}_{l''m''},\\
{}^{(1)}\Gamma^{lm}_{l'm'l''m''} &= - \frac{\lambda_1'\lambda_1''}{2}r^{(1)}_{l''m''}\left(\sigma_+h'_{r+}-i\sigma_- h'_{r-}\right),\\
{}^{(2)}\Gamma^{lm+}_{l'm'l''m''} &= \frac{\lambda_2'}{2\lambda_2}r^{(1)}_{l''m''}\left(\sigma_+ \partial_rh'_{+}-i\sigma_-\partial_rh'_{-}\right),\\
{}^{(1)}\Gamma^{lm+}_{l'm'l''m''} &= \frac{\lambda_1'\lambda_1''}{\lambda_2}r^{(1)}_{l''m''}\left(\sigma_+ h'_{r+}-i\sigma_-h'_{r-}\right),\\
{}^{(s)}\Gamma^{lm-}_{l'm'l''m''} &= {}^{(s)}\Gamma^{lm+}_{l'm'l''m''} \text{ with }\sigma_\pm\to -i \sigma_\mp.
\end{align}
\end{subequations}
Here we use the compact notation described in Appendix~\ref{sec:second_order_expressions}. These formulas will substantially collapse in Sec.~\ref{sec:horizon locking}.

\subsection{Null basis vectors}

On each slice ${\cal H}_v$ we introduce a pair of future-directed null vectors $\k^\alpha$ and $\n^\alpha$. Both are orthogonal to ${\cal H}_v$, satisfying
\begin{align}
	 \g_{\alpha \beta}\k^\alpha  \e_A ^\beta &= 0 , \label{basis_1} \\
	\g_{\alpha \beta} \n^\alpha \e_A ^\beta &= 0. \label{basis_2}
\end{align}
$\k^\alpha$ is chosen to point outward from ${\cal H}_v$, toward the horizon's exterior, and $\n^\alpha$ to point inward, into the black hole's interior. Together with $\e_A^\alpha$, these vectors form a basis for 4-vectors at points on ${\cal H}_v$. In the case of the event horizon, $\k^\alpha$ will be the tangent vector of the horizon generators; in the case of the apparent horizon, $\k^\alpha$ will not, generically, be tangent to the surface ${\cal H}$.

In both cases, we normalize $\k^\alpha$ such that $\k^v = 1$, and we scale $\n^\alpha$ such that it satisfies 
\begin{align}
	\g_{\alpha \beta}\k^\alpha \n^\beta   &= -1. \label{basis_3}
\end{align}
The metric can then be written as
\beq
\g_{\alpha\beta} = -\n_\alpha\k_\beta - \k_\alpha n_\beta + \gamma_{AB}\e_\alpha^A \e_\beta^B,
\eeq
where $e^A_\alpha := \gamma^{AB}\g_{\alpha\beta}e^\beta_B$, and the metric's inverse can be written as
\beq\label{ginverse in basis}
\g^{\alpha\beta} = -\n^\alpha\k^\beta - \k^\alpha n^\beta + \gamma^{AB}\e^\alpha_A \e^\beta_B.
\eeq

Given the normalization $k^v=1$, the orthonormality conditions~\eqref{basis_1}--\eqref{basis_3} uniquely determine $\k^\alpha$ and $\n^\alpha$ in terms of $r_\H$, $D_A r_\H$, and $\Omega_{AB}$. Assuming expansions of the form
\beq
\k^\alpha = \k^\alpha_{(0)} + \eps \k^\alpha_{(1)}+\eps^2\k^\alpha_{(2)} + O(\eps^3),\label{k expansion on H}
\eeq
and analogous for $\n^\alpha$, we can solve the orthonormality equations for the coefficients at each order. At leading order,
\begin{align}
\k^\alpha_{(0)}\partial_\alpha &= \partial_v +\frac{1}{2}\left[f(r_\H) - r_\H^{-2}\Omega^{AB}D_A r_\H D_B r_\H\right]\partial_r \nonumber\\
&\quad - r_\H^{-2}D_A r_\H \Omega^{AB}\partial_B,\\ 
\n^\alpha_{(0)}\partial_\alpha &= -\partial_r.
\end{align}

We again elide explicit expressions for $\k^\alpha_{(n)}$ and $\n^\alpha_{(n)}$ to higher order, instead presenting results for the re-expansions around the background horizon,
\begin{align}
    \k^\alpha &= \breve\k^\alpha_{(0)} + \eps\breve\k^\alpha_{(1)} + \eps^2 \breve\k^\alpha_{(2)} + O(\eps^3),\label{k expansion around H0}\\ 
    \n^\alpha &= \breve\n^\alpha_{(0)} + \eps\breve\n^\alpha_{(1)} + \eps^2 \breve\n^\alpha_{(2)} + O(\eps^3).
\end{align}
The leading terms in these expansions are
\begin{align}
    \breve\k^\alpha_{(0)}\partial_\alpha &= \partial_v,\\
    \breve\n^\alpha_{(0)}\partial_\alpha &= -\partial_r.
\end{align}
The first subleading terms are
\begin{align}
	\breve{\k}^\alpha_{(1)}\partial_\alpha  &= - \frac{1}{4 M} \Big( 2 M h^{(1)}_{vv} - r^{(1)} \Big)\partial_r \nonumber\\
    &\quad -\frac{1}{4M^2} \Big(h^{(1)}_{vA} + D_A r^{(1)} \Big)\Omega^{AB}\partial_B, \label{k1}\\
    \breve{\n}^\alpha_{(1)}\partial_\alpha &= \frac{1}{2}h^{(1)}_{rr}\partial_v + h^{(1)}_{vr}\partial_r +\frac{1}{4M^2} h^{(1)}_{rA}\Omega^{AB}\partial_B,\label{n1}
\end{align}
where all terms on the right are evaluated at $(v,2M,\theta^A)$.
The second-order terms in $k^\alpha$ are
\begin{subequations}\label{k2}
\begin{align}
    \breve{k}^v_{(2)} &=0,\\
	\breve{k}_{(2)}^r &= \frac{1}{8 M^2} \Big[2 M r^{(2)}  - 4 M^2 h^{(2)}_{vv}+\Omega^{AB}h^{(1)}_{vA}h^{(1)}_{vB}\nonumber\\ &\qquad  + 4 M^2 h^{(1)}_{vv} h^{(1)}_{vr}  - 4 M^2 \partial_{r} h^{(1)}_{vv} r^{(1)} \nonumber\\ 
	&\qquad- 2 M h^{(1)}_{vr} r^{(1)} -  (r^{(1)})^2  -  h^{(1)}_{vA} D^Ar^{(1)} \nonumber\\ 
	&\qquad + h^{(1)}_{vA} D^Ar^{(1)} -  D_Ar^{(1)}D^Ar^{(1)} \Big] , \label{k2r}\\
	\breve{k}_{(2)}^A &= \frac{1}{16 M^4} \Big(\Omega^{AC}\Omega^{BD}h^{(1)}_{BC} h^{(1)}_{vD} + \Omega^{AB}h^{(1)}_{rB}\breve{k}^r_{(1)}\nonumber\\ 
	&\qquad - 4 M^2\Omega^{AB} h^{(2)}_{vB} - 4 M^2\Omega^{AB} \partial_{r} h^{(1)}_{vB} r^{(1)}\nonumber\\ 
	&\qquad + 4 M\Omega^{AB} h^{(1)}_{vB} r^{(1)}   - 4 M^2 h^{(1)}_{vr} D^Ar^{(1)} \nonumber\\ 
	&\qquad+ 4 M r^{(1)}D^A r^{(1)} - 4 M^2D^A r^{(2)} \nonumber\\ 
	&\qquad+\Omega^{AC} h^{(1)}_{BC}D^B r^{(1)} \Big) . \label{k2A}
\end{align}
\end{subequations}
In our analysis we will not require the explicit expressions for $\breve{\n}^\alpha_{(n)}$ beyond $n=0$, but we use them as a consistency check in some of our calculations. We include them here for completeness:
\begin{subequations}\label{n2}
\begin{align}
\breve{\n}_{(2)}^v &= \frac{1}{8M^2}\Big(4 M^2 h^{(2)}_{rr} - h^{(1)}_{rA} h^{(1)}_{rB}\Omega^{AB} - 8 M^2 h^{(1)}_{vr} h^{(1)}_{rr} \nonumber\\
&\qquad\qquad + 4 M^2 \partial_{r}h^{(1)}_{rr} r^{(1)}\Big),\\
\breve{\n}_{(2)}^r &= \frac{1}{8M^2}\Big(8 M^2 h^{(2)}_{vr}-2 h^{(1)}_{vA} h^{(1)}_{rB}\Omega^{AB} - 6 M^2 h^{(1)}_{vv} h^{(1)}_{rr} \nonumber\\
&\qquad\qquad + 8 M^2 \partial_{r}h^{(1)}_{vr} r^{(1)} + 3 M
h^{(1)}_{rr} r^{(1)} \nonumber\\
&\qquad\qquad - 8 M^2 h^{(1)}_{vr}h^{(1)}_{vr}\Big),\\
\breve{\n}_{(2)}^A &= \frac{\Omega^{AB}}{16M^4}\Big(4 M^2 h^{(2)}_{rB}-4 M^2 h^{(1)}_{vr} h^{(1)}_{rB} -  h^{(1)}_{BC}h^{(1)}_{rD}\Omega^{CD} \nonumber\\
&\qquad\qquad- 2 M^2 h^{(1)}_{vB} h^{(1)}_{rr}  + 4 M^2 \partial_{r}h^{(1)}_{rB} r^{(1)} \nonumber\\
&\qquad\qquad- 4 M h^{(1)}_{rB} r^{(1)} + 2 M^2 h^{(1)}_{rr}D_{B}r^{(1)}\Big).
\end{align}
\end{subequations}

\subsection{Surface area and mass}

The surface area of the slice ${\cal H}_v$ is given by the integral 
\begin{equation}\label{area}
	A= \int_{{\cal H}_v}\!\! \sqrt{ \gamma }\, d^2\theta,
\end{equation}
where $\gamma$ is the determinant of the 2-metric $\gamma_{AB}$. By substituting the expansion~\eqref{intrinsic_metric}, we can write the surface element as an expansion 
\beq
\sqrt{\gamma}\,d^2\theta  = 4 M^2 \left[1 + \epsilon {\sqrt{\gamma}}^{(1)} + \epsilon^2  {\sqrt{\gamma}}^{(2)}+O(\eps^3)\right]d\Omega,\label{dS expansion}
\eeq
where the subleading terms are
\begin{align}
{\sqrt{\gamma}}^{(1)} &= \frac{\breve\gamma^{(1)}_\circ}{4M^2}\label{integrand_first},\\
{\sqrt{\gamma}}^{(2)} &= \frac{\breve\gamma^{(2)}_\circ}{4M^2}-\frac{\Omega^{AC}\Omega^{BD}\breve\gamma^{(1)}_{\langle AB\rangle}\breve\gamma^{(1)}_{\langle CD\rangle}}{64M^4}.\label{integrand_second}
\end{align}
We omit breves on the quantities ${\sqrt{\gamma}}^{(n)}$ for notational simplicity.

To evaluate the integral, we appeal to the harmonic expansions~\eqref{gamma harmonic expansions}, orthogonality 
relations \eqref{orthogonality-first}-\eqref{orthogonality-last}, and the identity $\int Y^{00}d\Omega = \sqrt{4\pi}$. The result is
\beq\label{area_second_order}
A = 16\pi M^2 + \eps A^{(1)} + \eps^2 A^{(2)} + O(\eps^3), 
\eeq
where
\begin{align}
A^{(1)} &= \sqrt{4\pi} \breve\gamma^{(100)}_{\circ},\label{A1}\\
A^{(2)}	&=\sqrt{4\pi}\breve\gamma^{(200)}_{\circ} \nonumber\\
&\quad - \sum_{lm}\frac{\lambda_2^2}{32M^2}\left(|\breve\gamma^{(1lm)}_{+}|^2+ |\breve\gamma^{(1lm)}_{-}|^2\right).\label{A2}
\end{align}

The area of the horizon provides a measure of the black hole's \textit{irreducible mass}, 
\begin{equation}\label{irreducible_mass}
	M_{\textup{irr}} := \sqrt{\frac{A}{16 \pi}},
\end{equation}
sometimes called the Christodoulou mass. 
As its name suggests, the irreducible mass cannot be lowered by any (classical) physical process. Historically, this definition arose from the case of a Kerr black hole~\cite{Christodoulou:70}, from which some amount of energy can be extracted via the Penrose process~\cite{Penrose_1971}. After substitution of the expansion~\eqref{area_second_order}, the irreducible mass reads
\begin{equation}\label{irreducible_mass_second}
	M_{\text{irr}} = M \bigg( 1+ \epsilon \frac{A^{(1)}}{2 A^{(0)}}+ \epsilon^2 \frac{4 A^{(0)} A^{(2)} - (A^{(1)})^2 }{8 (A^{(0)})^2} \bigg),
\end{equation}
where $A^{(0)}=16\pi M^2$.

The irreducible mass is closely related to another quasilocal measure of mass: the Hawking mass~\cite{Hayward:1993ph,Jaramillo:2010ay},
\beq\label{Hawking mass}
M_{\rm H} := M_{\rm irr}\left(1+\frac{1}{8\pi}\int_{{\cal H}_v} \vartheta_+ \vartheta_- \sqrt{\gamma}d^2\theta\right).
\eeq
Here $\vartheta_- := \gamma^{AB}e^\alpha_A e^\beta_B \nabla_\alpha \n_\beta$ and $\vartheta_+ := \gamma^{AB}e^\alpha_A e^\beta_B \nabla_\alpha \k_\beta$ are the expansion scalars associated with $\n^\alpha$ and $\k^\alpha$, respectively. Unlike irreducible mass, which simply measures the area of a surface, the Hawking mass directly involves the gravitational pull at the surface, as characterized by the expansion or contraction of the two null congruences.  In our case, $\vartheta_-$ will always be negative, while $\theta_+$ will be either zero or positive. An apparent horizon is defined by $\theta_+=0$, meaning $M_{\rm H}=M_{\rm irr}$ for an apparent horizon. But for the event horizon $\theta_+>0$, meaning $M_{\rm H}$ provides a simple alternative measure of the mass within the event horizon.

We return to these quantities in later sections.

\subsection{Intrinsic curvature}\label{sec:scalar curvature}

Since ${\cal H}_v$ is a 2-surface, its intrinsic curvature tensor can be written in terms of its Ricci scalar as
\beq
{\cal R}_{ABCD} = \frac{1}{2}{\cal R}(\gamma_{AC}\gamma_{BD}-\gamma_{AD}\gamma_{BC}),
\eeq
where we use a calligraphic ${\cal R}$ to avoid confusion with the curvature tensor of the unit two-sphere. In this section, we derive a perturbative formula for ${\cal R}$ through second order in $\eps$,
\beq
{\cal R} = \frac{1}{2M^2}+\eps \breve{\cal R}^{(1)}(v,\theta^A)+\eps^2 \breve{\cal R}^{(2)}(v,\theta^A) + O(\eps^3).\label{Ricci expansion}
\eeq

To carry out the expansion, we consider the metric $\gamma_{AB} = \breve{\gamma}^{(0)}_{AB} + \breve{\gamma}_{AB}$ with $\breve{\gamma}^{(0)}_{AB}:=4M^2\Omega_{AB}$. The  scalar curvature of this metric can be expanded in powers of $\breve{\gamma}_{AB}$ as
\beq
{\cal R}[\gamma] = {\cal R}[\breve{\gamma}^{(0)}] + \delta {\cal R}[\breve\gamma] + \delta^2{\cal R}[\breve\gamma] + O[(\breve\gamma)^3],
\eeq
following the notation in Appendix~\ref{sec:curvature tensors}. Explicitly, Eqs.~\eqref{dR} and \eqref{ddR} (with ${\cal R}_{AB}[\breve\gamma^{(0)}] =\Omega_{AB}$) reduce to
\begin{align}
\d{\cal R}[\breve\c] &= 
\frac{1}{16M^4}[D^AD^B \breve{\c}_{\langle AB\rangle}-2\breve{\c}_\circ-D^2\breve{\c}_\circ]\\
\d^2{\cal R}[\breve\c] &=
\frac{1}{64M^6}\Big[\tfrac{3}{4}D^E\breve{\c}_{CD}D_E\breve{\c}_{AB} -\tfrac{1}{2}D^E\breve{\c}_{CD}D_A\breve{\c}_{EB}\nonumber\\
&\quad - D_B\breve{\c}_{CD}^* D^E \breve{\c}^*_{  AE  } - \breve{\c}_{CD} \big( D_BD^E\breve{\c}_{AE} \nonumber \\
&\quad + D^ED_B\breve{\c}_{AE} - 2D_AD_B\breve{\c}_\circ - D^2\breve{\c}_{  AB }\big) \nonumber \\
&\quad +\breve{\c}_{AB}\breve{\c}_{CD}\Big]\Omega^{AC}\Omega^{BD},
\end{align}
where $\breve{\c}^*_{AB}=\breve{\c}_{AB} - 2\Omega_{AB}\breve\gamma_\circ$. 
Letting $\breve\gamma_{AB} = \eps\breve\gamma^{(1)}_{AB}+\eps^2\breve\gamma^{(2)}_{AB}+O(\eps^3)$, we then have
\begin{align}
\breve{\cal R}^{(1)} &= \delta{\cal R}[\breve\gamma^{(1)}],\\
\breve{\cal R}^{(2)} &= \delta{\cal R}[\breve\gamma^{(2)}] + \delta^2{\cal R}[\breve\gamma^{(1)}].    
\end{align}

Decomposing these quantities into harmonics, we find at first order
\begin{align}\label{R1lm}
{\cal R}^{(1)}_{lm} 
&= \frac{1}{16M^4}\left[\mu^2\breve\c^{(1lm)}_\circ + \frac{1}{2}\lambda_2^2\breve\c^{(1lm)}_+\right],
\end{align}
where $\mu^2:=(l+2)(l-1)$. This agrees with the result from Vega et al.~\cite{Vega_2011}. Decomposing the quadratic quantity $\delta^2{\cal R}[\breve\gamma^{(1)}]$ requires decompositions of products of angular functions into scalar harmonics. We perform that decomposition following the method outlined in Appendix~\ref{sec:coupling}, eventually arriving at
\begin{align}
{\cal R}^{(2)}_{lm} &= \frac{1}{16M^4}\left[\mu^2\breve\c^{(2lm)}_\circ + \frac{1}{2}(\lambda_2)^2\breve\c^{(2lm)}_+\right]\nonumber\\
&\quad+\sum_{\substack{l'm'\\l''m''}}\left[{}^{(3)}{\cal R}^{lm}_{l'm'l''m''}C^{lm0}_{l'm'3l''m''-3}\right.\nonumber\\
&\qquad\qquad
+{}^{(2)}{\cal R}^{lm}_{l'm'l''m''}C^{lm0}_{l'm'2l''m''-2}\nonumber\\
&\qquad\qquad
+{}^{(1)}{\cal R}^{lm}_{l'm'l''m''}C^{lm0}_{l'm'1l''m''-1}\nonumber\\
&\qquad\qquad
\left.+{}^{(0)}{\cal R}^{lm}_{l'm'l''m''}C^{lm0}_{l'm'0l''m''0},
\right],\label{R2lm}
\end{align}
where the $C$ symbols are given by Eq.~\eqref{coupling} and the functions ${}^{(s)}{\cal R}^{lm}_{l'm'l''m''}(v)$ are given in Eq.~\eqref{R2lm coefficients}.

\subsection{Gauss-Bonnet theorem}

For any closed two-dimensional Riemannian surface  $\mathcal{S}$, the Gauss-Bonnet theorem states that the surface's total curvature is related to its Euler characteristic $\chi({\cal S})$ according to
\beq
\int_{\cal S}{\cal R} \, dS = 4\pi\chi(\mathcal{S}),
\eeq
where $dS$ is the area element on $\mathcal{S}$. Applied to our case, where the surface ${\cal H}_v$ has the topology of a 2-sphere, the equality becomes
\beq\label{Gauss-Bonnet}
\int_{{\cal H}_v}\!{\cal R}\, \sqrt{\gamma}d^2\theta = 8\pi.
\eeq
In this section, we use this identity as a consistency check of our results for the surface area and intrinsic curvature.

Substituting the expansions~\eqref{dS expansion} and \eqref{Ricci expansion} into the left-hand side of the identity, we obtain
\begin{align}
\int_{{\cal H}_v}\!{\cal R}\, \sqrt{\gamma}d^2\theta &= 8\pi +4M^2\eps\left(\int\breve{\cal R}^{(1)}d\Omega +\frac{A^{(1)}}{8M^4}\right)\nonumber\\
&\quad +4M^2\eps^2\left[\int\left(\breve{\cal R}^{(2)}+{\cal R}^{(1)}\sqrt{\gamma}^{(1)}\right)d\Omega\right.\nonumber\\
&\qquad\qquad\qquad\left.+\frac{A^{(2)}}{8M^4}\right]+O(\eps^3).
\end{align}
Here we have used $4M^2\int\sqrt{\gamma}^{(n)}d\Omega=A^{(n)}$. Equating this expansion to the right-hand side of Eq.~\eqref{Gauss-Bonnet} yields an equation at each order in $\eps$,
\begin{align}
    -\frac{A^{(1)}}{8M^4} &= \int\breve{\cal R}^{(1)}d\Omega,\label{Gauss-Bonnet 1}\\
    -\frac{A^{(2)}}{8M^4} &= \int\left(\breve{\cal R}^{(2)}+\breve{\cal R}^{(1)}\sqrt{\gamma}^{(1)}\right)d\Omega.\label{Gauss-Bonnet 2}
\end{align}

The right-hand side of Eq.~\eqref{Gauss-Bonnet 1} evaluates to
\begin{subequations}
\begin{align}
 \int\breve{\cal R}^{(1)}d\Omega &= \sqrt{4\pi}\breve{\cal R}^{(1)}_{00} \\
&= -\frac{\sqrt{4\pi}\breve{\c}^{(100)}_\circ}{8M^4},
\end{align}
\end{subequations}
where we have appealed to Eq.~\eqref{R1lm}. Comparing this result to Eq.~\eqref{A1} for $A^{(1)}$, we see that Eq.~\eqref{Gauss-Bonnet 1} is satisfied.

Next, the right-hand side of Eq.~\eqref{Gauss-Bonnet 2} evaluates to
\begin{subequations}\label{Gauss-Bonnet 2 RHS}
\begin{align}
 \int\big(\breve{\cal R}^{(2)}+&\breve{\cal R}^{(1)}\sqrt{\gamma}^{(1)}\big)d\Omega \nonumber\\
 &=\sqrt{4\pi}\breve{\cal R}^{(2)}_{00}+\sum_{lm}\bar{\cal R}^{(1)}_{lm}\sqrt{\gamma}^{(1)}_{lm}\\
 &= -\frac{\sqrt{4\pi}\breve{\c}^{(200)}_\circ}{8M^4}+\sqrt{4\pi}\d^2{\cal R}_{00}[\breve\gamma^{(1)}]\nonumber\\
 &\quad +\frac{1}{128M^6}\sum_{lm}\left(2\mu^2|\breve{\gamma}^{(1lm)}_\circ|^2\right.\nonumber\\
 &\qquad\qquad\qquad \left.+\lambda_2^2 \bar\gamma^{(1lm)}_+\breve{\gamma}^{(1lm)}_\circ\right).
\end{align}
\end{subequations}
To avoid stacking bars on top of breves, here we let an overbar denote the complex conjugate of a quantity that otherwise would carry a breve. The monopole mode of the quadratic term in Eq.~\eqref{R2lm} can be simplified to 
\begin{align}
\d^2{\cal R}_{00}[\breve\gamma^{(1)}] &=\frac{1}{\sqrt{4\pi}256M^6}\sum_{lm}\bigg[-4\mu^2 |\breve\c^{(1lm)}_o|^2 \nonumber\\
&\quad+ \lambda_2^2\left(|\breve\c^{(1lm)}_-|^2+|\breve\c^{(1lm)}_+|^2\right) \nonumber\\
&\qquad\qquad- 2\lambda_2^2\bar\c^{(1lm) }_+\breve\c^{(1lm)}_\circ \bigg].
\end{align}
Substituting this into Eq.~\eqref{Gauss-Bonnet 2 RHS} and comparing the result to Eq.~\eqref{A2} for $A^{(2)}$, we find that Eq.~\eqref{Gauss-Bonnet 2} is satisfied.


\section{Apparent horizon}\label{sec:apparent horizon}

We now consider the apparent horizon. In this section, we obtain a perturbative description of the horizon's radial profile in terms of the metric perturbation. The horizon's area, mass, and curvature are then given by the generic formulas of the previous section. Our notation and conventions largely follow the textbook of Poisson~\cite{PoissonBook}.

\subsection{Specification of the horizon}

We first foliate the spacetime with surfaces $\Sigma_v$ of constant $v$. On each $\Sigma_v$, the apparent horizon ${\cal A}_v$ is a closed spatial 2-surface; this now plays the role of ${\cal H}_v$ from our generic treatment.\footnote{An apparent horizon is more commonly described as a 2-surface embedded in a spatial hypersurface. One can always find some foliation into spacelike 3-surfaces $\Sigma_{\tau}$ for some time function $\tau$ such that the apparent horizon ${\cal A}_\tau$ in $\Sigma_\tau$ is identical to ${\cal A}_v$. The construction in this section is indifferent to which of these submanifolds the apparent horizon is embedded into, but the foliation into surfaces labelled by $v$ is most natural in our perturbative context.} On this surface, we have the two future-directed null vector fields, $\k^\alpha$ and $\n^\alpha$, that are orthogonal to ${\cal A}_v$. To illuminate the definition of the apparent horizon, we extend $\k^\alpha$ and $\n^\alpha$ off of ${\cal A}_v$ by taking them to be the tangent vector fields to congruences of null curves, ${\cal C}_\k$ and ${\cal C}_\n$, respectively. The choice of these congruences is arbitrary; the curves making up ${\cal C}_\k$ (${\cal C}_\n$) may be accelerating and nonaffinely parametrized, for example, so long as they are tangent to $\k^\alpha$ ($\n^\alpha$) at ${\cal A}_v$. The congruences' kinematics at ${\cal A}_v$ are described by the 2-tensors
\begin{align}
B^{+}_{AB} &:= \e^\alpha_A \e^\beta_B\nabla_\alpha\k_\beta,\label{B+}\\
B^{-}_{AB} &:= \e^\alpha_A \e^\beta_B\nabla_\alpha\n_\beta,
\end{align}
and their expansion scalars are
\begin{align}
\vartheta_{+} := \gamma^{AB}B^{+}_{AB},\label{k expansion}\\
\vartheta_{-} := \gamma^{AB}B^{-}_{AB}.
\end{align}

A spatial 2-surface in $\Sigma_v$ is a \textit{trapped surface} if $\vartheta_{-} <0$ and $\vartheta_{+}<0$, and it is \textit{marginally trapped} if $\vartheta_{-} <0$ and $\vartheta_{+} = 0$. The apparent horizon ${\cal A}_v$ is the outermost marginally trapped surface in $\Sigma_v$. The collection of apparent horizons forms a 3-surface ${\cal A}:=\cup_v {\cal A}_v$, which we also refer to as the apparent horizon. This plays the role of ${\cal H}$ from our generic treatment. ${\cal A}$ is spacelike in dynamical regions of spacetime; it is then a {\em dynamical horizon} in the sense of Ashtekar~\cite{Ashtekar:2004cn}. It is null in stationary regions of spacetime; it is then an {\em isolated horizon}~\cite{Ashtekar:2004cn}.

$\vartheta_-$ will always be negative when $\vartheta_+$ vanishes, allowing us to calculate only $\vartheta_+$. The equation determining the apparent horizon's location is then 
\begin{equation}\label{expansion_2}
	\vartheta_+ = 0.
\end{equation}
One might imagine there being multiple solutions to this equation, requiring us to find the outermost one. (${\cal A}$ would represent a dynamical or isolated horizon in any case, but not an apparent horizon.) However, in our context of a perturbed Schwarzschild black hole, we will find that Eq.~ \eqref{expansion_2} specifies a unique surface near the background horizon.

Before calculating the expansion, we note that $\vartheta_{+}$ and $\vartheta_{-}$ only involve derivatives of $\k_\alpha$ and $\n_\alpha$ {\em within} ${\cal A}_v$. Therefore although it can be helpful to think of the expansion scalars in terms of congruences of curves, they are actually fully specified by the basis vectors $\k^\alpha$ and $\n^\alpha$ on ${\cal A}_v$. They do not depend on the complete congruences, nor do they depend on the vectors being tangent to geodesics. In the literature one more commonly sees the simplified form $\vartheta_+=\nabla_\alpha\k^\alpha$, which holds for an affinely parametrized congruence. Generically, for accelerating, nonaffinely parameterized curves, the curves in ${\cal C}_\k$ satisfy 
\beq
\k^\beta\nabla_\beta\k^\alpha = \kappa \k^\alpha + \kappa^A e^\alpha_A
\eeq
on ${\cal A}_v$, for some $\kappa$ and $\kappa^A$. [There is no component along $\n^\alpha$ because the contraction with $\k_\alpha$ must vanish: $\k_\alpha\k^\beta\nabla_\beta\k^\alpha=\frac{1}{2}\k^\beta\nabla_\beta(\k_\alpha\k^\alpha)=0$.] Using Eq.~\eqref{ginverse in basis}, we can therefore write the expansion as
\beq
\vartheta_+ = \nabla_{\alpha}\k^\alpha - \kappa.
\eeq
If the curves are affinely parametrized, regardless of whether they are geodesics or accelerated, this simplifies to the standard expression $\vartheta_+=\nabla_\alpha\k^\alpha$. However, because these expressions require an extension off of ${\cal A}_v$, we instead exclusively use Eq.~\eqref{k expansion}.

\subsection{Radial profile of the horizon}

In this section, we solve Eq.~\eqref{expansion_2} to find the  horizon's radial profile. 

We first write Eq.~\eqref{k expansion} in terms of background quantities and perturbative quantities. The tensor $B^+_{AB}$ defined in Eq.~\eqref{B+} can be written as
\beq\label{B+ perturbative}
B^{+}_{AB} = e^\alpha_A e^\beta_B\Big(g^{(0)}_{\beta\gamma}+h_{\beta\gamma}\Big)\Big({}^{(0)}\nabla_\alpha k^\gamma + C^\gamma_{\alpha\beta}[h]\k^\beta\Big).
\eeq
Here $C^\gamma_{\alpha\beta}$ is given by Eq.~\eqref{C tensor}. In evaluating Eq.~\eqref{B+ perturbative}, we first take derivatives at the coordinate location of the perturbed horizon, $x^\mu_\H$, using expansions of the form~\eqref{k expansion on H} for the null vector and applying radial derivatives as $\partial_r=\partial_{r_\H}$. After that, we carry out expansions of the form~\eqref{expansions around H0}, using Eqs.~\eqref{e expansion around H0} and \eqref{k expansion around H0}. Finally, when evaluating the contraction $\gamma^{AB}B^+_{AB}$, we use Eq.~\eqref{gamma inverse}. 

As a check of our calculation, we have also performed the operations in an alternative order, leaving $B^{+}_{AB}$ and $\gamma^{AB}$ at $r_\H$, using the form Eq.~\eqref{gamma expansion on H} for $\gamma^{AB}$, and then performing the expansion~\eqref{expansions around H0} at the level of the scalar quantity $\vartheta_+$. As an additional check, we have also used the alternative form $\vartheta_+ = (g^{\alpha\beta}+\n^\alpha \k^\beta + \k^\alpha \n^\beta)\nabla_\alpha\k_\beta$ to obtain the same result.

In all variations of the calculation, we arrive at
\beq\label{theta expansion}
\vartheta_+ = \eps \vartheta^{(1)}_+(v,\theta^A) + \eps^2 \vartheta^{(2)}_+(v,\theta^A) + O(\eps^3).
\eeq
The zeroth-order term identically vanishes because the background horizon has zero expansion in the background spacetime. The first-order term is
\begin{multline}
	\vartheta^{(1)}_+ = - \frac{1}{4 M^2} \Big[2 M h^{(1)}_{vv} - \partial_v h^{(1)}_{\circ} +D^Ah^{(1)}_{vA}\\   + (D^2-1) r^{(1)} \Big] .
\end{multline}
The second-order term is given in Eq.~\eqref{theta2}.

To solve the equations $\vartheta^{(n)}_+=0$, we expand all quantities in tensorial spherical harmonics using Eqs.~\eqref{scalar_metric_harm}--\eqref{metric_harm} and \eqref{r harmonic expansion} and then decompose $\vartheta^{(n)}_{+}$ into scalar-harmonic modes $\vartheta^{(n)}_{+,lm}$. After appealing to the identity~\eqref{eigenvalue} and the definitions \eqref{YA} and \eqref{XA}, we immediately find 
\begin{multline}
	\vartheta^{(1)}_{+,lm} = - \frac{1}{4 M^2} \Big[2 M h^{(1lm)}_{vv} - \partial_v h^{(1lm)}_{\circ} -\lambda_1^2 h^{(1lm)}_{v+}\\   - (1+\lambda_1^2) r^{(1)}_{lm} \Big].\label{expansion1lm}
\end{multline}
The solution to $\vartheta^{(1)}_{+,lm}=0$ is therefore
\begin{equation}\label{first_rad_ah}
	r^{(1)}_{lm}(v)= \frac{2Mh^{(1lm)}_{vv} -  \lambda_{1}^2 h_{v+}^{(1lm)} -  \partial_v h^{(1lm)}_\circ}{1 + \lambda_{1}^2}. 
\end{equation}
All quantities on the right are evaluated at $(v,r=2M)$.

Finding $\vartheta^{(2)}_{+,lm}$ requires decomposing products of tensorial harmonics into $Y_{lm}$ modes. As in our decompositions in Sec.~\ref{sec:geometry}, our method of decomposing such products is described in Appendix~\ref{sec:coupling}. Our calculation in this section in particular utilizes the identities~\eqref{first_C_int}--\eqref{last_C_int}. The result is an equation of the form
\begin{align}
	\vartheta^{(2)}_{+,lm} &= - \frac{1}{4 M^2} \Bigg\{2 M h^{(2lm)}_{vv} - \partial_v h^{(2lm)}_{\circ} -\lambda_1^2 h^{(2lm)}_{v+}\nonumber\\  &\qquad\quad +\sum_{\substack{l'm'\\l''m''}}\Big[ {}^{(2)}\Theta_{l'm'l''m''}^{lm}(v) C_{l' m' 2 l'' m'' - 2}^{lm0} \nonumber\\  &\qquad\qquad\quad + {}^{(1)}\Theta_{l'm'l''m''}^{lm}(v) C_{l' m' 1 l'' m'' -1}^{lm0}  \nonumber\\  &\qquad\qquad\quad + {}^{(0)}\Theta_{l'm'l''m''}^{lm}(v) C_{l' m' 0 l'' m'' 0}^{lm0}\Big] \nonumber\\  &\qquad\quad- (1+\lambda_1^2) r^{(2)}_{lm}\Bigg\},\label{theta2lm}
\end{align}
where the $C$ symbols are defined in Eq.~\eqref{coupling} and the functions ${}^{(s')}\Theta_{l'm'l''m''}^{lm}$ are given in Eq.~\eqref{theta2lm coefficients}.
The solution to $\vartheta^{(2)}_{+,lm}=0$ is therefore
\begin{align}
	r^{(2)}_{lm}(v)  &= \frac{2Mh^{(2lm)}_{vv} -  \lambda_{1}^2 h_{v+}^{(2lm)} -  \partial_v h^{(2lm)}_\circ}{1 + \lambda_{1}^2} \nonumber \\
	&\quad + \frac{1}{1 + \lambda_{1}^2}\sum_{\substack{l'm'\\l''m''}}\Big[ {}^{(2)}\Theta_{l'm'l''m''}^{lm}(v) C_{l' m' 2 l'' m'' - 2}^{lm0} \nonumber\\  &\qquad\qquad + {}^{(1)}\Theta_{l'm'l''m''}^{lm}(v) C_{l' m' 1 l'' m'' -1}^{lm0}  \nonumber\\  &\qquad\qquad + {}^{(0)}\Theta_{l'm'l''m''}^{lm}(v) C_{l' m' 0 l'' m'' 0}^{lm0}\Big]. \label{r2lm AH}
\end{align}

\subsection{Surface area, mass, and intrinsic curvature}

Given the perturbations~\eqref{first_rad_ah} and \eqref{r2lm AH} to the horizon's radial profile, we can compute the intrinsic metric~\eqref{intrinsic_metric} using \eqref{gamma harmonic expansions} with Eqs.~\eqref{gamma1 harmonic expansions} and \eqref{gamma2 harmonic expansions}. From the intrinsic metric we can then compute the surface area, mass, and intrinsic curvature using Eq.~\eqref{area_second_order}, Eq.~\eqref{irreducible_mass_second}, and Eq.~\eqref{Ricci expansion} [with Eqs.~\eqref{R1lm} and \eqref{R2lm}]. Because $\vartheta_+ = 0$ on the apparent horizon, the horizon's Hawking mass~\eqref{Hawking mass} is identical to its irreducible mass.

The surface area and the mass both require the monopole mode of $r^{(n)}$. For $l=0$, Eqs.~\eqref{first_rad_ah} and \eqref{r2lm AH} reduce to
\begin{align}
	r^{(1)}_{00}  &= 2Mh^{(100)}_{vv} -  \partial_v h^{(100)}_\circ,\label{r100 AH}\\
	r^{(2)}_{00}  &= 2Mh^{(200)}_{vv} -  \partial_v h^{(200)}_\circ \nonumber \\
	&\quad + \sum_{\substack{lm}}\frac{(-1)^m}{\sqrt{4\pi}}\Big( {}^{(2)}\Theta_{lml,-m}^{00} \nonumber\\  &\qquad\qquad - {}^{(1)}\Theta_{lml,-m}^{00}  + {}^{(0)}\Theta_{lml,-m}^{00}\Big). \label{r200 AH}
\end{align}
The explicit expression for $r^{(2)}_{00}$ is given in Eq.~\eqref{r200 AH explicit}. 

The surface area and mass at first order are easily evaluated. Note that an $l=0$ vacuum perturbation is necessarily a perturbation toward another Schwarzschild solution, which (with an appropriate choice of gauge) we can write as
\beq
\delta M\frac{\partial g^{(0)}_{\alpha\beta}}{\partial M}dx^\alpha dx^\beta = \frac{2\delta M}{r}dv^2,\label{EF mass perturbation}
\eeq
for some $\delta M$. Therefore the $l=0$ correction to the horizon radius, as given in Eq.~\eqref{r100 AH}, is
\beq\label{r100 AH dM}
r^{(1)}_{00} = \sqrt{16\pi}\delta M.
\eeq
With Eqs.~\eqref{A1} and \eqref{gamma1 harmonic expansions}, this implies that the correction to the surface area is
\beq\label{A1 AH}
A^{(1)} = 32\pi M\delta M,
\eeq
and the correction to the black hole mass, given in terms of the surface area in Eq.~\eqref{irreducible_mass_second}, is
\beq\label{M1 AH}
M^{(1)}_{\rm irr} = M^{(1)}_{\rm H} = \delta M
\eeq
(noting again that the Hawking and irreducible mass are necessarily identical for the apparent horizon). $\delta M$ can also be invariantly defined as the Abbott-Deser mass contained in ${\cal A}_v$~\cite{Abbott:1982jh,Dolan:2012jg}; for linear perturbations of Schwarzschild spacetime, all sensible definitions of mass agree.

In the context of a binary inspiral, where the two-timescale expansion~\eqref{metric_expansion_two-timescale} applies, all the same results hold true except that (i) the black hole's gradual absorption of energy requires that $\delta M$ becomes a function of $\tilde v$~\cite{miller2020twotimescale}, and (ii) $r^{(2)}_{lm}$ picks up an additional slow-time derivative term,
\beq\label{delta r2lm AH}
\delta r^{(2)}_{lm} = -\frac{\partial_{\tilde v}h^{(1lm)}_\circ}{1+\lambda_1^2},
\eeq
coming from the chain rule~\eqref{two_derivative} applied to the $v$ derivative in Eq.~\eqref{first_rad_ah}. After this correction is accounted for, all $v$ derivatives are then to be interpreted as the fast-time derivative $\Omega_i\frac{\partial}{\partial\varphi_i}$. The correction~\eqref{delta r2lm AH} plays an important role in our comparison of the apparent and event horizons in later sections.

We defer any evaluation of the scalar curvature or the second-order expressions to Sec.~\ref{sec:horizon locking}, where they will be substantially simplified.


\section{Event horizon}\label{sec:event horizon}

In this section, we obtain a perturbative description of the event  horizon's radial profile in terms of the metric perturbation, following the method of Refs.~\cite{Poisson:04,Vega_2011}. We then show that in the context of a two-timescale expansion, the event horizon is effectively localized in time. At the end of the section we begin our comparison of the two horizons.

\subsection{Specification of the horizon}

A black hole ${\cal B}$ is intrinsically a nonlocal object. It is formally defined as a region that is causally disconnected from future null infinity $\mathscr{I}^+$:
\begin{equation}\label{BH}
	{\cal B} = \mathcal{M} - J^-(\mathscr{I}^+),
\end{equation}
where $J^-(\mathscr{I}^+)$ is the causal past of $\mathscr{I}^+$. Its event horizon is the boundary of this region,
\begin{equation}\label{EH}
	{\cal H}^+= \partial {\cal B}.
\end{equation}
We denote the horizon as ${\cal H}^+$ to indicate that it is perturbatively close to the background spacetime's future horizon rather than its past horizon. ${\cal H}^+$ here plays the role of our generic surface ${\cal H}$ from Sec.~\ref{sec:geometry}, and cuts of constant $v$, ${\cal H}_v^+$, play the role of ${\cal H}_v$. The definition \eqref{EH} implies that the location of the event horizon at a given advanced time $v$ depends on the entire future history of the spacetime. 

To locate the horizon in practice, we first note that because it is a null surface, it is necessarily generated by a family of null geodesics. Each of these geodesics can be parametrized with advanced time $v$, such that it has coordinates $x^\alpha_\G(v)$. Since the curve must be within the horizon surface, which we parametrized as $x^\alpha_\H(v,\theta^A)$, we can write $x^\alpha_\G(v)$ as
\beq
x^\alpha_\G(v) = x^\alpha_\H(v,\theta^A_\G(v)). 
\eeq
Its tangent vector $k^\alpha = \frac{dx^\alpha_\G}{dv}$ is then given by
\begin{equation}\label{k_EH}
k^\alpha = \frac{\partial x^\alpha_\H }{\partial v} + \dot\theta^A_\G \e^\alpha_A ,
\end{equation}
where 
\begin{align}
\frac{\partial x^\alpha_\H}{\partial v}\partial_\alpha &= \partial_v + \frac{\partial r_\H}{\partial v}\partial_r, \\
\dot\theta^A_\G e^\alpha_A\partial_\alpha &= \dot\theta^A_\G\frac{\partial r_\H}{\partial \theta^A}\partial_r + \dot\theta^A_\G\partial_A,
\end{align}
and $\dot\theta^A_\G:=d\theta^A_\G/dv$. This implies $k^v=1$, $k^A = \dot\theta^A_\G$ and
\beq
k^r= \frac{\partial r_\H}{\partial v} + \dot\theta^A_\G D_A r_\H. \label{kr EH}
\eeq

Because it is tangent to the null surface's generators, $k^\alpha$ must also be orthogonal to the horizon. This means $k^\alpha$ satisfies all the same orthonormality conditions as in our generic treatment in  Sec.~\ref{sec:geometry}: $k_\alpha k^\alpha=0$, \eqref{basis_1}, and \eqref{basis_3}. Therefore on each cut of the horizon, ${\cal H}^+_v$, $k^\alpha$ is uniquely determined as a function of $r_\H$ and $h_{\alpha\beta}$ exactly as in the generic treatment. So in particular, $k^A$ (or equivalently, $\dot\theta^A_\G$) is given in terms of $k^r$ by Eq.~\eqref{k expansion around H0} with the $A$ component of Eq.~\eqref{k1} and with \eqref{k2A}. However, rather than using Eqs.~\eqref{k1} and \eqref{k2r} for $k^r$, we can instead use the form~\eqref{kr EH}. The null condition $k_\alpha k^\alpha=0$ then becomes a first-order differential equation for the radial profile $r_\H$ as a function of $v$. 

So far, this description only specifies that ${\cal H}^+$ is a null surface. To specify {\em which} null surface it is, we need to impose \textit{teleological} boundary conditions. We assume that in the distant future, the spacetime settles to a stationary, Kerr black hole. In that stationary state, the event horizon has the standard radial profile of the Kerr horizon. Our teleological condition on $r_\H(v,\theta)$ is that when $v\to\infty$, $r_\H$ reduces to the Kerr horizon's radial profile. In other words, the integral curves of $k^\alpha$ must start in the distant future as the generators of the Kerr horizon and then be evolved backward in time from that end state. In Sec.~\ref{sec:localization} we comment on some subtleties in this condition.

\subsection{Radial profile of the horizon}

We obtain evolution equations for $r_\H$ by expanding Eq.~\eqref{kr EH} in the form~\eqref{k expansion around H0}, which implies 
\begin{align}
\breve{k}_{(1)}^r &= \frac{\partial r^{(1)}}{\partial v}, \\
\breve{k}_{(2)}^r &= \frac{\partial r^{(2)}}{\partial v} + k_{(1)}^A  D_A r^{(1)}. \label{last_k_EH}
\end{align}
The other components are described above.

Substituting our expansion of $k^\alpha$ into $(g^{(0)}_{\alpha\beta}+h_{\alpha\beta})k^\alpha k^\beta=0$, we find
\begin{equation}\label{first_rad_eh}
  r^{(1)} - 4M\frac{\partial r^{(1)}}{\partial v} = 2M h^{(1)}_{vv}
\end{equation}
and
\begin{align}\label{second_rad_eh}
	r^{(2)} - 4M \frac{\partial r^{(2)}}{\partial v} &= 2M h^{(2)}_{vv} + \frac{( r^{(1)} )^2}{2 M}  + 4M h^{(1)}_{vr} \frac{\partial r^{(1)}}{\partial v}  \nonumber \\  
	&\quad - \frac{\Omega^{AB}h^{(1)}_{vA} h^{(1)}_{vB}}{2M} - \frac{h^{(1)}_{vA} D^Ar^{(1)}}{M} \nonumber \\
	&\quad -  \frac{D_Ar^{(1)}D^A r^{(1)}}{2 M}  + r^{(1)} \partial_r h^{(1)}_{vv},
\end{align}
where all quantities on the right are evaluated at $(v,2M,\theta^A)$. 

We can write these equations in the common form
\beq
\frac{\partial r^{(n)}}{\partial v}	-\kappa_0 r^{(n)} = -\kappa_0 F^{(n)}(v,\theta^A),
\eeq
where we have introduced $\kappa_0:= 1/(4M)$, the surface gravity of the unperturbed Schwarzschild black hole. Expanding the equations in scalar spherical harmonics, we obtain
\begin{equation}\label{generic_eq_rad}
	\frac{dr^{(n)}_{lm}}{dv} - \kappa_0 r^{(n)}_{lm}(v) =  -\kappa_0 F^{(n)}_{lm}(v).
\end{equation}
The first-order driving term is immediately found to be
\begin{equation}\label{first_rad_lm_eh}
	  F^{(1)}_{lm}(v) = 2M h^{(1lm)}_{vv}(v,2M).
\end{equation}
Calculating the $lm$ modes of the second-order driving term requires decomposing products of vector harmonics into single scalar harmonics; as stated previously, this is described in Appendix~\ref{sec:coupling}. The result is
\begin{align}
	 F^{(2)}_{lm}(v)  &=  2M h^{(2lm)}_{vv}(v,2M) \nonumber \\
	 &\quad +\sum_{\substack{l'm'\\ l'' m''}}  \big[ {}^{(1)}H^{lm}_{l'm'l''m''}(v) C_{l' m' 1 l'' m'' -1}^{lm0} \nonumber \\
	 &\qquad\quad + {}^{(0)}H^{lm}_{l'm'l''m''}(v) C_{l' m' 0 l'' m'' 0}^{lm0}\big] , \label{compact_full_EH_radius}
\end{align}
with 
\begin{align}\label{second_rad_full_lm}
{}^{(1)}H^{lm}_{l'm'l''m''} &= \frac{\lambda_1' \lambda_1''}{ 8 M \sqrt{\pi}}  \left[ \sigma_\otimes\Bigl(h'_{v-} h''_{v-}+h'_{v+}h''_{v+}\Bigr)\right. \nonumber\\
&\quad + 2 \Bigl(\sigma_+  h'_{v+} - i \sigma_- h'_{v-}\Bigr) r^{(1)}_{l'' m''}  \nonumber\\
&\quad \left.+ \sigma_+ r^{(1)}_{l' m'} r^{(1)}_{l'' m''}  \right], \\
{}^{(0)}H^{lm}_{l'm'l''m''} &= \frac{1}{4 M \sqrt{\pi}} \left(2 M h'_{vr} r^{(1)}_{l'' m''}  - 4 M^2 h'_{vv} h''_{vr}\right.\nonumber \\
&\quad \left.+ r^{(1)}_{l' m'} r^{(1)}_{l'' m''} + 4 M^2 r^{(1)}_{l' m'} \partial_r h''_{vv} \right).
\end{align}
Here we use the compact notation described in Appendix~\ref{sec:second_order_expressions}, and we additionally define $\sigma_\otimes:=\sigma_+-i\sigma_-$.

The differential equations~\eqref{generic_eq_rad} have the teleological solutions
\begin{align}\label{teleological}
	r^{(n)}_{lm}(v) &= \kappa_0 \int_{v}^{\infty} e^{-\kappa_0 (v' - v)} F^{(n)}_{lm}(v') dv'.
\end{align}
Under the assumption that $F^{(n)}_{lm}$ can be treated as effectively constant after some late time $T$, the solution~\eqref{teleological} correctly becomes the constant $r^{(n)}_{lm}=F^{(n)}_{lm}(T)$ for all $v>T$. If we adopted a causal solution instead, the solution would grow exponentially as $e^{\kappa_0 v}$ at late times. 

To exactly evaluate the integral~\eqref{teleological} at a given time $v$, we need to know $F^{(n)}_{lm}$ for all $v'>v$. This requires simulating the spacetime's entire evolution, allowing it to settle to a stationary state, and only then finding the location of the horizon at earlier times.

\subsection{Timescales and temporal localization on the horizon(s)}\label{sec:localization}

In Refs.~\cite{Poisson:04,Vega_2011}, Poisson and collaborators show that under certain circumstances, the teleological effects on the horizon's location are strongly suppressed, and the horizon is effectively localized in time. In this section, we review that argument, recall why it does not apply to binaries, and then apply a variant of it to show that in a small-mass-ratio inspiral, the timescales are such that the horizon is always temporally localized except in an interval of advanced time around the final plunge.

The essential idea of the localization is that the exponential factor in Eq.~\eqref{teleological} exponentially suppresses the effects of the distant future. Integrating by parts, as we did to obtain Eq.~\eqref{integral_approximation1}, we can express Eq.~\eqref{teleological} as
\begin{subequations}\label{Poisson localization}
\begin{align}
	r^{(n)}_{lm}(v) &= F^{(n)}_{lm}(v) + \int_{v}^{\infty} e^{-\kappa_0 (v' - v)} \frac{dF^{(n)}_{lm}}{dv'} dv'\\
	&\ \,\,\vdots \nonumber\\
	&= F^{(n)}_{lm}(v) +\frac{1}{\kappa_0}\frac{dF^{(n)}_{lm}}{dv} + \frac{1}{\kappa_0^2}\frac{d^2F^{(n)}_{lm}}{dv^2} +\ldots
\end{align}
\end{subequations}
This is a sensible approximation if the characteristic frequency of $F^{(n)}_{lm}$ is much smaller than $\kappa_0$.

In a binary inspiral, the frequencies $\Omega_{\bm{k}}$ can be much larger than $\kappa_0$, meaning this approximation is inappropriate. However, we can nevertheless develop a similar localization approximation. We do this in two steps: First, we show that the far future has negligible impact on the location of the horizon at times significantly before plunge. Second, we show that during the inspiral phase, the approximations~\eqref{integral_approximation1} and \eqref{Poisson localization} can be combined to localize the horizon.

Generically, a small-mass-ratio binary evolves through four phases~\cite{Taracchini:2014zpa}: the slow inspiral; the transition to plunge when the companion approaches the innermost stable circular orbit (or more generically, when it approaches the separatrix between stable and plunge orbits~\cite{Stein:2019buj}); the plunge itself; and finally, the post-merger ringdown, when the black hole settles to a stationary, Kerr state. Each of these phases has an associated evolution timescale. The inspiral is characterized by the orbital timescale $\sim 1/\Omega_{\bm{k}}$ and the long radiation-reaction time $t_{\rm rr}\sim M/\eps$. The transition to plunge is characterized again by the orbital timescale, but also by the transition timescale $\sim M/\eps^{1/5}$~\cite{Ori_2000}, as the orbit more rapidly evolves during the transition phase. The plunge itself occurs rapidly, on the timescale $\sim M$. Finally, the ringdown phase itself has two types of evolution: the rapid exponential decay of the black hole's quasinormal modes~\cite{Berti:2009kk}, and the late-time tails~\cite{Price:1971fb} that decay with a power law $v^{-2l-3}$ along the horizon~\cite{Barack:1999ya}.

We now consider a time $v$ in the inspiral phase. Our first goal is to show that all the later phases have negligible impact on the location of the horizon at time $v$. In the process, we find an estimate of the cutoff time where our treatment breaks down. 

We first split the integral~\eqref{teleological} into four segments corresponding to the different phases of the system, $\int_v^{\infty} = \int_v^{v_T} + \int_{v_T}^{v_P} + \int_{v_P}^{v_R} + \int_{v_R}^{\infty}$, where the labels refer respectively to transition, $T$, to plunge, $P$, and to ringdown, $R$. We then consider these one by one, starting with the integral over the transition regime, which we write as
\begin{equation}
I_T = \kappa_0 e^{ \kappa_0 (v - v_T)}  \int_{v_T}^{v_P}  e^{- \kappa_0 (v' - v_T)} F^{(n)}_{lm}(v') dv' . 
\end{equation}
To ensure that we can neglect this integral, we require it to be much smaller than $\epsilon$, such that the contribution for $n=1$ is negligible compared to the $O(\eps^2)$ effects that we calculate. 

Although the transition is rapid compared to the inspiral, it is slow compared to the orbital period, and throughout the transition we can adopt an adapted two-timescale form for the forcing function, $F^{(n)}(v) = \sum_{\bm{k}}F^{(n)}_{\bm{k}}(\epsilon^{1/5}v)e^{-i\varphi_{\bm{k}}(v,\eps)}$, where $\frac{d\varphi_{\bm{k}}}{dv}=\Omega_{\bm{k}}(\eps^{1/5}v)$. (We suppress $lm$ indices for the remainder of this discussion; the approximations can be carried out at the level of the sum of $lm$ modes or at the level of individual modes). For each ${\bm{k}}$ mode, the integral then has the form
\beq
I_T^{\bm{k}}=\kappa_0 e^{ \kappa_0 (v - v_T)}  \int_{v_T}^{v_P}  e^{-\psi^T_{\bm{k}}(v',\eps)} F^{(n)}_{\bm{k}}(\eps^{1/5}v') dv', 
\eeq
where $\psi^T_{\bm{k}}(v',\eps):= \kappa_0 (v' - v_T)+i\varphi_{\bm{k}}(v',\eps)$. This function  has the properties
\begin{align}
\psi^T_{\bm{k}}(v_T,\eps) &= i\varphi_{\bm{k}}(v_T,\eps),\\
\frac{d\psi^T_{\bm{k}}}{dv'} &:=\omega^T_{\bm{k}}(\eps^{1/5}v')= \kappa_0+i\Omega_{\bm{k}}(\eps^{1/5}v'). 
\end{align}
Repeatedly integrating by parts, we obtain
\begin{multline}
I_T^{\bm{k}} = 	\frac{\kappa_0 e^{\kappa_0(v-v_T)}}{\omega^T_{\bm{k}}} \left[e^{-\psi^T_{\bm{k}}(v_T,\eps)}F^{(n)}_{\bm{k}}(\eps^{1/5} v_T) \right.\\
\left.- e^{-\psi^T_{\bm{k}}(v_P,\eps)}F^{(n)}_{\bm{k}}(\eps^{1/5} v_P)+ O(\eps^{1/5})\right].
\end{multline}
The quantity in brackets is order 1. Therefore, for $I_T$ to be much smaller than $\eps$, we need $e^{ \kappa_0 (v - v_T)} \ll \epsilon$. This relation translates into a relation for $(v-v_T)$:
\begin{equation}\label{cutoff}
\kappa_0 | v - v_T | \gg | \ln{\epsilon} |.
\end{equation} 
So we conclude that during the inspiral phase, we may neglect the impact of the transition phase so long as we restrict ourselves to advanced times satisfying Eq.~\eqref{cutoff}. Note that this is a far smaller time interval than the radiation-reaction time, implying that the companion can get very near to the transition before the transition's influence on the horizon is felt.

Next, we consider the integral over the plunge phase, which can be written as
\begin{equation}
I_P = \kappa_0e^{ \kappa_0 (v - v_P)} \int_{v_P}^{v_R}e^{- \kappa_0 (v' - v_P)} F^{(n)}_{lm}(v') dv' .
\end{equation}
During the plunge, the timescale on which $F(v)$ varies is $\sim M$, implying the contribution from the integral is of order 1, excluding the exponential factor outside it. Therefore the cutoff~ \eqref{cutoff} ensures that $I_P$ is much smaller than $\eps$. 

The same argument applies to the integral over the ringdown phase,
\begin{equation}
I_R=\kappa_0e^{ \kappa_0 (v - v_R)}\int_{v_R}^{\infty}e^{- \kappa_0 (v' - v_R)} F^{(n)}_{lm}(v') dv'.
\end{equation}
The cutoff~\eqref{cutoff} ensures that $I_R$ is much smaller than $\eps$.

We have now shown that all the future phases have negligible impact on the horizon during the inspiral phase. We have only one remaining integral,
\beq
r^{(n)}_{lm} = \kappa_0\int_v^{v_T}e^{-\kappa_0(v'-v)}F^{(n)}_{lm}(v')dv',
\eeq
which we are able to localize in time using the now-familiar integration by parts. If we expand the forcing function in two-timescale form, as $F^{(n)} = F^{(n)}_{\bm{k}}(\tilde v)e^{-i\varphi_{\bm{k}}(v,\eps)}$, then for each mode $\bm{k}$, we have an integral
\beq
r^{(n)}_{\bm{k}} = \kappa_0 \int_{v}^{v_T} e^{- \psi_{\bm{k}}(v',v,\eps)}  F^{(n)}_{\bm{k}}(\eps v') dv',
\eeq
where we have defined
\beq
\psi_{\bm{k}}(v',v,\eps):=\kappa_0(v'-v)+i\varphi_{\bm{k}}(v',\eps).
\eeq
This function  has the properties
\begin{align}
    \psi_{\bm{k}}(v,v,\eps) &= i \varphi_{\bm{k}}(v,\eps),\\
    \frac{d\psi_{\bm{k}}}{dv'} &:=\omega_{\bm{k}}(\eps v')= \kappa_0+i\Omega_{\bm{k}}(\eps v').\label{omega_k}
\end{align}
If we now repeatedly integrate by parts while appealing to these properties and the cutoff~\eqref{cutoff}, we obtain the approximation
\begin{align}\label{integral_approximation2}
\int_v^{v_T}\!\! F^{(n)}_{\bm{k}}&(\eps v')e^{-\psi_{\bm{k}}(v',v,\eps)}dv' \nonumber\\ &= e^{-i \varphi_{\bm{k}}(v,\eps)}\left[ \frac{ F^{(n)}_{\bm{k}}(\tilde v)}{\omega_{\bm{k}}(\tilde v)}+\frac{\eps}{\omega_{\bm{k}}}\frac{d}{d\tilde v}\frac{ F^{(n)}_{\bm{k}}(\tilde v)}{\omega_{\bm{k}}(\tilde v)} \right]\!\nonumber\\
&\qquad+o(\eps).
\end{align}
As promised, although neither of the approximations~\eqref{integral_approximation1} or \eqref{Poisson localization} is alone accurate, a variant which combines them does provide an accurate approximation that localizes the teleological integral. The oscillations in the two-timescale approximation approximately average out over the integration domain, while the exponential decay eliminates the impact of the very far future. In the case of quasistationary modes, with $\bm{k}=0$, the approximation~\eqref{integral_approximation2} simply reduces to Eq.~\eqref{Poisson localization}, but for these modes $F^{(n)}$ is the slowly varying function $F^{(n)}_{\bm{0}}(\tilde v)$, meaning each derivative in Eq.~\eqref{Poisson localization} comes with a power of $\eps$.

By substituting Eq.~\eqref{first_rad_lm_eh} for $F^{(1)}$, we now obtain the temporally localized perturbations to the event horizon's radius:
\beq
r^{(1)}_{\bm{k}} = \frac{2M\kappa_0 h^{(1,\bm{k})}_{vv}}{\omega_{\bm{k}}},\label{r1k EH}
\eeq
and
\begin{align}
r^{(2)}_{\bm{k}} &= \frac{\kappa_0 F^{(2)}_{\bm{k}}}{\omega_{\bm{k}}}+\frac{2M\kappa_0 \partial_{\tilde v}h^{(1,\bm{k})}_{vv}}{\omega^2_{\bm{k}}}\nonumber\\
&\qquad- \frac{2Mi\kappa_0 h^{(1,\bm{k})}_{vv}}{\omega^3_{\bm{k}}}\frac{d\Omega_{\bm{k}}}{d\tilde v}.\label{r2k EH}
\end{align}
These formulas can be made more explicit using
\beq
\frac{\kappa_0}{\omega_{\bm{k}}} = \frac{1}{1+4iM\Omega_{\bm{k}}}.
\eeq
For quasistationary modes, these results give the slowly evolving {\em average} corrections to the horizon radius:
\begin{align}
\left\langle r^{(1)}\right\rangle &= 2M\left\langle h^{(1)}_{vv}\right\rangle,
\label{<r1> EH}\\
\left\langle r^{(2)}\right\rangle &= \left\langle F^{(2)}\right\rangle +8M^2\frac{d}{d\tilde v}\left\langle h^{(1)}_{vv}\right\rangle, \label{<r2> EH}
\end{align}
where we recall the definition~\eqref{average_definition}  of the average of a two-timescale function.

Before moving to the next section, we comment on how applicable our results are to other scenarios and phases. In this section we have focused on the location of the horizon in the inspiral phase. The same arguments apply to the transition phase, and we expect the event horizon to remain localized for a significant portion of the transition (although, to our knowledge, there has not yet been a complete multiscale treatment of the transition phase, including the metric perturbation in addition to the companion's trajectory). 

In the plunge phase, the horizon is not localizable, and moreover, much of our analysis throughout this paper breaks down. As the companion approaches the black hole, additional generators join the horizon~\cite{Hamerly-Chen:10,Emparan-Martinez:16,Hussain-Booth:17,Emparan-Martinez-Zilhao:17}. The caustics they form create a cusp on the horizon. Although our treatment of individual generators remains valid in that case, the horizon does not have a smooth induced metric, and the generic treatment in Sec.~\ref{sec:geometry} becomes invalid. The additional generators that join the horizon can also begin from a great distance away from it. The cusp extends into the infinite past on the horizon, suggesting one should worry that this spoils our treatment of the inspiral phase as well. However, the cusp is exponentially suppressed in the same way as other effects discussed in this section, and we can safely ignore it.

Finally, in the ringdown phase, our generic treatment of the horizons apply perfectly well. But since quasinormal modes have frequencies and decay rates comparable to $\kappa_0$, there is no temporal localization in the early stage of the ringdown. At late times, if the power-law tails die out on a much larger timescale than $1/\kappa_0$, then Eq.~\eqref{Poisson localization} should apply.

\subsection{Surface area, mass, and intrinsic curvature}

Just as in the case of the apparent horizon, given the perturbations~\eqref{teleological} to the horizon's radial profile, or \eqref{r1k EH} and \eqref{r2k EH} in an inspiral, we can compute the intrinsic metric~\eqref{intrinsic_metric} using \eqref{gamma harmonic expansions} with Eqs.~\eqref{gamma1 harmonic expansions} and \eqref{gamma2 harmonic expansions}. From the intrinsic metric we can then compute the surface area, irreducible and Hawking mass, and intrinsic curvature using Eq.~\eqref{area_second_order}, Eqs.~\eqref{irreducible_mass_second} and \eqref{Hawking mass}, and Eq.~\eqref{Ricci expansion} [with Eqs.~\eqref{R1lm} and \eqref{R2lm}].

The surface area and masses require the $l=0$ term in the horizon radius, $r^{(n)}_{00}$. These are obtained from the $l=0$ forcing functions in Eqs.~\eqref{first_rad_lm_eh} and \eqref{second_rad_full_lm}, which read
\begin{align}
    F^{(1)}_{00} &= 2 M h^{(100)}_{vv},\\
	F^{(2)}_{00} &= 2 M h^{(200)}_{vv}
	+  \frac{1}{4M^2\sqrt{\pi}} \sum_{lm} \Big[ 2Mh^{(1lm)}_{vr} \bar r^{(1)}_{lm}  \nonumber  \\ 
	&\quad - 2\lambda_{1}^2 h_{v+}^{(1lm)} \bar r^{(1)}_{lm} -4M^2 h^{(1lm)}_{vv}\bar h^{(1lm)}_{vr} \nonumber  \\ 
	&\quad - (\lambda_{1}^2-1)|r^{(1)}_{lm}|^2 - \lambda_{1}^2(|h_{v-}^{(1lm)}|^2 +
		|h_{v+}^{(1lm)}|^2) \nonumber  \\ 
	&\quad +4M^2 r^{(1)}_{lm} \partial_r \bar h^{(1lm)}_{vv} \Big].\label{second_rad_00_eh}
\end{align}

Like we did for the apparent horizon, we can quickly obtain the surface area and mass at first order. Using Eq.~\eqref{EF mass perturbation} for $h^{(100)}_{vv}$ in Eq.~\eqref{teleological}, we recover $r^{(1)}_{00} = \sqrt{16\pi}\delta M$; this is identical to the result~\eqref{r100 AH dM} for the apparent horizon. The same can also be obtained from Eq.~\eqref{<r1> EH}, noting that a vacuum monopole perturbation can always be written in a gauge in which it is a pure ${\bm{k}}=0$ mode. Equations~\eqref{A1} and \eqref{irreducible_mass_second} then imply $A^{(1)} = 32\pi M\delta M$ and $M^{(1)}_{\rm irr} = \delta M$, just as for the apparent horizon.

Unlike the apparent horizon, the event horizon's Hawking mass~\eqref{Hawking mass} differs from its irreducible mass. However, the difference only enters at second order. At first order, the expansion $\vartheta_+$ vanishes~\cite{Poisson:04,Vega_2011}; we reproduce this result in the next section. We therefore have $\vartheta_+ = \eps^2\vartheta^{(2)}_+ +O(\eps^3)$. The expansion scalar for the ingoing null vector is easily calculated to be $\vartheta_- = -\frac{1}{M}+O(\eps)$. Therefore the Hawking mass of the event horizon is
\beq\label{M_H EH}
M_{\rm H} = M_{\rm irr} - \frac{\eps^2M^2}{2\pi}\int \vartheta^{(2)}_+ d\Omega +O(\eps^3).
\eeq
We simplify this formula in the next section. There, we make a thorough comparison of the two horizons. To preface that comparison, we note that because $\vartheta_+=O(\eps^2)$, the event horizon is in fact identical to the apparent horizon at first order. As alluded to in the Introduction, they only begin to differ at second order.


\section{Gauge fixing and invariant properties of the horizons}\label{sec:horizon locking}

So far our calculations have not specified a choice of gauge. At first order, it is straightforward to show that quantities such as the black hole mass are invariant under the linear transformation~\eqref{Delta h1}. Moreover, the horizons themselves are invariant 3-surfaces. However, their foliations into surfaces of constant $v$ are inherently gauge dependent. Even given some foliation, a transformation within each 2-surface does not leave all our quantities invariant. As a simple example, consider the horizon's scalar curvature. Under a gauge transformation generated by a vector $\xi^\alpha$ that is tangent to ${\cal H}_v$, the first-order correction to the scalar curvature, ${\cal R}^{(1)}$, transforms as ${\cal R}^{(1)}\to {\cal R}^{(1)} + {\cal L}_\xi {\cal R}^{(0)}$. Since the zeroth-order curvature is constant on the horizon, this implies that ${\cal R}^{(1)}$ is invariant under these transformations. However, ${\cal R}^{(2)}$ then transforms as ${\cal R}^{(2)}\to {\cal R}^{(2)} + {\cal L}_\xi {\cal R}^{(1)}$. Since ${\cal R}^{(1)}$ is not constant, ${\cal R}^{(2)}$ is not invariant.

Quantities such as the horizon area and mass of ${\cal H}_v$ {\em are} invariant under transformations within ${\cal H}_v$, since they are defined as integrals over ${\cal H}_v$. But they are not invariant under a transformation that alters the foliation. And analyzing how they transform is nontrivial because we have described the horizons' locations using gauge-dependent parametrizations $x^\alpha_{\cal H}$.

In this section, we construct invariant quantities associated with the curvature, area, and mass of the horizons. Our procedure for constructing these invariant quantities is based on gauge fixing: we write all variables in terms of the transformation to a preferred, fully fixed gauge in which the perturbed event horizon remains at the coordinate location $r=2M$. Such a gauge is described as {\em horizon locking}~\cite{Vega_2011}. Our procedure also gives invariant geometrical meaning to our foliation and to the apparent horizon's location relative to the event horizon. At the end of the section, we are able to isolate the precise differences between the two horizons.

Since we will be comparing between quantities on ${\cal A}$ and ${\cal H}^+$, in this section we explicitly add a ``${\cal A}$" or ``${\cal H}^+$" label to quantities such as $r^{(n)}$, $M_{\rm irr}$, and ${\cal R}$.

\subsection{Gauge-fixed metric perturbations}

Referring to Eqs.~\eqref{Delta h1} and \eqref{Delta h2}, we define the gauge-fixed metric perturbations to be
 \begin{align}
 	\hat{h}^{(1)}_{\alpha \beta} &= h^{(1)}_{\alpha \beta} + \mathcal{L}_{\zeta_{(1)}} g^{(0)}_{\alpha \beta}, \label{first_order_gauge_metric} \\
 	\hat{h}^{(2)}_{\alpha \beta} &= h^{(2)}_{\alpha \beta} + \mathcal{L}_{\zeta_{(2)}} g^{(0)}_{\alpha \beta} + \frac{1}{2} \mathcal{L}^2_{\zeta_{(1)}} g^{(0)}_{\alpha \beta} + \mathcal{L}_{\zeta_{(1)}} h^{(1)}_{\alpha \beta}. \label{second_order_gauge_metric}
\end{align}
In terms of Eddington-Finkelstein components, these equations read
\begin{subequations}\label{h1hat components}
\begin{align}
	\hat{h}^{(1)}_{vv} &= h^{(1)}_{vv}  -\frac{2M}{r^2}\zeta_{(1)}^{r} + 2\partial_v \zeta_{(1)}^{r}-2f\partial_v\zeta^v_{(1)}, \label{hhat1vv} \\
	\hat{h}^{(1)}_{vr} &= h^{(1)}_{vr} +\partial_v \zeta_{(1)}^v + \partial_r \zeta_{(1)}^{r}-f\partial_r\zeta^v_{(1)}, \label{hhat1vr}\\
	\hat{h}^{(1)}_{vA} &= h^{(1)}_{vA}+r^2\Omega_{AB} \partial_v \zeta_{(1)}^B -fD_A\zeta^v_{(1)}\nonumber\\
	&\quad + D_A  \zeta_{(1)}^r \label{hhat1vA},\\
	 \hat{h}^{(1)}_{rr} &= h^{(1)}_{rr} + 2\partial_r\zeta^v_{(1)}\label{hhat1rr},\\
	 \hat{h}^{(1)}_{rA} &= h^{(1)}_{rA} + r^2\Omega_{AB}\partial_r\zeta^B_{(1)}+D_A\zeta^v_{(1)},\label{hhat1rA}\\
	 \hat{h}^{(1)}_{\circ} &= h^{(1)}_\circ+2r\zeta^r_{(1)} + r^2D_{A}\zeta^A_{(1)},\label{hhat1 trace}\\
	 \hat{h}^{(1)}_{\langle AB\rangle} &= h^{(1)}_{\langle AB\rangle} + 2r^2\Omega_{C\langle A}D_{B\rangle}\zeta^C_{(1)}.\label{hhat1 tracefree}
\end{align}
\end{subequations}
The components of $\hat h^{(2)}_{\alpha\beta}$ are given by the same equations with the replacements $\zeta^\alpha_{(1)}\to\zeta^\alpha_{(2)}$ and $h^{(1)}_{\alpha\beta}\to H^{(2)}_{\alpha \beta}$, where
\beq\label{H}
H^{(2)}_{\alpha \beta}:=  h^{(2)}_{\alpha \beta} + \frac{1}{2} \mathcal{L}^2_{\zeta_{(1)}} g^{(0)}_{\alpha \beta} + \mathcal{L}_{\zeta_{(1)}} h^{(1)}_{\alpha \beta}. 
\eeq

In these expressions, $\zeta^\alpha_{(n)}$ is the unique vector that transforms from the ``user gauge" (whichever gauge one happens to use to solve the field equations) to the fixed gauge. By imposing geometrical conditions on $\hat h_{\alpha\beta}^{(n)}$, we will express $\zeta^\alpha_{(n)}$ explicitly in terms of $h^{(n)}_{\alpha\beta}$. This process will be expedited by introducing the tensor-harmonic expansion
\begin{align}
	\zeta_{(n)}^a &= \sum_{lm} \zeta_{(nlm)}^a Y^{lm} , \\
	\zeta_{(n)}^A &= \sum_{lm}\left(\zeta_{(nlm)}^{+} Y_{lm}^A + \zeta_{(nlm)}^{-} X_{lm}^{A}\right). 
\end{align}
Once $\zeta^{\alpha}_{(n)}$ is determined, the quantities $\hat h^{(n)}_{\alpha\beta}$ will then be given by simple formulas in terms of $h^{(n)}_{\alpha\beta}$. These formulas will be gauge invariant: $\hat h^{(n)}_{\alpha\beta}$ takes the same value regardless of the gauge that $h^{(n)}_{\alpha\beta}$ is in. The horizons' surface area, curvature, and mass will then be written in terms of manifestly invariant quantities with clear geometrical meanings.

Finally, to facilitate this gauge-fixing procedure in the case of a binary inspiral, we introduce a division of each field into its quasistationary and oscillatory pieces,
\begin{align}
h^{(n)}_{\alpha\beta} &= \left\langle h^{(n)}_{\alpha\beta}\right\rangle + j^{(n)}_{\alpha\beta},\label{hn division}\\
\hat h^{(n)}_{\alpha\beta} &= \left\langle \hat h^{(n)}_{\alpha\beta}\right\rangle + \hat j^{(n)}_{\alpha\beta},\label{hnhat division}\\
H^{(2)}_{\alpha\beta} &= \left\langle H^{(2)}_{\alpha\beta}\right\rangle +  J^{(2)}_{\alpha\beta},\label{H division}\\
\zeta^\alpha_{(n)} &= \left\langle \zeta_{(n)}^{\alpha}\right\rangle + \eta^\alpha_{(n)}.\label{zeta division}
\end{align}
Each of the oscillatory quantities has an expansion of the form
\beq
j^{(n)}_{\alpha\beta} = \sum_{\bm{k}\neq\bm{0}}j^{(n\bm{k})}_{\alpha\beta}(\tilde v,r,\theta^A)e^{-i\varphi_{\bm{k}}},
\eeq
which excludes the quasistationary, $\bm{k}=0$ terms from expansions of the form~\eqref{metric_expansion_two-timescale}. When substituting these two-timescale forms into Eqs.~\eqref{first_order_gauge_metric}--\eqref{second_order_gauge_metric}, the first-order equation becomes ~\eqref{h1hat components} with $\partial_v\to \Omega_i\partial_{\varphi_i}$. A $v$ derivative acting on $\tilde v$ dependence is demoted to second order. We can absorb those terms into a redefinition of $H^{(2)}_{\alpha\beta}$:  
\begin{subequations}\label{H slow-time terms}
\begin{align}
H^{(2)}_{vv}&\to \tilde H^{(2)}_{vv} := H^{(2)}_{vv} +2\partial_{\tilde v}\zeta^r_{(1)}-2f\partial_{\tilde v}\zeta^v_{(1)},\\
H^{(2)}_{vr}&\to\tilde H^{(2)}_{vr}:= H^{(2)}_{vr} +\partial_{\tilde v}\zeta^v_{(1)},\\
H^{(2)}_{vA}&\to\tilde H^{(2)}_{vA}:= H^{(2)}_{vA} +r^2\Omega_{AB}\partial_{\tilde v}\zeta^B_{(1)},
\end{align}
\end{subequations}
with other components unchanged. All other $v$ derivatives in the second-order expressions are then replaced with $\Omega_i\partial_{\varphi_i}$.

\subsection{Gauge fixing}

\subsubsection{Horizon locking}

We first impose the condition that the event horizon of the perturbed spacetime lies at the coordinate radius $r=2M$. An example of a gauge condition that enforces this is the Killing gauge~\cite{Vega_2011}, defined by $h^{(n)}_{v\alpha}=0$ for all $\alpha$. For reasons we describe momentarily, we impose a slight variant of that condition. We first impose
\begin{align}
\hat{h}^{(1)}_{vv}\big|_{{\cal H}^0} &= 0,\label{horizon locking 1}\\
\hat{h}^{(2)}_{vv}\big|_{{\cal H}^0} & = \frac{\Omega^{AB}\hat{h}^{(1)}_{vA}\hat{h}^{(1)}_{vB}\big|_{{\cal H}^0}}{4M^2}.\label{horizon locking 2}
\end{align}
Here $\big|_{{\cal H}^0}$ indicates evaluation at $(v,2M,\theta^A)$; we do not require these conditions to hold at any points away from $r=2M$. Examining Eq.~\eqref{first_rad_eh}, we see that in a gauge satisfying Eq.~\eqref{horizon locking 1}, the leading perturbation to the event horizon's radius, $r^{(1)}_{{\cal H}^+}$, vanishes. Given this, examining Eq.~\eqref{second_rad_eh}, we see that in a gauge satisfying both Eqs.~\eqref{horizon locking 1} and \eqref{horizon locking 2}, the second-order perturbation $r^{(2)}_{{\cal H}^+}$ likewise vanishes. In both cases, we assume the teleological solution for $r^{(n)}_{{\cal H}^+}$, which rules out the nontrivial homogeneous solutions.

By combining Eq.~\eqref{hhat1vv} with Eqs.~\eqref{horizon locking 1} and \eqref{horizon locking 2}, we find
\begin{align}
    \partial_v \zeta_{(1)}^{r} - \kappa_0 \zeta_{(1)}^r  &= -\frac{1}{2} h^{(1)}_{vv}, \label{zeta1_r_eq}\\
      \partial_v \zeta_{(2)}^{r} - \kappa_0 \zeta_{(2)}^r  &= -\frac{1}{2}H^{(2)}_{vv} +\frac{\Omega^{AB}\hat{h}^{(1)}_{vA}\hat{h}^{(1)}_{vB}}{8M^2}. \label{zeta2_r_eq}  
\end{align}
It should be understood that all quantities are evaluated at $r=2M$ in these expressions. 
These equations are identical in form to the equations \eqref{first_rad_eh} and \eqref{second_rad_eh} for the perturbations to the event horizon's radius. So they also have the teleological solution provided by the formula \eqref{teleological},
\begin{align}
	\zeta_{(1)}^r\big|_{{\cal H}^0} &= \frac{1}{2} \int_{v}^{\infty} e^{- \kappa_0 (v' - v)}  h^{(1)}_{vv}\big|_{{\cal H}^0} dv',\label{zeta1r}\\
	\zeta_{(2)}^r\big|_{{\cal H}^0} &= \frac{1}{2} \int_{v}^{\infty} e^{-\kappa_0 (v' - v)} \nonumber\\&\qquad\times \left(H^{(2)}_{vv} -\frac{\Omega^{AB}\hat{h}^{(1)}_{vA}\hat{h}^{(1)}_{vB}}{4M^2}\right)\bigg|_{{\cal H}^0}dv'.\label{zeta2r}
\end{align}

Note that the conditions~\eqref{horizon locking 1} and \eqref{horizon locking 2} do not dictate that the perturbed horizon's generators have the same coordinate description as the background horizon's. The generators all lie within the surface $r=2M$, but the perturbed generators do not correspond to lines of constant $\theta^A$ within that surface. In many (but not all) cases, we can freely enforce that they {\em are} lines of constant $\theta^A$ by demanding
\beq\label{horizon generator condition}
\hat h^{(n)}_{vA}\big|_{{\cal H}^0} =0. 
\eeq
From Eq.~\eqref{hhat1vA}, this implies 
\begin{align}
\zeta^+_{(1)}\big|_{{\cal H}^0} &=\zeta^+_{(1)}\big|_{{\cal S}^0}-\frac{1}{4M^2}\int_{-\infty}^v \left(h^{(1)}_{v+}+\zeta^r_{(1)}\right)\big|_{{\cal H}^0}dv',\label{zeta+ r=2M}\\
\zeta^-_{(1)}\big|_{{\cal H}^0} &=\zeta^-_{(1)}\big|_{{\cal S}^0} -\frac{1}{4M^2}\int_{-\infty}^v h^{(1)}_{v-}\big|_{{\cal H}^0}dv',\label{zeta- r=2M}
\end{align}
where ${\cal S}^0$ is the bifurcation sphere, $(r=2M,v=-\infty)$, and we have omitted $lm$ indices for visual simplicity. The same equations apply at second order with the replacements $\zeta^\alpha_{(1)}\to \zeta^\alpha_{(2)}$ and $h_{v\pm}^{(1)}\to H_{v\pm}^{(2)}$.

We can see from Eqs.~\eqref{k1} and \eqref{k2} that with the above conditions, the normal vector to the event horizon (and therefore the tangent to the event horizon generators) now has the same components as the background normal vector,
\beq
\hat k^\alpha_{{\cal H}^+}\partial_\alpha = \partial_v.\label{locked horizon generator}
\eeq

However, we cannot make the simplifications~\eqref{horizon generator condition} and \eqref{locked horizon generator} in the case of a binary inspiral. In an inspiral, the condition~\eqref{horizon generator condition} leads to a large, order-$\eps^{-1}$ vector field, 
\beq
\left\langle\zeta^-_{(1lm)}\right\rangle\Big|_{{\cal H}^0} = -\frac{1}{4M^2\eps}\int_{-\infty}^{\tilde v} \left\langle h^{(1lm)}_{v-}\right\rangle\Big|_{{\cal H}^0} d\tilde v',
\eeq
invalidating our asymptotic expansions. This will occur, for example, due to the black hole's slowly varying spin, which accumulates over the course of the inspiral as the black hole absorbs gravitational waves. 

To adapt the condition~\eqref{horizon generator condition} to an inspiral, we split our fields into quasistationary and oscillatory terms, following Eqs.~\eqref{hn division}--\eqref{zeta division}. We then impose
\beq\label{fast horizon generator condition}
\hat j^{(n)}_{vA}\big|_{{\cal H}^0} = 0,
\eeq
implying the analogues of Eqs.~\eqref{zeta+ r=2M} and \eqref{zeta- r=2M},
\begin{align}
\eta^+_{(1\bm{k})}\big|_{{\cal H}^0} &= \frac{1}{4iM^2\Omega_{\bm{k}}} \left(j^{(1\bm{k})}_{v+}+\eta^r_{(1\bm{k})}\right)\big|_{{\cal H}^0},\label{eta+ r=2M}\\
\eta^-_{(1\bm{k})}\big|_{{\cal H}^0} &= \frac{1}{4iM^2\Omega_{\bm{k}}} j^{(1\bm{k})}_{v-}\big|_{{\cal H}^0}.\label{eta- r=2M}
\end{align}
Here we have again omitted $lm$ indices. At second order the same equations hold with the replacements $\eta^\alpha_{(1)}\to \eta^\alpha_{(2)}$ and $j_{v\pm}^{(1)}\to \tilde J_{v\pm}^{(2)}$, where $\tilde J_{v\pm}^{(2)}$ includes the slow-time derivative terms from Eq.~\eqref{H slow-time terms}. Referring to Eqs.~\eqref{k1} and \eqref{k2}, we see that these conditions enforce
\beq
\hat k^\alpha_{{\cal H}^+}\partial_\alpha = \partial_v +  \hat k^A_{{\cal H}^+}\partial_A,\label{locked horizon generator two-time}
\eeq
where 
\beq
\hat k^A_{{\cal H}^+} = -\frac{\eps}{4M^2}\Omega^{AB}\<\hat h^{(1)}_{vB}\>\big|_{{\cal H}^0}+O(\eps^2)
\eeq
is (at least at leading nonzero order) a slowly varying angular component. The horizon generators in this case wrap around the horizon slowly, with a small, slowly varying frequency $\frac{d\theta^A}{dv} = \hat k^A_{{\cal H}^+}$.

There is a straightforward analogy between our gauge fixing in the two cases. In the case that we do not have quasistationary effects, Eqs.~\eqref{zeta+ r=2M} and \eqref{zeta- r=2M} leave the constants $\zeta^\pm_{(n)}\big|_{{\cal S}^0}$ unspecified; this is an incomplete gauge fixing on the horizon. In the case of an inspiral, we instead have the slowly varying vector fields $\<\zeta^\pm_{(n)}\>\big|_{{\cal H}^0}$ unspecified. Similarly, we can immediately relate the time-varying pieces of Eqs.~\eqref{zeta+ r=2M} and \eqref{zeta- r=2M} to Eqs.~\eqref{eta+ r=2M} and \eqref{eta- r=2M} using the time-localizing approximation~\eqref{integral_approximation1}. Moreover, Eqs.~\eqref{horizon locking 1}--\eqref{zeta2r} apply in both cases. 

To bring all the equations into two-timescale form, we can localize Eqs.~\eqref{zeta1r} and \eqref{zeta2r} using the approximation~\eqref{integral_approximation2}. One can use only the leading term in that approximation and then make the adjustments in Eq.~\eqref{H slow-time terms}, or equivalently, one can include the subleading term in the approximation and omit the adjustments in Eq.~\eqref{H slow-time terms}. The result is
\begin{align}
    \zeta^r_{(1\bm{k})}\big|_{{\cal H}^0} &= \frac{h^{(1\bm{k})}_{vv}}{2\omega_{\bm{k}}},\label{zeta1r two-timescale}\\
    \zeta^r_{(2\bm{k})}\big|_{{\cal H}^0} &= \frac{\tilde H^{(2\bm{k})}_{vv}}{2\omega_{\bm{k}}} - \frac{\Omega^{AB}\big\langle \hat  h^{(1)}_{vA}\big\rangle\big\langle \hat  h^{(1)}_{vB}\big\rangle\delta_{\bm{k},\bm{0}}}{2M}
    ,\label{zeta2r two-timescale}
\end{align}
where all quantities are evaluated at $r=2M$, and we have used Eq.~\eqref{fast horizon generator condition}.

\subsubsection{Foliation locking}

Having locked the event horizon to $r=2M$ and its generators to lines of constant (or slowly varying) $\theta^A$, we now impose the foliation-locking condition
\beq
\hat g^{\alpha\beta}\partial_\alpha v\,\partial_\beta v = 0.\label{foliation locking ghat}
\eeq
This condition enforces that a surface of constant $v$ is a null surface. It does not enable us to uniquely identify a specific cut of the horizon, but it does enforce that our cuts of constant $v$ correspond to a foliation defined by the intersections of the horizon with a family of ingoing null surfaces $\Sigma_v$. 

There are several ways to satisfy Eq.~\eqref{foliation locking ghat}. A straightforward method  is to impose
\begin{align}
    \hat h^{(n)}_{r\alpha} &= 0.\label{foliation locking}
\end{align}
This is referred to as a lightcone gauge condition~\cite{Preston:2006ze}. If we additionally enforced $\hat h^{(n)}_\circ=0$, it would put the metric in the standard radiation gauge, in which $\hat h^{(n)}_{\<AB\>}$ represents the transverse-tracefree ingoing gravitational waves at the horizon. However, we will not impose this additional condition. The condition~\eqref{foliation locking} alone enforces not only that $\Sigma_v$ is a null surface, but also (through $\hat h^{(n)}_{rA}=0$) that the generators of $\Sigma_v$ are lines of constant $\theta^A$, and that their tangent vector (or equivalently, the normal to the surface) has components
\beq
 \hat n^\alpha_{{\cal H}^+}\partial_\alpha = -\partial_r\label{nhat}
\eeq
[which is consistent with Eqs.~\eqref{n1} and \eqref{n2}]. As a consequence, $r$ is an affine parameter along the surface's generators. We will see that Eq.~\eqref{nhat} also applies for the ingoing null vector orthogonal to ${\cal A}_v$ (through second order in $\eps$).

By combining $\hat h^{(n)}_{rr}=0$ with Eq.~\eqref{hhat1rr}, we find
\begin{align}
\zeta^v_{(1)} &= \zeta^v_{(1)}\big|_{{\cal H}^0} -\frac{1}{2}\int_{2M}^r h^{(1)}_{rr}dr'.\label{zeta1v r>2M}
\end{align}
Here, and in all cases in this section, the same equation applies at second order with the replacements $\zeta^\alpha_{(1)}\to\zeta^\alpha_{(2)}$ and $h^{(1)}_{\alpha\beta}\to H^{(2)}_{\alpha \beta}$. These equations uniquely determine $\zeta^v_{(n)}$ up to its value at the horizon. Similarly, combining $\hat h^{(n)}_{vr}=0$ with Eq.~\eqref{hhat1vr}, we find
\begin{align}
    \zeta^r_{(1)} &= \zeta^r_{(1)}\big|_{{\cal H}^0}-\int_{2M}^r \left(h^{(1)}_{vr}+\partial_v\zeta^v_{(1)}+\frac{f(r')}{2}h^{(1)}_{rr}\right) dr'.\label{zeta1r r>2M}
\end{align}    
Equation~\eqref{zeta1r r>2M} with Eqs.~\eqref{zeta1v r>2M} and \eqref{zeta1r} (with their second-order analogues) fully determine $\zeta^r_{(n)}$.

Next, $\hat h^{(n)}_{rA} = 0$ combined with Eq.~\eqref{hhat1rA} implies
\begin{align}
\zeta^+_{(1)} &= \zeta^+_{(1)}\big|_{{\cal H}^0} - \int_{2M}^r\frac{h^{(1)}_{r+} +\zeta^v}{r'^2}dr',\label{zeta+ r>2M}\\
\zeta^-_{(1)} &= \zeta^-_{(1)}\big|_{{\cal H}^0} - \int_{2M}^r\frac{h^{(1)}_{r-}}{r'^2} dr'.\label{zeta- r>2M}
\end{align}
These relations are in terms of individual $lm$ modes, but for visual simplicity we have suppressed $lm$ indices on all quantities. Equations~\eqref{zeta+ r>2M} and \eqref{zeta- r>2M} and their second-order analogues, with Eqs.~\eqref{zeta1v r>2M}, \eqref{zeta+ r=2M}, and \eqref{zeta- r=2M} [or Eqs.~\eqref{eta+ r=2M} and \eqref{eta- r=2M}], fully specify $\zeta^\pm_{(n)}$ up to the constants $\zeta^\pm_{(n)}\Big|_{{\cal S}^0}$ or up to the slowly varying quantities $\<\zeta^\pm_{(n)}\>\big|_{{\cal H}^0}$.

We have now fully fixed the gauge up to $\zeta^v_{(n)}\big|_{{\cal H}^0}$ and $\zeta^\pm_{(n)}\big|_{{\cal S}^0}$. $\zeta^\pm_{(n)}\big|_{{\cal S}^0}$ corresponds to a transformation within ${\cal H}_v^+$, while $\zeta^v_{(n)}\big|_{{\cal H}^0}$ directly corresponds to a specification of the cut ${\cal H}^+_v$; given this ${\cal H}^+_v$, $\Sigma_v$ is then the specific null surface that intersects ${\cal H}^+$ at that cut. To fix the choice of cut, we impose the even-parity part of the condition~\eqref{horizon generator condition} to linear order in distance from the horizon, in the sense that
\beq\label{dhv+=0}
\partial_r\hat{h}^{(nlm)}_{v+}\big|_{{\cal H}^0} = 0. 
\eeq
Geometrically, this condition, together with our other gauge-fixing conditions, enforces that the 2-vector 
\beq
\hat\omega_A := - \hat n^{{\cal H}^+}_\alpha (\hat e_{{\cal H}^+})^\beta_A\hat\nabla_\beta \hat k^\alpha_{{\cal H}^+}
\eeq
has no even-parity piece (the odd-parity piece of this vector is gauge invariant~\cite{Vega_2011}, at least at first order). When combined with Eq.~\eqref{hhat1vA}, Eq.~\eqref{dhv+=0} implies
\begin{multline}\label{zeta1v eqn r=2M}
    2(\partial_v+\kappa_0)\zeta^v_{(1)} = \partial_r h^{(1)}_{v+}-\partial_v h^{(1)}_{r+}-h^{(1)}_{vr}\\-\frac{1}{M}\left(\zeta^r_{(1)}+h^{(1)}_{v+}\right)\quad (l>0),
\end{multline}
where all fields are evaluated at $r=2M$, and we have suppressed $lm$ indices. To obtain this form of the equation, we have substituted Eq.~\eqref{zeta+ r>2M} for $\partial_r\zeta^+_{(1)}$ and Eq.~\eqref{zeta1r r>2M} for $\partial_r\zeta^r_{(1)}$. The well-behaved solution to Eq.~\eqref{zeta1v eqn r=2M} is
\begin{multline}
    \zeta^v_{(1)}\Big|_{{\cal H}^0} = \frac{1}{2}\int_{-\infty}^v e^{\kappa_0(v'-v)}\Big[\partial_r h^{(1)}_{v+}-\partial_v h^{(1)}_{r+}-h^{(1)}_{vr}\\-\frac{1}{M}\left(\zeta^r_{(1)}+h^{(1)}_{v+}\right)\Big]\Big|_{{\cal H}^0}dv'\ \  (l>0).\label{zeta1v r=2M}
\end{multline}
If we mirror the derivation of Eq.~\eqref{integral_approximation2}, we can expand this in the two-timescale form\footnote{The subleading term, which will contribute to the second-order vector in the two-timescale expansion, is straightforwardly obtained in analogy with Eq.~\eqref{integral_approximation2}.}
\begin{align}\label{zeta1v r=2M two-timescale}
\zeta^v_{(1\bm{k})} &= \frac{\partial_r h^{(1\bm{k})}_{v+}+i\Omega_{\bm{k}}h^{(1\bm{k})}_{r+}-h^{(1\bm{k})}_{vr}-\frac{1}{M}\left(\zeta^r_{(1\bm{k})}+h^{(1\bm{k})}_{v+}\right)}{2(\kappa_0-i\Omega_{\bm{k}})} \nonumber\\
&\quad+O(\eps),
\end{align}
where all quantities are evaluated at $r=2M$. 
These equations, with Eq.~\eqref{zeta1v r>2M} (and their second-order analogues), now determine $\zeta^v_{(n)}$ for all $r$ and all $l>0$.

To partially fix the $l=0$ mode, we impose the conditions~\eqref{horizon locking 1} and \eqref{horizon locking 2} to linear order in distance,
\begin{align}
\partial_r\hat{h}^{(n00)}_{vv}\big|_{{\cal H}^0} &= 0,\label{dhvv=0}
\end{align}
noting that for $n=2$ this is only applied outside the two-timescale context, where $\hat h^{(n)}_{vA}=0$. The condition~\eqref{dhvv=0}, together with our other gauge-fixing conditions, enforces that the average surface gravity on the horizon remains equal to its background value,
\beq
\frac{1}{4\pi}\int_{{\cal H}^+_v} \hat\kappa\, d\Omega = \kappa_0 + O(\eps^3),
\eeq
where $\hat\kappa$ is defined from $\hat k^\beta_{{\cal H}^+}\hat\nabla_\beta\hat k^\alpha_{{\cal H}^+} = \hat\kappa \hat k^\alpha_{{\cal H}^+}$.

Taking a derivative of Eq.~\eqref{hhat1vv} and applying these conditions, we find 
\begin{align}
    \zeta^v_{(1)}\big|_{{\cal H}^0} &=  \zeta^v_{(1)}\big|_{{\cal S}^0}+ \int\displaylimits^v_{-\infty}\int\displaylimits^{v'}_{-\infty}\!e^{\kappa_0(v''-v')} \bigg(\frac{1}{2}\partial_r h^{(1)}_{vv}-\partial_v h^{(1)}_{vr}\nonumber\\
    &\quad+\frac{\kappa_0}{M}\zeta^r_{(1)}+\kappa_0 h^{(1)}_{vr}\Big)\!\bigg|_{{\cal H}^0}dv''dv'\quad  (l=0)\label{zeta1v l=0}
\end{align}
and analogous at second order. This  solution to Eq.~\eqref{dhvv=0} is well behaved in the infinite past. However, there is no solution that is well behaved in the infinite future: the vector field grows linearly with $v$ if the metric perturbation becomes stationary at late times. We should therefore consider Eq.~\eqref{zeta1v l=0} as fixing any nonstationary part of the metric perturbation, taking that part to vanish at late times. 

This division is clearer in a two-timescale expansion, where in place of Eq.~\eqref{dhvv=0} we can impose a condition on the purely oscillatory part of the perturbation,
\beq
\partial_r\hat{j}^{(n00)}_{vv}\big|_{{\cal H}^0} = 0.\label{dj1vv=0}
\eeq
This fixes the oscillatory part of the vector field:
\beq
\eta^v_{(1\bm{k})}\big|_{{\cal H}^0} = \frac{\eta^r_{(1\bm{k})}+2M^2\partial_r j^{(1\bm{k})}_{vv}+4M^2\omega_{\bm{k}}j^{(1\bm{k})}_{vr}}{4M^2i\Omega_{\bm{k}}(i\Omega_{\bm{k}}-\kappa_0)}\bigg|_{{\cal H}^0}\ (l=0),\label{eta1v l=0 r=2M}
\eeq  
where $\omega_{\bm{k}}$ was defined in Eq.~\eqref{omega_k}, and at second order the same holds with $\eta^\alpha_{(1)}\to\eta^\alpha_{(2)}$ and $j^{(1)}_{\alpha\beta}\to\tilde  J^{(2)}_{\alpha\beta}$. Equation~\eqref{dj1vv=0} enforces that the average surface gravity contains no oscillatory part:
\beq
\frac{1}{4\pi}\int_{{\cal H}^+_v} \hat\kappa\, d\Omega = \frac{1}{4\pi}\int_{{\cal H}^+_v}\< \hat\kappa\> d\Omega.
\eeq

The final, unspecified freedom in the foliation is the choice of $\<\zeta^v_{(n00)}\>\Big|_{{\cal H}^0}$ (or $\zeta^v_{(n00)}\big|_{{\cal S}^0}$, outside a two-timescale context), which corresponds to a slowly varying (or constant), uniform shift in time along the horizon. We return to this freedom below.

\subsubsection{Euclidean radius locking}\label{sec:embedding locking}

Having locked the horizon and specified the foliation (up to uniform time translations), we now fix the gauge {\em within} each cut. We specify $\zeta^A_{(n)}\Big|_{{\cal S}^0}$ by imposing
\beq\label{pure trace}
 \hat{h}^{(n)}_{\langle AB\rangle}\big|_{{\cal S}^0} = 0.
\eeq
The analogous condition is more impactful in the case of an inspiral, where we  impose
\beq\label{pure trace average}
 \<\hat{h}^{(n)}_{\langle AB\rangle}\>\Big|_{{\cal H}^0} = 0.
\eeq
This puts the time-averaged induced metric on ${\cal H}_v^+$ in a ``pure trace" form,
\begin{multline}
\<\hat\gamma^{{\cal H}^+}_{AB}\> = \Big[4M^2 + \eps\<\hat\gamma^{(1)}_\circ\>\\+ \eps^2\<\hat\gamma^{(2)}_\circ\>+O(\eps^3)\Big]\Omega_{AB},\label{gamma pure trace}
\end{multline}
giving it the same appearance as the metric on a closed 2-surface with radial profile
\begin{multline}
\<\hat r_{\rm E}(\theta^A)\> = 2M\bigg[1+\frac{\eps}{4M^2}\<\hat\gamma^{(1)}_\circ\>\\
+ \frac{\eps^2}{4M^2}\<\hat\gamma^{(2)}_\circ\>+O(\eps^3)\bigg]^{1/2}
\end{multline}
in flat, Euclidean 3-space. There is then a one-to-one association between the metric's trace and a Euclidean radius. We expand on this association in Sec.~\ref{visualizing embedding}.

Combining Eq.~\eqref{hhat1 tracefree} with Eq.~\eqref{pure trace average}, we find
\begin{align}
\<\zeta^{\pm}_{(1lm)}\>\Big|_{{\cal H}^0} &= -\frac{1}{8M^2}\<h^{(1lm)}_\pm\>\Big|_{{\cal H}^0} \quad(l>1),\label{zeta1pm av}\\
\<\zeta^{\pm}_{(2lm)}\>\Big|_{{\cal H}^0} &= -\frac{1}{8M^2}\<H^{(2lm)}_\pm\>\Big|_{{\cal H}^0} \quad(l>1).\label{zeta2pm av}
\end{align}
Outside the two-timescale context, these equations instead apply for $\zeta^\pm_{(n)}\big|_{{\cal S}^0}$ (after removing the angular brackets). With this, we have fully fixed the $l>1$ modes of the vector field. 

To fix the even-parity dipole mode of  $\<\zeta^A_{(n)}\>\Big|_{{\cal H}^0}$, we impose that
\beq\label{gamma no dipole}
\<\hat h^{(n1m)}_\circ\>\Big|_{{\cal H}^0} = 0.
\eeq
This enforces that the intrinsic metric~\eqref{gamma pure trace} has no $l=1$ contribution, meaning that the geometry is manifestly round for $l=0,1$. The fact that we can eliminate the $l=1$ contribution corresponds to the fact that a spatial translation cannot affect the intrinsic curvature of a surface in Euclidean space. Combining Eq.~\eqref{hhat1 trace} with Eq.~\eqref{gamma no dipole}, we find
\begin{align}
\<\zeta^{+}_{(11m)}\>\Big|_{{\cal H}^0} &= \frac{\<h^{(11m)}_\circ\>\big|_{{\cal H}^0}}{8M^2}+\frac{\<\zeta^r_{(11m)}\>\big|_{{\cal H}^0}}{2M},\label{zeta1+ av dipole}\\
\<\zeta^{+}_{(21m)}\>\Big|_{{\cal H}^0} &= \frac{\<H^{(21m)}_\circ\>\big|_{{\cal H}^0}}{8M^2}+\frac{\<\zeta^r_{(21m)}\>\big|_{{\cal H}^0}}{2M}.\label{zeta2+ av dipole}
\end{align}
We return to this type of gauge fixing of the even-parity dipole in Sec.~\ref{visualizing embedding}.

\subsubsection{Residual Killing fields}

We are now left with two unrestricted modes: the $l=0$ piece of $\zeta^v_{(n)}\big|_{{\cal S}^0}$ and the $l=1$, odd-parity piece of $\zeta^A_{(n)}\big|_{{\cal S}^0}$. These pieces of the vector field cannot be fixed by imposing conditions on $\hat h^{(1)}_{\alpha\beta}$. They correspond to the timelike and rotational Killing fields of the background spacetime, which trivially contribute nothing in Eq.~\eqref{first_order_gauge_metric}. If we were only concerned with first-order perturbations, we could entirely ignore these pieces, as they would have no impact on the metric perturbation. However, by leaving the first-order vectors $\zeta^v_{(100)}\big|_{{\cal S}^0}$ and $\zeta^-_{(11m)}\big|_{{\cal S}^0}$ unspecified, we leave our {\em second}-order metric perturbations unfixed.

More concretely, consider a gauge transformation generated by a vector $\xi^\alpha_{(1)}$, as in Eq.~\eqref{Delta h1}. If we let $h^{(1)}_{\alpha\beta}\to h^{(1)}_{\alpha\beta}+{\cal L}_{\xi_{(1)}}g^{(0)}_{\alpha\beta}$ and $ \zeta^\alpha_{(1)} \to  \zeta^\alpha_{(1)} +\Delta \zeta^\alpha_{(1)}$ in Eq.~\eqref{first_order_gauge_metric}, we find that the transformation induces a change 
\beq
\Delta \hat h^{(1)}_{\alpha\beta} = {\cal L}_{\Delta\zeta_{(1)}} g^{(0)}_{\alpha\beta} + {\cal L}_{\xi_{(1)}} g^{(0)}_{\alpha\beta}.\label{Dhhat1}
\eeq
If we substitute the transformation~\eqref{Delta h1} into our equations for $\zeta^\alpha_{(1)}$, we find that $\zeta^\alpha_{(1)}$ transforms as
\beq
\Delta \zeta^\alpha_{(1)} = -\xi^\alpha_{(1)} +\Xi^\alpha_{(1)},\label{zeta1 transformation}
\eeq
where
\beq
\Xi^\alpha_{(1)}\partial_\alpha := \xi^v_{(100)}\big|_{{\cal S}^0}Y^{00}\partial_v + \sum_m \xi^-_{(11m)}\big|_{{\cal S}^0}X^A_{1m}\partial_A
\eeq
is a linear combination of Killing fields. These Killing fields arise in $\Delta \zeta^\alpha_{(1)}$ from letting $h^{(1)}_{\alpha\beta}\to h^{(1)}_{\alpha\beta}+{\cal L}_{\xi_{(1)}}g^{(0)}_{\alpha\beta}$ in Eqs.~\eqref{zeta1v l=0} and \eqref{zeta- r=2M}; the integral in each case introduces a mode of $\xi^\alpha_{(1)}$ at ${\cal S}^0$. Returning to Eq.~\eqref{Dhhat1}, we now see
\beq
\Delta\hat h^{(1)}_{\alpha\beta} = 0.
\eeq
In words, $\hat h^{(1)}_{\alpha\beta}$ is invariant, as we would expect.

However, if we next consider a second-order gauge transformation generated by vectors $\xi^\alpha_{(1)}$ and $\xi^\alpha_{(2)}$, as in Eq.~\eqref{Delta h2}, then a short calculation starting from Eq.~\eqref{second_order_gauge_metric} reveals that 
\beq
\Delta \hat h^{(2)}_{\alpha\beta} = {\cal L}_\Upsilon g^{(0)}_{\alpha\beta} + {\cal L}_{\Xi_{(1)}} h^{(1)}_{\alpha\beta},\label{Dhhat2}
\eeq
where
\beq
\Upsilon^\alpha := \Delta\zeta^\alpha_{(2)} + \xi^\alpha_{(2)} + \frac{1}{2}\!\left[\zeta_{(1)}+\Xi_{(1)},\xi_{(1)}\right]^\alpha.
\eeq
Here $[\cdot,\cdot]$ denotes the commutator. It is easy to check that $\hat h^{(2)}_{\alpha\beta}$ is {\em not} invariant because $\zeta^\alpha_{(1)}$ is not fully fixed. To show this, note that by construction, $\hat h^{(2)}_{v\alpha}\big|_{{\cal H}^0}=0$ in all gauges, implying  $\Delta\hat h^{(2)}_{v\alpha}\big|_{{\cal H}^0}=0$. Substituting this condition in Eq.~\eqref{Dhhat2} and solving for $\Upsilon^\alpha$, we find
\begin{align}
    \Upsilon^r\big|_{{\cal H}^0} &= \int_v^\infty e^{-\kappa_0(v'-v)}{\cal L}_{\Xi_{(1)}} h^{(1)}_{vv}\big|_{{\cal H}^0}dv',\label{Upsilonr}\\
    \Upsilon^A\big|_{{\cal H}^0} &= \Upsilon^A\big|_{{\cal S}^0} \nonumber\\
    &\quad - \frac{\Omega^{AB}}{4M^2}\int_{-\infty}^v \!\!\left(D_B\Upsilon^r+{\cal L}_{\Xi_{(1)}}h^{(1)}_{vB}\right)\!\big|_{{\cal H}^0} dv'.\label{UpsilonA}
\end{align}
If we now examine $\Delta \hat h^{(2)}_\circ=\frac{1}{2}\Omega^{AB}\Delta \hat h^{(2)}_{AB}$, for example, substituting Eqs.~\eqref{Upsilonr} and \eqref{UpsilonA} into Eq.~\eqref{Dhhat2}, we find that
\beq
\Delta \hat h^{(2)}_\circ = 2r \Upsilon^r + r^2 D_A \Upsilon^A + \Xi^\alpha_{(1)}\partial_\alpha h^{(1)}_{\circ}\label{Dhhat2 trace}
\eeq
is a manifestly nonzero quantity. We then conclude that $\hat h^{(2)}_\circ$ is prevented from being invariant by the fact that $\Xi^\alpha_{(1)}$ is not a Killing field of the perturbed spacetime.

To fix $\zeta^v_{(100)}\big|_{{\cal S}^0}$ and $\zeta^-_{(11m)}\big|_{{\cal S}^0}$, we can impose additional conditions on $\hat h^{(2)}_{\alpha\beta}$. 
However, we will be satisfied with the fact that some choice can be made; we will not make any particular choice. 

We equivocate in this way because the quantity we calculate explicitly in Sec.~\ref{sec:quasicircular_binaries} is invariant  (within a large class of gauges) under the residual gauge freedom. The quantity we calculate is the surface area of the horizon, which transforms as
\beq
\Delta \hat A = \sqrt{4\pi}\eps^2\Delta \hat h^{(200)}_\circ\big|_{{\cal H}^0} +O(\eps^3)
\eeq
under the residual gauge freedom. This follows from Eqs.~\eqref{A1 EH} and \eqref{A2 EH} in the case of the event horizon. [Equation~\eqref{A2hat AH} with Eq.~\eqref{theta200 EH gauge fixed} implies that an additional term appears for the apparent horizon, but it is also proportional to $\Delta \hat h^{(200)}_\circ$.] Appealing to Eq.~\eqref{Dhhat2 trace}, we see that
\beq
\Delta \hat A = \sqrt{4\pi}\eps^2\!\left(\!4M \Upsilon^r_{00} + \Xi^v_{(1)}\partial_v h^{(100)}_\circ\right)\!\big|_{{\cal H}^0} +O(\eps^3).
\eeq
Both terms on the right vanish in the case that $h^{(100)}_{vv}$ and $h^{(100)}_\circ$ are independent of $v$. Since there always exist gauges in which a spherically symmetric vacuum perturbation is stationary, we conclude that $\hat A$ is invariant within that class of gauges even if we do not fix the residual Killing degrees of freedom.

Here we have focused on the case of regular perturbation theory. The story is very much the same in a two-timescale expansion, except that the residual freedoms $\<\zeta^v_{(100)}\>\Big|_{{\cal H}^0}$ and $\<\zeta^{+}_{(11m)}\>\Big|_{{\cal H}^0}$ appear in $\hat h^{(2)}_{\alpha\beta}$ in two ways: through the Lie derivatives ${\cal L}_{\zeta_{(1)}}^2g^{(0)}_{\alpha\beta}$ and ${\cal L}_{\zeta_{(1)}}h^{(1)}_{\alpha\beta}$ and through slow-time derivatives $\partial_{\tilde v}\<\zeta^v_{(100)}\>\Big|_{{\cal H}^0}$ and $\partial_{\tilde v}\<\zeta^{+}_{(11m)}\>\Big|_{{\cal H}^0}$.

\subsubsection{Summary}\label{sec:gauge fixing summary}

In summary, our gauge-fixing procedure has accomplished the following:
\begin{enumerate}
    \item fixed the coordinate radius of the event horizon ${\cal H}^+$ to be $r=2M$
    \item fixed surfaces of constant $v$, $\Sigma_v$, to be null
    \item fixed the foliation into 2-surfaces ${\cal H}_v^+={\cal H}^+\cap\Sigma_v$, to enforce the conditions~\eqref{dhv+=0} and \eqref{dhvv=0} [or~\eqref{dj1vv=0}] 
    \item fixed the coordinates $(r,\theta^A)$ such that the generators of $\Sigma_v$ are given by lines of constant $\theta^A$, with $r$ an affine parameter
    \item fixed the angular coordinates $\theta^A$ on the horizon such that the horizon generators are lines of constant $\theta^A$ (or, in an inspiral, such that they wrap around the horizon slowly, on the radiation-reaction timescale)
    \item further fixed the angular coordinates on the horizon such that the slowly varying part of the induced metric on ${\cal H}_v$ has the same form as a 2-surface embedded (with the natural  identification of angular coordinates) in Euclidean 3-space.
\end{enumerate}

If we use regular perturbation theory (i.e., with no two-timescale expansion), our choices put the gauge-fixed metric perturbation in the form
\begin{multline}
\hat h_{\alpha\beta}dx^\alpha dx^\beta = [O(f)]dv^2+2[O(f)]_Advd\theta^A\\+
\hat h_{AB}d\theta^A d\theta^B,
\end{multline}
where the individual components are given by Eq.~\eqref{h1hat components}. Here we use $f=1-2M/r$ as a measure of coordinate distance from the horizon. At $r=2M$ the metric perturbation reduces to 
\beq
\hat h_{\alpha\beta}dx^\alpha dx^\beta = \hat h_{AB}d\theta^A d\theta^B.
\eeq
The $lm$ modes of $h^{(1)}_{AB}$ at $r=2M$ are given by
\begin{align}
\hat h^{(1)}_{\circ} &= h^{(1)}_\circ + 4M\zeta^r_{(1)} -4M^2\lambda^2_1\zeta^+_{(1)},\label{hhatABreg trace}\\
\hat h^{(1)}_{\pm} &= h^{(1)}_\pm + 8M^2\zeta^\pm_{(1)} \label{hhatABreg STF}
\end{align}
(suppressing $lm$ labels) with  $\zeta^r_{(1)}$ given by Eqs.~\eqref{zeta1r} and $\zeta^\pm_{(1)}$ by Eqs.~\eqref{zeta+ r=2M} and \eqref{zeta- r=2M} [with $\zeta^\pm_{(1)}\big|_{{\cal S}^0}$ given by Eqs.~\eqref{zeta1pm av} and \eqref{zeta1+ av dipole}]. The modes of $\hat h^{(2)}_{AB}$ are given by the same formulas with $\zeta^\alpha_{(1)}\to\zeta^\alpha_{(2)}$ and $h^{(1)}_{AB}\to H^{(2)}_{AB}$.

In a two-timescale expansion, the gauge-fixed perturbation is more complicated due to the slow evolution of the black hole. The components $\hat h_{vv}$ and $\hat h_{vA}$ have slowly varying pieces that do not scale with $f$, and the metric perturbations at $r=2M$ are instead
\begin{align}
\hat h^{(1)}_{\alpha\beta}dx^\alpha dx^\beta &= 2\<\hat h^{(1)}_{vA}\>dv\, d\theta^A+\hat h^{(1)}_{AB}d\theta^A d\theta^B,\\
\hat h^{(2)}_{\alpha\beta}dx^\alpha dx^\beta &= \<\hat h^{(2)}_{vv}\>dv^2 +2\<\hat h^{(2)}_{vA}\>dv\, d\theta^A \nonumber\\
&\quad+ \hat h^{(2)}_{AB}d\theta^A d\theta^B.
\end{align}
According to Eqs.~\eqref{hhat1vA} and \eqref{horizon locking 2}, the slowly varying $v\alpha$ components on ${\cal H}^0$ are given by
\begin{align}
\<\hat h^{(1)}_{vA}\> &= \<h^{(1)}_{vA} + 2MD_Ah^{(1)}_{vv}\>,\\
\<\hat h^{(2)}_{vv}\> &= \frac{\Omega^{AB}\<\hat h^{(1)}_{vA}\>\<\hat h^{(1)}_{vB}\>}{4M^2},\\
\<\hat h^{(2)}_{vA}\> &= \< \tilde H^{(2)}_{vA} 
+ D_A \zeta^r_{(2)}\>,
\end{align}
with $\zeta^r_{(2)}$ given by Eq.~\eqref{zeta2r two-timescale}.

The metric perturbation on a constant-$v$ cut of ${\cal H}_0$ divides into a purely oscillatory trace-free piece plus a trace piece that contains both slowly varying and oscillatory contributions: 
\begin{align}
\hat h^{(n)}_{AB} &= \hat j^{(n)}_{\<AB\>} + \left( \<\hat h^{(n)}_\circ\>+\hat j^{(n)}_\circ\right)\Omega_{AB}.\label{hhatAB}
\end{align}
At first order, the oscillatory contributions are
\begin{subequations}\label{jhat1 modes}
\begin{align}
\hat j^{(1\bm{k})}_+ &= j^{(1\bm{k})}_{+} +\frac{2}{i\Omega_{\bm{k}}}\left(j^{(1\bm{k})}_{v+}+\frac{j^{(1\bm{k})}_{vv}}{2\omega_{\bm{k}}}\right),\\
\hat j^{(1\bm{k})}_- &= j^{(1\bm{k})}_{-} +\frac{2j^{(1\bm{k})}_{v-}}{i\Omega_{\bm{k}}},\\
\hat j^{(1\bm{k})}_\circ &= j^{(1\bm{k})}_\circ +\frac{2M}{\omega_{\bm{k}}}j^{(1\bm{k})}_{vv}-\frac{\lambda^2_1}{i\Omega_{\bm{k}}}\left(j^{(1\bm{k})}_{v+}+\frac{j^{(1\bm{k})}_{vv}}{2\omega_{\bm{k}}}\right),
\end{align}
\end{subequations}
and the slowly varying piece is
\beq\label{<h1_circ> gauge fixed}
\<\hat h^{(1)}_\circ\> = \begin{cases} \<h^{(1)}_\circ + 8M^2h^{(1)}_{vv}+\frac{\lambda^2_1}{2}h^{(1)}_+\> & (l\neq1),\\
0 & (l=1).
\end{cases}
\eeq
At second order, the same equations apply for $\hat j^{(2)}_\pm$, $\hat j^{(2)}_\circ$, and $\<\hat h^{(2)}_\circ\>$ with the replacements $j^{(1)}_{\alpha\beta}\to \tilde J^{(2)}_{\alpha\beta}$ and $h^{(1)}_{\alpha\beta}\to \tilde H^{(2)}_{\alpha\beta}$.

\subsection{Expansion scalar}

In the next two sections, we express the properties of the event and apparent horizon in terms of our gauge-fixed perturbations. To aid that analysis, we derive here simplified expressions for the expansion scalar of either horizon. 

In gauge-fixed form, for either horizon, the expansion scalar~\eqref{expansion1lm} is
\begin{multline}
	\hat\vartheta^{(1)}_{+,lm} = \frac{1}{4 M^2} \Big[\partial_v \hat h^{(1lm)}_{\circ} +\lambda_1^2 \hat h^{(1lm)}_{v+}\\+ (1+\lambda_1^2) \hat r^{(1)}_{lm} \Big].\label{theta1lm gauge fixed}
\end{multline}
It is straightforward to show that the linearized vacuum Einstein equation at $r=2M$ implies
\beq
\hat{\delta R}_{vv}\big|_{{\cal H}^0} = -\frac{(\partial_v-\kappa_0)\left(\partial_v\hat h^{(1)}_\circ+\lambda_1^2\hat h^{(1)}_{v+}\right)}{4 M^2} = 0. 
\eeq
This has the now-familiar form~\eqref{generic_eq_rad}, implying the unique well-behaved solution 
\beq\label{EFE soln}
\partial_v\hat h^{(1)}_\circ+\lambda_1^2\hat h^{(1)}_{v+} = 0
\eeq
or equivalently, $\partial_v\hat h^{(1)}_\circ = D^A\hat h^{(1)}_{vA}$. Therefore
\beq
\hat\vartheta^{(1)}_{+,lm} = \frac{(1+\lambda_1^2) \hat r^{(1)}_{lm}}{4 M^2}.
\eeq

For the event horizon, where $\hat r^{(1)}_{{\cal H}^+} = 0$, this implies
\beq
\vartheta^{(1)}_{{\cal H}^+} =0.
\eeq
(Here we omit the subscript $+$, but it is understood that $\hat\vartheta_{{\cal H}^+}$ refers to the expansion scalar associated with $\hat k^\alpha$.) 
For the apparent horizon, where $\vartheta^{(1)}_{\cal A} = 0$, it instead implies
\beq
\hat r^{(1)}_{\cal A} = 0.\label{rhat1A}
\eeq
In words, the event and apparent horizon are identical at first order. The fact that the event horizon's expansion vanishes at this order is well known~\cite{Poisson:04}, although we are unaware of its implications having been spelled out.

Given the above, on both horizons the expansion reduces to
\beq
\vartheta_+ = \eps^2\hat\vartheta^{(2)}_+ + O(\eps^3),
\eeq
with the expression~\eqref{theta2} collapsing to
\begin{multline} 
	\hat\vartheta^{(2)}_+ = \frac{1}{32 M^4} \Big[ 8 M^2(1-D^2) \hat r^{(2)} + 8 M^2\partial_v \hat h^{(2)}_{\circ} \\ - \frac{1}{2}\partial_v\!\left(\Omega^{AC}\Omega^{BD} \hat h^{(1)}_{\<AB\>} \hat h^{(1)}_{\<CD\>}\right) \\ + 2D^B\!\left(\hat h^{(1)}_{vA}\Omega^{AC}\hat h^{(1)}_{\<BC\>}\right)\Big].\label{theta2-horizon-locked} 
\end{multline}
To obtain this form we have written $\hat h^{(1)}_{vA}D^A\hat h^{(1)}_\circ$ as a total divergence minus $D^A\hat h^{(1)}_{vA}\hat h^{(1)}_\circ$ and then appealed to Eq.~\eqref{EFE soln}. Equation~\eqref{theta2-horizon-locked} simplifies further in the case that $\hat h^{(1)}_{vA}=0$, but the important features are present in either case: every term involving the metric perturbation is either a total divergence (which vanishes upon integration over the horizon) or a total time derivative (which vanishes upon averaging over fast time in a two-timescale expansion).

\subsection{Invariant properties of the event horizon}

By construction, our gauge fixing ensures that
\beq\label{rn EH horizon locked}
\hat r^{(n)}_{{\cal H}^+} = 0.
\eeq
This implies that the gauge-fixed metric on ${\cal H}_v^+$, from Eq.~\eqref{intrinsic_metric}, is 
\beq
\hat\gamma^{{\cal H}^+}_{AB} = 4M^2\Omega_{AB} + \eps \hat h^{(1)}_{AB} + \eps^2 \hat h^{(2)}_{AB} + O(\eps^3),
\eeq
where $\hat h^{(n)}_{AB}$ is given in Eqs.~\eqref{hhatABreg trace}--\eqref{hhatABreg STF} or \eqref{hhatAB}. The null vectors $\hat k^\alpha_{{\cal H}^+}$ and $\hat n^\alpha_{{\cal H}^+}$ are given by Eqs.~\eqref{locked horizon generator} [or \eqref{locked horizon generator two-time}] and \eqref{nhat}, and the tangent vectors $(\hat e_{{\cal H}^+}\!)^\alpha_A$ by $\delta^\alpha_A$.

Given this geometry, the scalar curvature of ${\cal H}_v^+$ is given by Eq.~\eqref{Ricci expansion} with the replacement $\breve{\gamma}^{(n)}_{AB}\to \hat{h}^{(n)}_{AB}$. Similarly, the surface area of ${\cal H}_v^+$, given by Eq.~\eqref{area_second_order}, is
\beq
\hat A_{{\cal H}^+} = 16\pi M^2 +\eps A^{(1)}_{{\cal H}^+} + \eps^2 A^{(2)}_{{\cal H}^+} + O(\eps^3),
\eeq
with
\begin{align}
\hat A^{(1)}_{{\cal H}^+} &= \sqrt{4\pi} \hat h^{(100)}_{\circ},\label{A1 EH}\\
\hat A^{(2)}_{{\cal H}^+}	&=\sqrt{4\pi}\hat h^{(200)}_{\circ}\nonumber\\
&\quad - \sum_{lm}\frac{\lambda_2^2}{32M^2}\left(|\hat h^{(1lm)}_{+}|^2+ |\hat h^{(1lm)}_{-}|^2\right).\label{A2 EH}
\end{align}

In order to calculate the mass of the horizon, we require the expansion scalar. Referring to Eq.~\eqref{theta2-horizon-locked}, we see that
\beq
\hat\vartheta_{{\cal H}^+} = \eps^2 \hat\vartheta^{(2)}_{{\cal H}^+} +O(\eps^3),
\eeq
with
\begin{multline} 
	\hat\vartheta^{(2)}_{{\cal H}^+} =\frac{1}{32 M^4} \Big[8 M^2\partial_v \hat h^{(2)}_{\circ}+ 2D^B\!\left(\hat h^{(1)}_{vA}\Omega^{AC}\hat h^{(1)}_{\<BC\>}\right) \\ - \frac{1}{2}\partial_v\!\left(\Omega^{AC}\Omega^{BD} \hat h^{(1)}_{\<AB\>} \hat h^{(1)}_{\<CD\>}\right)\Big].\label{theta2-horizon-locked EH} 
\end{multline}
Therefore the event horizon's Hawking mass~\eqref{M_H EH} reads
\beq\label{M_H EH horizon-locked}
\hat M^{{\cal H}^+}_{\rm H} = \hat M^{{\cal H}^+}_{\rm irr} - \frac{\eps^2M^2}{2\pi}\int \vartheta^{(2)}_{{\cal H}^+} d\Omega +O(\eps^3),
\eeq
where $\hat M^{{\cal H}^+}_{\rm irr}=\sqrt{\hat A_{{\cal H}^+}/(16\pi)}$ is the irreducible mass of the event horizon. It will also be helpful, when comparing to the apparent horizon's mass in the next section, to write this as
\beq\label{M_H EH horizon-locked v2}
\hat M^{{\cal H}^+}_{\rm H} = \hat M^{{\cal H}^+}_{\rm irr} - 2\eps^2M^2 \vartheta^{(2)}_{{\cal H}^+,00}Y^{00} +O(\eps^3).
\eeq
The monopolar piece of the expansion scalar simplifies to the total time derivative
\begin{multline}
	\vartheta^{(2)}_{{\cal H}^+,00}Y^{00} = \frac{1}{128M^4} \partial_v\Big[32M^2\hat h^{(200)}_\circ Y^{00} \\
	- \sum_{lm}\frac{\lambda_2^2}{4\pi}\left(|\hat h_{+}^{(1lm)}|^2+|\hat h_{-}^{(1lm)}|^2\right)\Big].\label{theta200 EH gauge fixed}
\end{multline}
Note that since the expansion is  nonnegative (by the area theorem), the Hawking mass is  less than or equal to the irreducible mass. We see that in stationary cases, the expansion vanishes, as we would expect, and the two types of mass become equal to one another.

So far in this section we have implicitly worked in regular perturbation theory. In the context of a two-timescale expansion, all of the above equations hold true except that $\hat\vartheta^{(2)}_{{\cal H}^+}$ picks up an additional term from the $v$ derivative in Eq.~\eqref{theta1lm gauge fixed},
\beq
\delta\hat\vartheta^{(2)}_{{\cal H}^+} = \frac{\partial_{\tilde v} \hat h^{(1)}_{\circ}}{4M^2}.
\eeq
We can use this to obtain a simplified relation between the slowly varying parts of the Hawking and irreducible masses, $\langle M^{{\cal H}^+}_{\rm H} \rangle$ and $\langle M^{{\cal H}^+}_{\rm irr} \rangle$. 
Since Eq.~\eqref{theta200 EH gauge fixed} is a total time derivative, it implies 
\beq
\<\hat\vartheta^{(2)}_{{\cal H}^+}\> = \<\delta\hat\vartheta^{(2)}_{{\cal H}^+}\> = \frac{1}{4M^2}\partial_{\tilde v}\< \hat h^{(1)}_{\circ}\>. 
\eeq
We then have 
\beq
\langle M^{{\cal H}^+}_{\rm H}\rangle = \langle \hat M^{{\cal H}^+}_{\rm irr}\rangle - \frac{\eps^2}{8\pi}\partial_{\tilde v}\int\hat h^{(1)}_{\circ} d\Omega +O(\eps^3).
\eeq
The integral picks out the spherically symmetric piece of $h^{(1)}_{\circ}$, and the slow-time derivative picks out the quasistationary piece. Such a piece is necessarily a perturbation toward another Schwarzschild black hole, as given in Eq.~\eqref{EF mass perturbation}, with $\delta M$ a function $\tilde v$. According to Eq.~\eqref{<h1_circ> gauge fixed}, this corresponds to a gauge-fixed perturbation
\beq
\hat h^{(1)}_{\circ} = 8M\delta M, 
\eeq
and so
\beq\label{<MH> vs <Mirr> EH}
\langle \hat M^{{\cal H}^+}_{\rm H}\rangle = \langle \hat M^{{\cal H}^+}_{\rm irr}\rangle -\frac{\eps^2}{\kappa_0}\frac{d\delta M}{d\tilde v} + O(\eps^3).
\eeq
We may also note, from Eq.~\eqref{M1 AH} and the surrounding discussion,  that 
\beq
\left\langle\frac{dM_{\rm irr}}{dv}\right\rangle = \eps^2 \frac{d\delta M}{d\tilde v} + O(\eps^3),
\eeq
regardless of which horizon $M_{\rm irr}$ refers to or whether we use its horizon-locked variant. Therefore Eq.~\eqref{<MH> vs <Mirr> EH} can be written as
\beq
\langle \hat M^{{\cal H}^+}_{\rm H}\rangle = \langle \hat M^{{\cal H}^+}_{\rm irr}\rangle -\frac{1}{\kappa_0}\left\langle\frac{d\hat M^{{\cal H}^+}_{\rm irr}}{dv}\right\rangle + O(\eps^3).\label{MHH vs MHirr}
\eeq
We recognize this as precisely the form of the equation governing the horizon location, Eq.~\eqref{generic_eq_rad}. 

In principle we can convert Eq.~\eqref{MHH vs MHirr} into a teleological integral relationship:
\beq
\langle\hat M^{{\cal H}^+}_{\rm irr}\rangle = \kappa_0 \int_v^\infty e^{-\kappa_0(v'-v)}\langle \hat M^{{\cal H}^+}_{\rm H}\rangle dv' +O(\eps^3).
\eeq
But it should be noted that our derivation only applies well before merger, in the sense described in Sec.~\ref{sec:localization}. So a cutoff should be imposed on the upper limit of integration in this relationship.

\subsection{Invariant properties of the apparent horizon}

Our horizon-locking condition does not place the apparent horizon at the coordinate location $r=2M$. Instead, with the event horizon at $r=2M$, the gauge-fixed radius of the apparent horizon is
\beq\label{rAH horizon locked}
\hat r_{\cal A} = 2M + \eps^2 \hat r^{(2)}_{\cal A} + O(\eps^3),
\eeq
where we have used the result~\eqref{rhat1A} to set $\hat r^{(1)}_{\cal A}=0$. We can also write this as
\beq\label{rAH vs rEH horizon locked}
\hat r_{\cal A} = \hat r_{{\cal H}^+} + \eps^2 \hat r^{(2)}_{\cal A} + O(\eps^3).
\eeq
We stress that the equality at linear order, $ r^{(1)}_{\cal A} = r^{(1)}_{{\cal H}^+}$, is an invariant statement. Since $r^{(1)}_{\cal A}$ and $r^{(1)}_{{\cal H}^+}$ transform in the same way under a gauge transformation, their equality in the horizon-locked gauge implies their equality in all gauges. The second-order term, $\hat r^{(2)}_{\cal A}$, then provides an invariant measure of the apparent horizon's displacement from the event horizon.

The second-order radial displacement follows from Eq.~\eqref{theta2-horizon-locked} with $\vartheta^{(2)}_{+,\cal A}=0$, or equivalently from Eq.~\eqref{r2lm AH}. We write the result in terms of the expansion scalar of the event horizon:
\begin{align}\label{rhat2lm AH}
	\hat r^{(2)}_{{\cal A},lm}(v) = -\frac{4M^2\hat\vartheta^{(2)}_{{\cal H}^+,lm}}{1+\lambda_1^2}.
\end{align}
The modes of the expansion are given by
\begin{align}
\hat\vartheta^{(2)}_{{\cal H}^+,lm} &=  \frac{\partial_v \hat h^{(2lm)}_\circ}{4M^2} \nonumber \\
	&\quad + \sum_{\substack{l'm'\\l''m''}}\Big[ {}^{(2)}\hat\Theta_{l'm'l''m''}^{lm}(v) C_{l' m' 2 l'' m'' - 2}^{lm0} \nonumber\\  &\qquad\qquad + {}^{(1)}\hat\Theta_{l'm'l''m''}^{lm}(v) C_{l' m' 1 l'' m'' -1}^{lm0}\Big]
\end{align}
with the explicit coefficients given in Eq.~\eqref{theta2lm coefficients gauge fixed}.
This formula for $\hat\vartheta^{(2)}_{{\cal H}^+,lm}$ can be obtained directly from Eq.~\eqref{theta2lm} or from Eq.~\eqref{theta2-horizon-locked EH}. Since the event horizon's expansion is nonnegative, these relationships suggest that $\hat r^{(2)}_{{\cal A},lm}$ is negative or zero; the apparent horizon lies inside the event horizon, as it should. 

Equation~\eqref{rAH horizon locked} implies that the gauge-fixed metric on ${\cal A}_v$, given by Eq.~\eqref{intrinsic_metric}, is 
\begin{multline}
\hat\gamma^{{\cal A}}_{AB} = 4M^2\Omega_{AB} + \eps \hat h^{(1)}_{AB} + \eps^2\left(\hat h^{(2)}_{AB}+4M\hat r^{(2)}_{\cal A}\Omega_{AB}\right)\\ + O(\eps^3).
\end{multline}
Or in terms of the metric on ${\cal H}_v^+$,
\beq
\hat\gamma^{{\cal A}}_{AB} = \hat\gamma^{{\cal H}^+}_{AB} +4M\eps^2\hat r^{(2)}_{\cal A}\Omega_{AB} + O(\eps^3).\label{gamma AH vs gamma EH gauge fixed}
\eeq
From Eqs.~\eqref{k1}--\eqref{n2} and \eqref{basis vectors}, the basis vectors on ${\cal A}_v$ are given by
\begin{align}
    \hat k^\alpha_{{\cal A}}\partial_\alpha &= \hat k^\alpha_{{\cal H}^+}\partial_\alpha +\eps^2 \frac{\hat r^{(2)}_{\cal A}}{4M}\partial_r+O(\eps^3),\\
   \hat n^\alpha_{{\cal A}}\partial_\alpha &= \hat n^\alpha_{{\cal H}^+}\partial_\alpha+O(\eps^3),\\
    (\hat e_{{\cal A}})^\alpha_A\,\partial_\alpha &= (\hat e_{{\cal H}^+})^\alpha_A\,\partial_\alpha + \eps^2 D_A \hat r^{(2)}_{\cal A}\partial_r +O(\eps^3).
\end{align}
Note that since these quantities have all been pulled back to ${\cal H}^0$, the relationships between them do not involve comparing tensors at different points.

This metric's intrinsic curvature, from Eq.~\eqref{Ricci expansion}, can be written in terms of the event horizon's curvature as
\beq
{\cal R}_{\cal A} = {\cal R}_{{\cal H}^+} + \frac{\eps^2}{4M^3}\sum_{lm}\mu^2 \hat r^{(2)}_{{\cal A},lm} + O(\eps^3),\label{R AH vs R EH gauge fixed}
\eeq
where, recall, $\mu^2:=(l+2)(l-1)$. Similarly, the surface area of ${\cal A}_v$, from Eq.~\eqref{area_second_order}, is
\begin{align}
\hat A_{\cal A} = \hat A_{{\cal H}^+} + 16\pi M\hat r^{(2)}_{{\cal A},00}Y^{00}.\label{A2hat AH}
\end{align}	
This involves the monopolar correction to the apparent horizon  radius, which reads 
\beq
	\hat r^{(2)}_{{\cal A},00} = -4M^2\hat\vartheta^{(2)}_{{\cal H}^+,00},
\eeq
with Eq.~\eqref{theta200 EH gauge fixed}.

From the surface area, with Eq.~\eqref{irreducible_mass_second}, we can read off the second-order contribution to the apparent horizon's irreducible mass:
\beq
\hat M^{(2)}_{{\rm irr},{\cal A}} = \hat M^{(2)}_{{\rm irr},{\cal H}^+}  - 2M^2\hat \vartheta^{(2)}_{{\cal H}^+,00}Y^{00}.\label{Mirr AH vs Mirr EH gauge fixed}
\eeq
Comparing this result for the irreducible mass to Eq.~\eqref{M_H EH horizon-locked v2}, we observe that it is precisely the second-order contribution to the Hawking mass of the event horizon. Since $\hat M^{\cal A}_{\rm irr} = \hat M^{\cal A}_{\rm H}$, this implies that the two horizons have identical Hawking masses through second order:
\beq\label{MH AH = MH EH}
\hat M^{\cal A}_{\rm H} = \hat M^{{\cal H}^+}_{\rm H}+O(\eps^3).
\eeq
Because the event horizon is larger than or equal to the apparent horizon, its irreducible mass is slightly larger (by an amount of order $\eps^2$) than the apparent horizon's. However, the event horizon's Hawking mass is slightly smaller than its irreducible mass, and here we see that the slight difference is precisely the same as the difference between the two horizons' irreducible mass. 

All of the above applies in regular perturbation theory. In the context of a two-timescale expansion, it all remains valid except that the apparent horizon radius picks up the additional term~\eqref{delta r2lm AH}, which we reproduce here in terms of the gauge-fixed metric perturbation,
\beq
\delta\hat r^{(2)}_{{\cal A},lm} = -\frac{\partial_{\tilde v} \hat h^{(1lm)}_\circ}{1+\lambda_1^2}.
\eeq
Since the contribution in Eq.~\eqref{rhat2lm AH} is a total fast-time derivative, $\delta\hat r^{(2)}_{{\cal A},lm}$ is the only contribution that does not average to zero. That is,
\beq
\< \hat r^{(2)}_{{\cal A},lm}\> = \<\delta\hat r^{(2)}_{{\cal A},lm}\> = -\frac{\partial_{\tilde v}\big\langle\hat h^{(1lm)}_\circ\big\rangle}{1+\lambda_1^2}.\label{<r AH> gauge fixed}
\eeq

The averaged horizon radius allows us to derive averaged equations for the apparent horizon area and mass. Following the same arguments as led to Eq.~\eqref{<MH> vs <Mirr> EH}, we find
\beq
\<\hat r^{(2)}_{{\cal A},00}\>Y^{00} = \< \delta \hat r^{(2)}_{{\cal A},00}\>Y^{00} = -8M\frac{d\delta M}{d\tilde v}.\label{<delta r200> gauge fixed}
\eeq
Equation~\eqref{A2hat AH} then tells us that the slowly varying piece of the horizon area is 
\beq
\left\langle\hat A_{\cal A}\right\rangle = \left\langle\hat A_{{\cal H}^+}\right\rangle -\frac{32M\pi \eps^2}{\kappa_0}\frac{d\delta M}{d\tilde v} +O(\eps^3),\label{<A AH> vs <A EH> gauge fixed}
\eeq
and the slowly varying irreducible mass is
\beq
\<\hat M^{\cal A}_{\rm irr}\> = \<\hat M^{{\cal H}^+}_{\rm irr}\> -\frac{1}{\kappa_0}\left\langle\frac{d\hat M^{{\cal H}^+}_{\rm irr}}{dv} \right\rangle + O(\eps^3).\label{<Mirr AH> vs <Mirr EH> gauge fixed}
\eeq
Of course, since Eq.~\eqref{MH AH = MH EH} is valid even without averaging, it is also valid on average:
\beq
\<\hat M^{\cal A}_{\rm H}\> = \<\hat M^{{\cal H}^+}_{\rm H}\>+O(\eps^3).
\eeq


\section{Case study: quasicircular binaries}\label{sec:quasicircular_binaries}

In this section, we apply our results in the particular case studied in Ref.~\cite{Pound_2020}: a quasicircular binary. We start by summarizing key properties of the metric in this scenario. We then describe Ref.~\cite{Pound_2020}'s calculation of the apparent horizon's irreducible mass, and we present new results for the numerical difference between the two horizons' irreducible masses. We conclude by presenting a visualization of the horizon at first order, specifically highlighting the motion of the black hole, which has not been accounted for in previous visualizations in the literature.

\subsection{Two-timescale character}

The two-timescale expansion for quasicircular binaries was detailed in Ref.~\cite{miller2020twotimescale}. The orbiting companion is placed on a quasicircular orbit 
\beq\label{xp}
x^\mu_p(t,\eps) = [t,r_p(\eps t,\eps),\pi/2,\phi_p(t,\eps)],
\eeq
where the subscript $p$ refers to ``particle"; at leading order in the system's mass ratio $m/M$, the companion can be represented as a point mass. In the parametrization~\eqref{xp}, the orbital radius is slowly decaying, and the orbital phase is the integral of a slowly increasing orbital frequency: \beq
\phi_p(t,\eps) = \int^t_0 \Omega(\eps t')dt'.
\eeq
On time scales much shorter than  $M/\sqrt{\eps}$, this phase can be approximated by the geodesic phase,
\beq\label{phi_p approx}
\phi_p(t,\eps) = \Omega_0 t + O(\eps t^2),
\eeq
where $\Omega_0 := \Omega(0) = \sqrt{M/r_p(0)^3}$. As a consequence, the time $M/\sqrt{\eps}$ is referred to as the dephasing time. In addition to determining the phase evolution, the orbital frequency provides a convenient parameterization of the orbital radius, and we can write $r_p=r_0(\Omega)+O(\eps)$, where $r_0(\Omega):=(M\Omega)^{2/3}$ is the geodesic relationship.

The quasicircular orbit is linked to a distinct two-timescale behavior of the metric perturbation. We can see how this comes about from the leading-order stress-energy tensor of the companion, which reads
\begin{equation}\label{stress_energy_quasicircular}
	T^{\alpha \beta} = \frac{m u^\alpha u^\beta}{u^t r_p^2} \delta(r-r_p) \delta\big(\theta- \pi/2\big) \delta(\phi - \phi_p).
\end{equation}
Here 
\beq
u^\alpha := \frac{dx^\alpha_p}{d\tau} = u^t \left(1,\eps \frac{dr_p}{d\tilde t},0,\Omega\right)
\eeq
is the four-velocity of the particle, normalized to satisfy $g^{(0)}_{\alpha\beta}u^\alpha u^\beta=-1$, and $\tilde t = \eps t$. The completeness relation $\delta(\theta-\pi/2)\delta(\phi-\phi_p)=\sum_{lm}Y^{lm}(\theta^A)\bar Y^{lm}(\pi/2,\phi_p)$, together with the fact that 
\beq
\bar Y^{lm}(\pi/2,\phi_p)\propto e^{-im\phi_p}, 
\eeq
immediately implies that each $lm$ mode of the stress-energy has the form of a slowly varying amplitude times a rapidly oscillating phase factor $e^{-im\phi_p}$. The Einstein equations ensure that the metric perturbation inherits this same form. In a neighbourhood of the event horizon, we then arrive at solutions of the form
\begin{subequations}\label{two-timescale hab quasicircular}
\begin{align}
    h^{(n)}_{ab} &= \sum_{lm} h^{(nml)}_{ab}Y^{lm}\,e^{-im\phi_p(v,\eps)},\\
	h^{(n)}_{\circ} &= \sum_{lm} h^{(nlm)}_{\circ} Y^{lm}\,e^{-im\phi_p(v,\eps)},\\
	h^{(n)}_{aA} &= \sum_{lm} \left(h^{(nlm)}_{a+} Y^{lm}_A + h^{(nlm)}_{a-} X^{lm}_A\right)e^{-im\phi_p(v,\eps)},\\
	h^{(n)}_{\langle AB\rangle } &= \sum_{lm}\left( h^{(nlm)}_{+} Y^{lm}_{AB} +h^{(nlm)}_{-} X^{lm}_{AB}\right)e^{-im\phi_p(v,\eps)},
\end{align}
\end{subequations}
where the coefficients are functions of $(r,\tilde v)$. This is the two-timescale expansion~\eqref{two-timescale h quasicircular} with the identification of the fast-time phase variable as $\varphi_\phi = \phi_p$. The periodic dependence on $\phi_p(t,\eps)$ at the particle translates into the same periodic dependence on $\phi_p(v,\eps)$ at the horizon. The slow-time dependence of the mode amplitudes, such as $h^{(nml)}_{ab}(r,\tilde v)$, represents a dependence on the orbital frequency $\Omega$, the perturbation $\delta M(\tilde v)$ to the black hole's mass, and an analogous perturbation $\delta J(\tilde v)$ to the black hole's spin. Following Ref.~\cite{Pound_2020}, in this paper we restrict ourselves to ``moments" of slow time at which $\delta J$ vanishes.

Beyond its general two-timescale character, the metric perturbation in this scenario has an important feature: it only depends on $\phi$ and $\phi_p$ in the combination $(\phi-\phi_p)$. This means that averages over the horizon surface ${\cal H}_v$ are also  averages over fast time. Quantities such as the horizon area and mass therefore have no oscillatory parts, meaning
\beq\label{A=<A>}
A = \<A\> \quad \text{and}\quad M_{\rm BH} = \<M_{\rm BH}\>. 
\eeq
Here ``$M_{\rm BH}$" can be either the irreducible or Hawking mass. Another way of saying this is that the black hole's area and mass are approximately conserved quantities: they only change very slowly, on a time scale much longer than the orbital period. We can view this as a consequence of the spacetime possessing an approximate Killing vector. Since the metric only depends on the combination $(\phi-\phi_p)$, the helical vector $K^\alpha\partial_\alpha = \partial_t + \Omega\, \partial_\phi$ satisfies ${\cal L}_\xi g_{\alpha\beta} = O(\eps^2)$. However, the event horizon is not a Killing horizon at this order, as $K^\alpha$ is not null at the horizon.

As discussed in Sec.~\ref{sec:localization}, our two-timescale approximation and our treatment of the event horizon both break down as the companion approaches the transition to plunge. For a quasicircular orbit, this implies we must cut off our treatment at a time before the companion reaches the innermost stable circular orbit (ISCO) at $r_0=6M$. 
Estimating an appropriate cutoff would require a careful matching between our two-timescale expansion and a transition-regime expansion near the ISCO. But we can surmise a rough estimate. In the transition regime, the orbital radius has the form $r_p\sim 6M + \eps^{2/5}r^{(1)}_p(\eps^{1/5}t)$~\cite{Ori_2000}, implying that we should switch from our two-timescale expansion to a transition-regime expansion at a radius $r_p(\tilde v_T)$ satisfying $M\gg (r_p-6M)\gtrsim\eps^{2/5}M$. We must then additionally satisfy the cutoff~\eqref{cutoff} to ensure that the transition regime does not significantly impact the event horizon. Since $\frac{dr_p}{dv}\sim \eps$ in the inspiral regime, the cutoff $|v-v_{T}|\gg |\ln\eps|$ translates to $[r_p(\tilde v)-r_p(\tilde v_T)]\gg |\eps \ln\eps|$. However, since $\eps^{2/5}\gg|\eps \ln\eps|$, this secondary restriction has negligible impact. 

In the following sections, we use data for the mode amplitudes $h^{(nml)}_{ab}$, $h^{(nml)}_{a\pm}$, $h^{(nml)}_{\circ}$, and $h^{(nml)}_{\pm}$ that were obtained as a part of the calculations in Ref.~\cite{Pound_2020}. The modes were computed in the Lorenz gauge by solving the two-timescale expansion of the field equations, as described in Ref.~\cite{miller2020twotimescale}. The first-order mode amplitudes satisfy the same field equations as they would in an ordinary frequency-domain treatment, and they were computed using the frequency-domain code described in Ref.~\cite{Akcay:2013wfa}; in this paper we use data up to $l=30$ for a sequence of values of $r_0$ from $6M$ to $25M$. The only second-order mode we use is the monopole mode, $l=m=0$. The explicit details of its computation will be presented elsewhere.

\subsection{Mass of the black hole}

We compute the mass of the black hole in several ways. We first summarize the calculation as performed in Ref.~\cite{Pound_2020}:
\begin{enumerate}
    \item Compute the $lm$ modes of the first-order perturbation of the horizon radius, $r^{(1)}_{{\cal A},lm}=r^{(1)}_{{\cal H}^+,lm}$, from either Eq.~\eqref{first_rad_ah} or \eqref{r1k EH}.\footnote{We have confirmed that the two formulas yield identical numerical results.} For quasicircular orbits, the latter reduces to
     \begin{equation}\label{r1lm EH and AH}
    	r^{(1)}_{{\cal A},lm} = r^{(1)}_{{\cal H}^+,lm} = \frac{2M h^{(1lm)}_{vv}}{1+4iM\Omega_m},
 	\end{equation}
 	with $\Omega_m:=m\Omega$. Here $r^{(n)}_{{\cal H},lm}(\tilde v)$ is the coefficient in the expansion \begin{align}
        r^{(n)}_{\cal H} =\sum_{lm} r^{(n)}_{{\cal H},lm}(\tilde v)e^{-im\phi_p(v,\eps)}Y^{lm}.
    \end{align}
	\item Compute the monopolar second-order correction to the apparent horizon radius, $r^{(2)}_{{\cal A},00}$, using Eq.~\eqref{r200 AH explicit}. In this formula, $\partial_v = \mp i\Omega_m$ or 0, depending on whether it acts on an unconjugated quantity ($-$), a conjugated quantity ($+$), or an absolute value ($0$). Note that since $r^{(2)}_{00}$ is an $m=0$ mode, it does not depend on the phase $\phi_p$; in Eq.~\eqref{r200 AH explicit}, all terms in the sum come as pairs $\chi_{lm} \bar\psi_{lm}$, meaning the fast-time phase factors $e^{\pm im\phi_p}$ cancel.
    \item Compute the correction $\delta r^{(2)}_{{\cal A},00}$ from Eq.~\eqref{delta r2lm AH}. In the Lorenz gauge, this correction can be found analytically to agree with the gauge-fixed formula \eqref{<delta r200> gauge fixed}: 
    \beq\label{delta r200 Lorenz}
	\delta r^{(2)}_{00}Y^{00} = -8M\frac{d\delta M}{d\tilde v}.
	\eeq
	This is not true in all gauges; Eq.~\eqref{<delta r200> gauge fixed} is gauge invariant, but Eq.~\eqref{delta r2lm AH} is not.
	\item Compute the intrinsic metric~\eqref{intrinsic_metric} on the apparent horizon using \eqref{gamma harmonic expansions} with ~\eqref{gamma1 harmonic expansions} and \eqref{gamma2 harmonic expansions}. Note that only the monopole, trace mode $\breve\gamma^{(200)}_\circ$ is required at second order.
	\item Compute the surface area and irreducible mass of the apparent horizon using Eqs.~\eqref{area_second_order} and \eqref{irreducible_mass_second}.
\end{enumerate}

Given our results in this paper, we can now carry out several alternative calculations. First, we can calculate the irreducible mass of the event horizon using identical steps, simply replacing $r^{(2)}_{{\cal A},00}$ with  $r^{(2)}_{{\cal H}^+,00}$, as given in Eq.~\eqref{r2k EH} with Eq.~\eqref{second_rad_eh}. Because $h^{(n00)}_{vv}$ vanishes at $r=2M$ in the Lorenz gauge, this formula reduces to
\beq
r^{(2)}_{{\cal H}^+,00} =  \frac{F^{(2)}_{00}}{1+4iM\Omega_m}
\eeq
with $F^{(2)}_{00}$ as given in Eq.~\eqref{second_rad_00_eh}. Note that we do not require an analogue of Eq.~\eqref{delta r200 Lorenz} in this case because Eq.~\eqref{r2k EH} already contains the complete second-order contribution to $r^{(2)}_{{\cal H}^+}$ in the two-timescale expansion.

Another alternative is to utilize our gauge-fixed expressions:
\begin{enumerate}
    \item Compute the gauge-fixed perturbations $\hat h^{(1lm)}_\circ$ and $\hat h^{(1lm)}_{\pm}$ on the horizon, from Eqs.~\eqref{jhat1 modes} and \eqref{<h1_circ> gauge fixed}. Here $j^{(1{\bm k})} \to h^{(1lm)}$ for $m\neq0$ modes and $\<h^{(1lm)}\> = h^{(1l0)} \delta_0^m$.
	\item Compute the monopole mode of the gauge-fixed second-order perturbation on the horizon,
	\beq\label{hhatcirc200}
	\hat h^{(200)}_\circ = H^{(200)}_\circ + 8M^2 H^{(200)}_{vv},
	\eeq
	with $H^{(200)}_\circ$ and $H^{(200)}_{vv}$ as given in Eqs.~\eqref{Hcirc200} and \eqref{Hvv200}.
    \item Compute the event horizon's area using Eq.~\eqref{A2 EH} and its irreducible mass $\hat M^{(2)}_{{\cal H}^+,\rm irr}$ using Eq.~\eqref{irreducible_mass_second}.
	\item If desired, add the correction in Eq.~\eqref{<MH> vs <Mirr> EH} to obtain the Hawking mass of the event horizon or, equivalently, the irreducible mass of the apparent horizon:
	\beq
	\hat M^{(2)}_{{\cal H}^+,\rm H}  = \hat M^{(2)}_{{\cal A},\rm irr} = \hat M^{(2)}_{{\cal H}^+,\rm irr} -4M\frac{d\delta M}{d\tilde v}.
	\eeq	
\end{enumerate}

\begin{figure}[t]
	\includegraphics[width=\columnwidth]{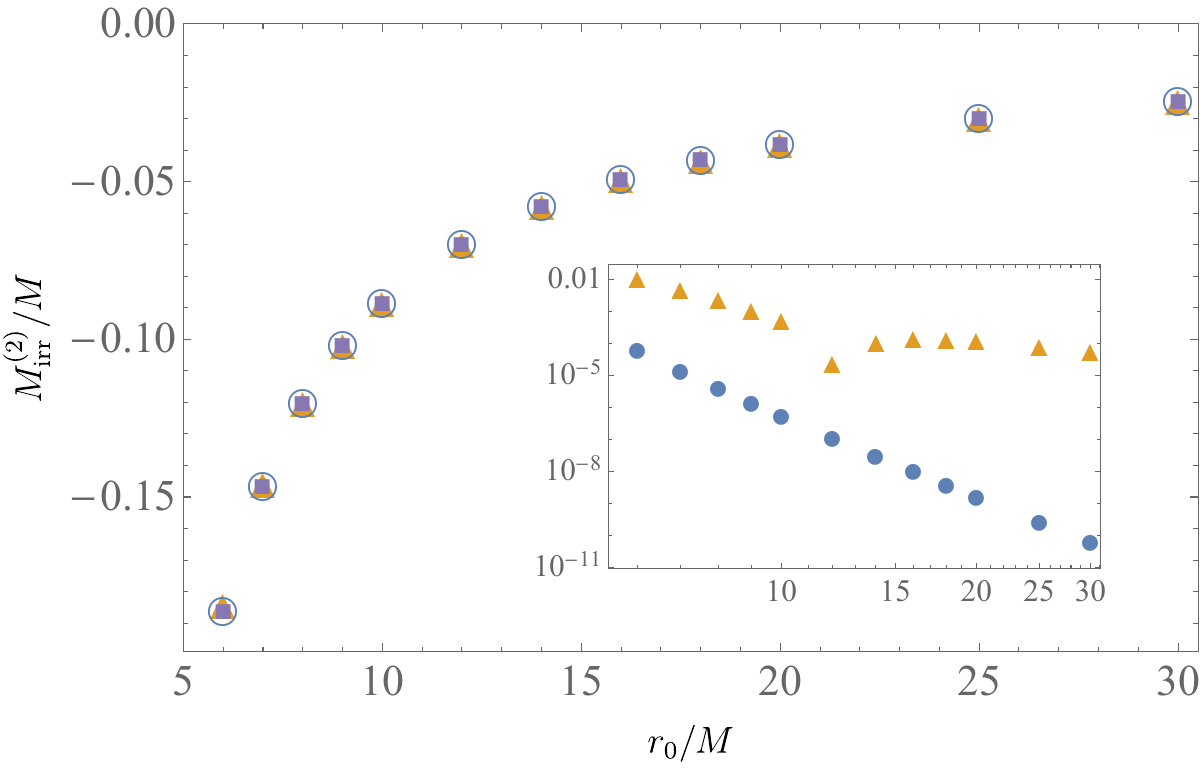}
	\caption{Second-order contribution to the black hole's irreducible mass as a function of orbital radius. The upper plot shows values as measured from the apparent horizon in the Lorenz gauge ($M^{(2)}_{{\cal A},\rm irr}$, solid purple squares), from the apparent horizon with gauge fixing ($\hat M^{(2)}_{{\cal A},\rm irr}$, solid orange triangles), and from the event horizon in the Lorenz gauge ($M^{(2)}_{{\cal H}^+,\rm irr}$, open blue circles). The inset shows the relative differences $\left|\left(\hat M^{(2)}_{{\cal A},\rm irr}- M^{(2)}_{{\cal A},\rm irr}\right)/M^{(2)}_{{\cal A},\rm irr}\right|$ (orange triangles) and $\left|\left(M^{(2)}_{{\cal H}^+,\rm irr}- M^{(2)}_{{\cal A},\rm irr}\right)/M^{(2)}_{{\cal A},\rm irr}\right|$ (blue circles).}
	\label{fig:irreducible mass}
	
\end{figure}

Figure~\ref{fig:irreducible mass} shows the result of all three calculations of $M^{(2)}_{{\cal A},\rm irr}$ and $M^{(2)}_{{\cal H}^+,\rm H}$ as a function of $r_0$. However, we must be careful in interpreting the results for two reasons. 

First, the data do not represent the evolution of a single system. The data at each orbital radius was obtained with the black hole's angular momentum set to zero. While we can freely choose to set it to zero at a single moment of an inspiral, we cannot set it to zero throughout the inspiral: as the system evolves, the black hole's spin changes due to gravitational-wave flux through the horizon, which is always nonzero. Hence, instead of representing snapshots over the course of an inspiral, the data from different orbital radii represent snapshots from {\em different} inspirals. In practice, the black hole's spin evolves by a very small amount over an inspiral, so our results are strongly indicative of the behavior in a true evolution, but the conceptual distinction should be kept in mind.

The second reason for caution is that these results on their own do not represent a meaningful correction to the black hole mass. This is because we can always freely add a stationary mass perturbation to the second-order metric perturbation (i.e., a perturbation toward another Schwarzschild solution). Since this perturbation is a vacuum solution, regular at the horizon and at infinity, there is no restriction on it. The method used to calculate $h^{(2)}_{\alpha\beta}$ in Ref.~\cite{Pound_2020}, which involves integrating the second-order source against a retarded Green's function, picks out a particular solution, but we do not have a specific measure of how much of that solution corresponds to a perturbation toward another Schwarzschild metric. A meaningful measurement would only come from measuring differences between masses.  
The calculation in Ref.~\cite{Pound_2020} provides an example in the calculation of the binary's gravitational binding energy. The binding energy is the difference between the system's Bondi mass and the two components' total rest mass. In that difference, perturbations toward another Schwarzschild solution cancel out, as they contribute an identical amount to $M_{\rm Bondi}$ as to $M_{\rm BH}$.

Nevertheless, the results in Fig.~\ref{fig:irreducible mass} do contain valuable information. They show that the gauge-fixed mass differs from the mass in the Lorenz gauge by a quite small amount, despite the fact that the mass is a foliation-dependent quantity. Likewise, they show that the irreducible mass of the event horizon is virtually indistinguishable from that of the apparent horizon. We must also keep in mind that the absolute differences between these quantities are multiplied by $\eps^2$, making them truly tiny in a realistic small-mass-ratio binary. Moreover, our calculations have established that the difference between the irreducible masses of the two horizons is invariant (given the choice of foliation in Sec.~\ref{sec:horizon locking}): according to Eq.~\eqref{<MH> vs <Mirr> EH} or \eqref{<Mirr AH> vs <Mirr EH> gauge fixed},  the (averaged) irreducible masses of the two horizons always differ by $4M\frac{d\<M_{\rm BH}\>}{dv} = 4M\eps^2{\cal F}_{\cal H}$, where ${\cal F}_{\cal H}$ is the dimensionless flux of energy into the black hole, normalized by $\eps^2$. ${\cal F}_{\cal H}$ is numerically small for all orbital configurations~\cite{Fujita:2009us}, suggesting that even aside from the suppression by $\eps^2$, the two measures of black hole mass always agree to several significant digits. While this is a generally accepted conclusion from full numerical relativity simulations, our results provide a complementary, sharper statement in the perturbative context.

This robustness of the black hole mass has practical implications. It implies that the binding energy computed in Ref.~\cite{Pound_2020} is largely insensitive to the definition of black hole mass. That in turn suggests that the waveforms currently being generated using the binding energy~\cite{Wardell-etal:2021} are similarly robust.

\subsection{Motion and teleology of the horizon} \label{visualizing embedding}

\begin{figure}[tb]
     \captionsetup[subfigure]{labelformat=empty}
     \begin{subfigure}[b]{0.23\textwidth}
         \centering
         \includegraphics[width=\textwidth]{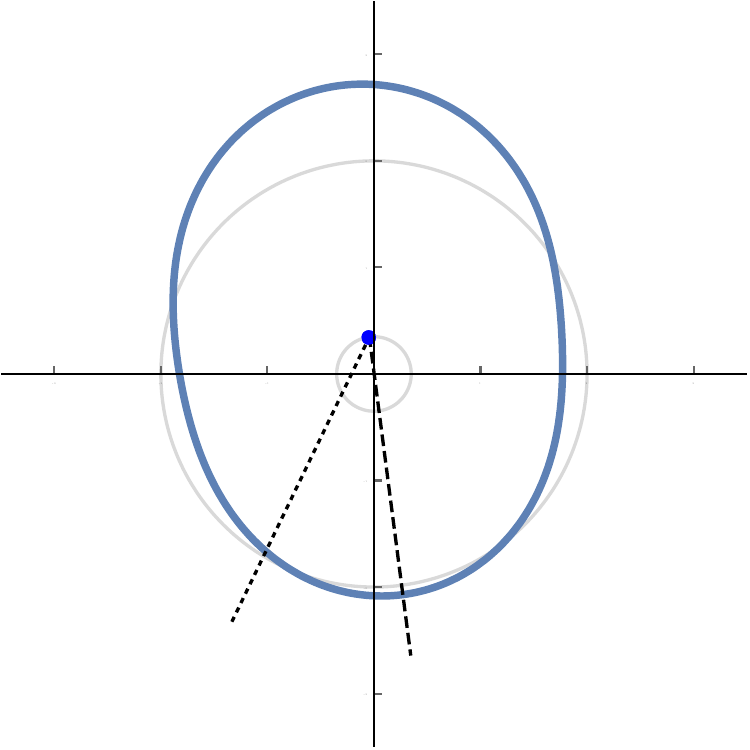}
         \caption{$v=3v_{\text{max}}/4$}
         \label{fig:r073vmax4}
     \end{subfigure}
     \hfill
     \begin{subfigure}[b]{0.23\textwidth}
         \centering
         \includegraphics[width=\textwidth]{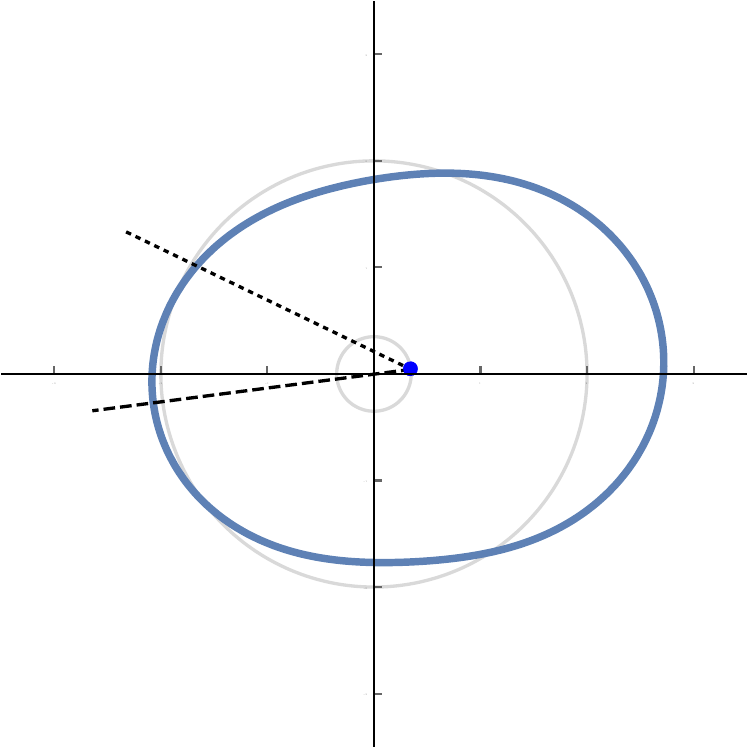}
         \caption{$v=v_{\text{max}}/2$}
         \label{fig:r072vmax4}
     \end{subfigure}\vspace{1em}
     \hfill
      \begin{subfigure}[b]{0.23\textwidth}
         \centering
         \includegraphics[width=\textwidth]{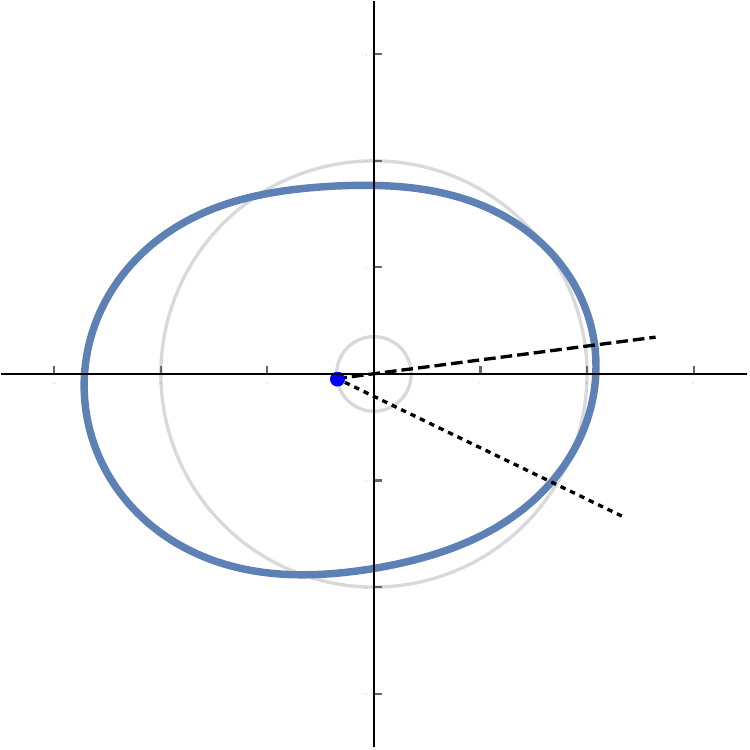}
         \caption{$v=0$}
         \label{fig:r07v0}
     \end{subfigure}
     \hfill
     \begin{subfigure}[b]{0.23\textwidth}
         \centering
         \includegraphics[width=\textwidth]{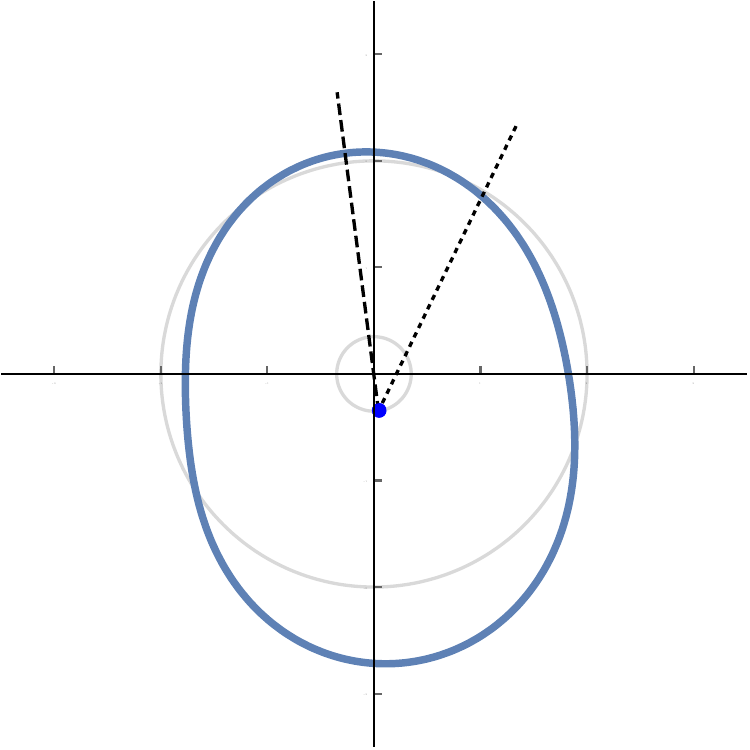}
         \caption{$v=v_{\text{max}}/4$}
         \label{fig:r07vmax4}
       \end{subfigure}
        \caption{Snapshots of the event horizon in the $x$--$y$ plane at four moments of advanced time, with deformations due to a particle in circular orbit at $r_0=7M$ and a mass ratio $\eps=0.1$. Here $v_{\rm max}=2\pi/\Omega$, such that the four snapshots show one complete orbital period. To make the distortion visible on the scale of the plot, we have omitted $m=0$ modes and multiplied $l>1$ modes by a factor of 200. The dotted line points toward the position of the particle at advanced time $v$, while the dashed line points toward the maximal deformation of the horizon (the ``tidal bulge"); the relative angle between them is $31.7^\circ$. The large reference circle represents the unperturbed horizon at $r=2M$. The blue dot indicates the coordinate ``center" of the black hole, which traces out the circular trajectory of radius $r_{\rm BH}=0.350M$ represented by the smaller reference circle.} 
        \label{fig:r07}
\end{figure}

\begin{figure}[tb]
     \centering
     \captionsetup[subfigure]{labelformat=empty}
     \begin{subfigure}[b]{0.23\textwidth}
         \centering
         \includegraphics[width=\textwidth]{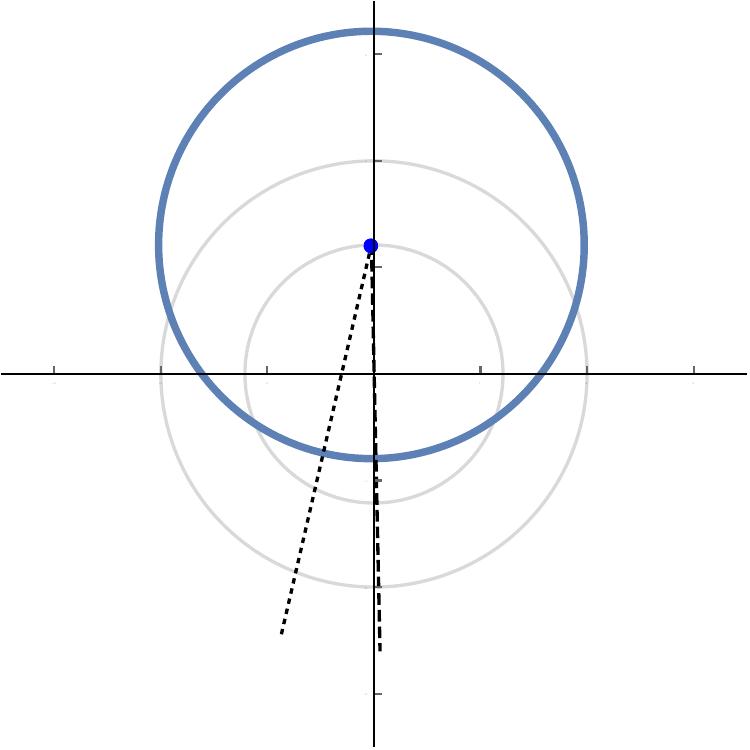}
         \caption{$v=3v_{\text{max}}/4$}
         \label{fig:r0253vmax4}
     \end{subfigure}
     \hfill
     \begin{subfigure}[b]{0.23\textwidth}
         \centering
         \includegraphics[width=\textwidth]{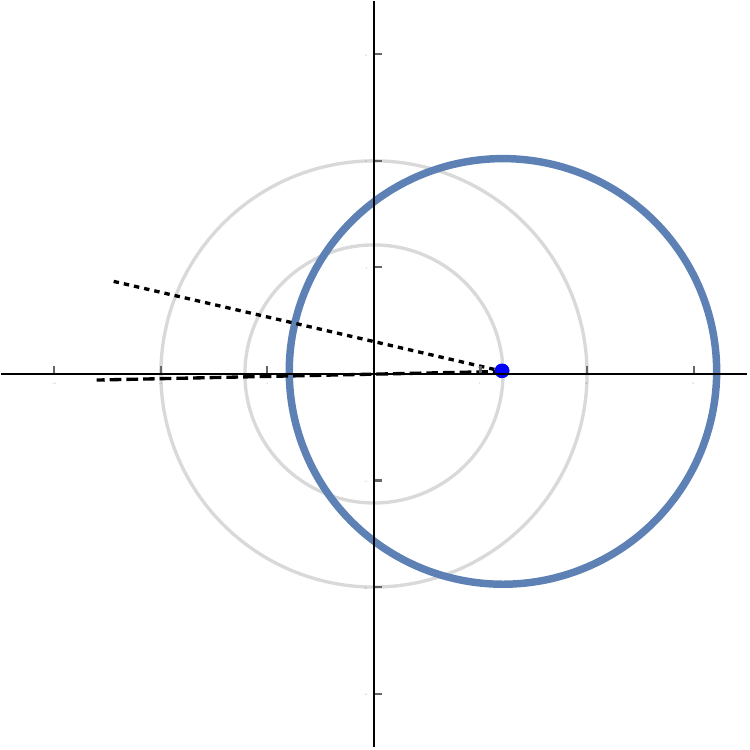}
         \caption{$v=v_{\text{max}}/2$}
         \label{fig:r0252vmax4}
     \end{subfigure}\vspace{1em}
     \hfill
      \begin{subfigure}[b]{0.23\textwidth}
         \centering
         \includegraphics[width=\textwidth]{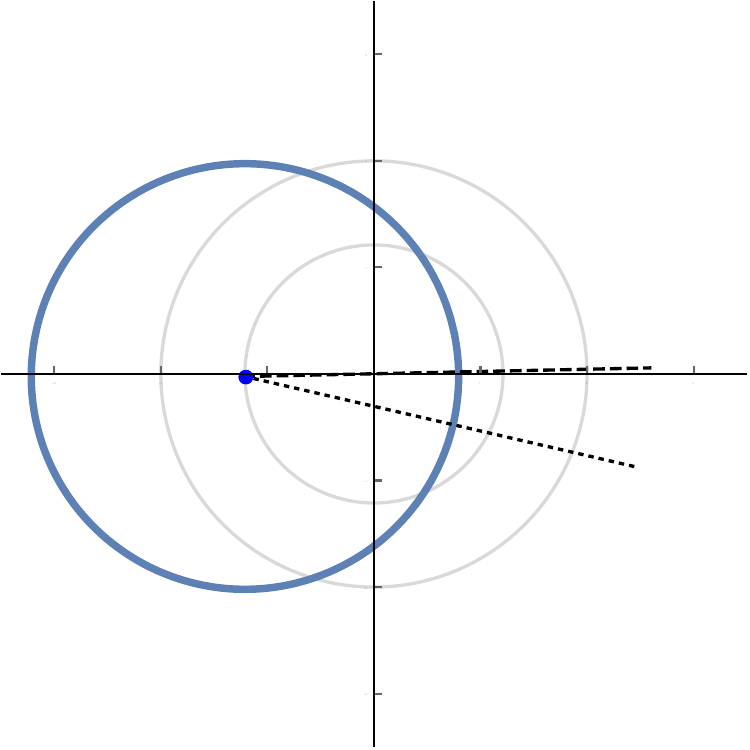}
         \caption{$v=0$}
         \label{fig:r025v0}
     \end{subfigure}
     \hfill
     \begin{subfigure}[b]{0.23\textwidth}
         \centering
         \includegraphics[width=\textwidth]{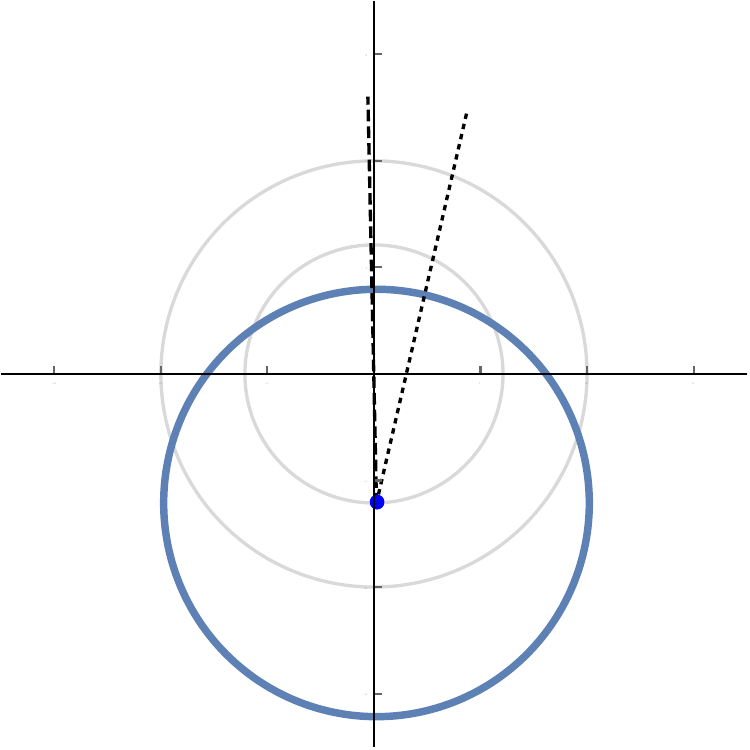}
         \caption{$v=v_{\text{max}}/4$}
         \label{fig:r025vmax4}
     \end{subfigure}
        \caption{Snapshots of the horizon for a particle in circular orbit at $r_0=25M$, with other details as described in the caption of Fig.~\ref{fig:r07}. The relative angle between the tidal bulge and the particle location is $14.5^\circ$, and the radius of the black hole's orbit is $r_{\rm BH}=1.21M$.}
        \label{fig:r025}
\end{figure}

 \begin{figure}[tb]
    \centering
     \begin{subfigure}[tb]{0.23\textwidth}
         \centering
         \includegraphics[width=\textwidth]{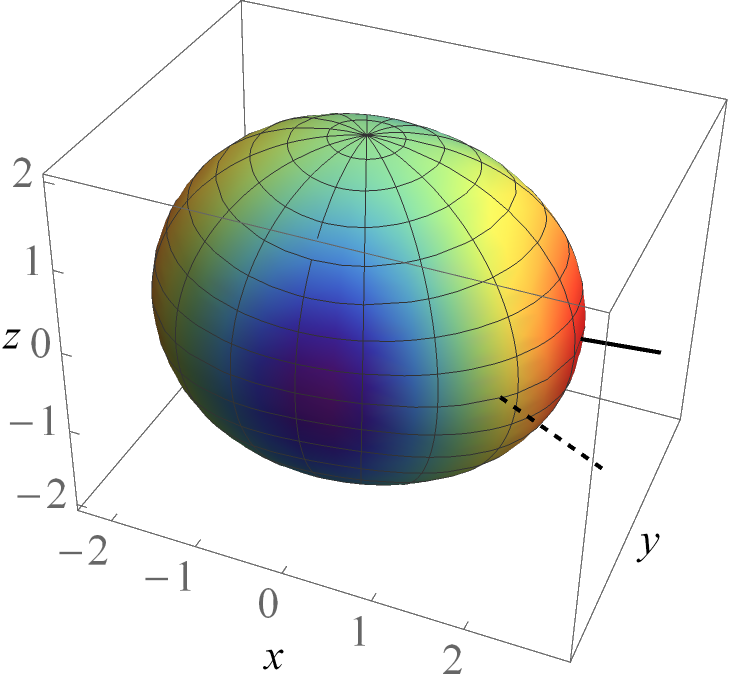}
         \label{fig:3D b}
     \end{subfigure}
     \hfill
     \begin{subfigure}[tb]{0.23\textwidth}
         \centering
         \includegraphics[width=\textwidth]{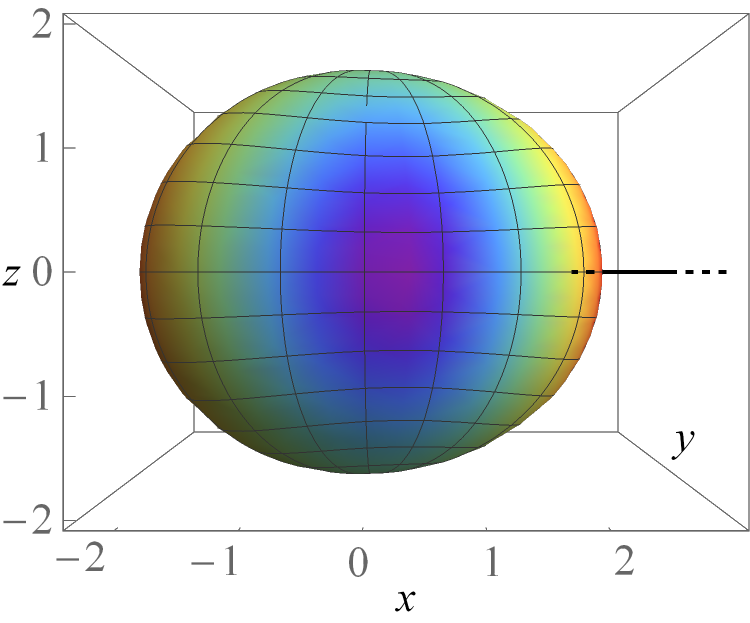}
         \label{fig:3D d}
     \end{subfigure}
        \caption{Three-dimensional visualizations of the horizon surface at $v=0$, with the same parameters as Fig.~\ref{fig:r07}. Here we use the embedding~\eqref{Euclidean embedding v1} rather than \eqref{Euclidean embedding v2}.}
        \label{fig:r07 3D}
\end{figure}

As a final element of our discussion, we present a visualization of the horizon. Unlike the rest of the paper, this section is necessarily restricted to linear perturbations. The reason is simple: the visualization requires modes with $l>0$, and no second-order data is yet available for these modes. Because of this restriction, our presentation here largely reproduces earlier literature. However, we highlight an aspect of the visualization that was omitted in earlier descriptions: the {\em motion} of the black hole. We also point out that features described as teleological in the literature cannot have a truly teleological origin, and we offer an alternative explanation for the behavior.

In this section we also depart from our treatment in other sections in that we specifically focus on behavior on the orbital timescale $1/\Omega$ rather than trying to maintain accuracy on both the orbital and radiation-reaction timescale. This means that we may make the replacement $\varphi_p(v,\eps)\to \Omega v$, as in Eq.~\eqref{phi_p approx}, which is consistent at first order in $\eps$ on the orbital time scale.

\subsubsection{Motion of the black hole}

We first review the visualization method outlined in Refs.~\cite{Vega_2011,Hughes1} (following earlier work in Refs.~\cite{Hartle:1973zz, Hartle:1974gy}). Since the horizon's coordinate radius is inherently gauge dependent, and actually identical to the background horizon's radius in a horizon-locking gauge, the method instead depicts the shape of the horizon based on its intrinsic curvature. To do so, one embeds the horizon  into 3-dimensional Euclidean space. 

The embedding is found by constructing a closed 2-surface in Euclidean space that has the same intrinsic curvature as the horizon. In Cartesian coordinates $x^i=(x,y,z)$, at a given value of $v$, such a surface can be parametrized as
\begin{align}\label{Euclidean embedding v1}
x^i_{\rm E}(\theta^A;v) = r_{\rm E}(\theta^A;v)\Omega^i(\theta^A), 
\end{align}
where $\Omega^i:=(\sin\theta \cos\phi,\sin\theta\sin\phi,\cos\theta)$ is the usual radial unit vector, which satisfies $\delta_{ij}\Omega^i\Omega^j=1$. We write the radius of the surface as
\begin{equation} 
r_{\rm E}(\theta^A;v) = 2M[1 + \rho(\theta^A;v)], \label{r_E}
\end{equation}
and we expand the small correction as
\begin{equation} \label{delta r_E}
\rho(\theta^A;v) = \sum_{lm}\rho_{lm}\,Y^{lm}(\theta^A)e^{-i\Omega_m v}.
\end{equation}

The metric on the surface, induced  from the Euclidean metric $ds^2=\delta_{ij}dx^idx^j$, is
\begin{align}\label{metric of embedded surface}
ds^2 = r^2_{\rm E}(\theta^A;v)d\Omega^2 
= 4M^2\left[1+2\rho(\theta^A;v)\right]d\Omega^2,
\end{align}
where we have discarded the term of order $\rho^2$. Noting that this metric is of the form~\eqref{gamma expansion on H}, we can read off its scalar curvature from Eqs.~\eqref{Ricci expansion} and \eqref{R1lm}, with $\breve{\c}^{(1lm)}_+\to0$ and $\breve{\c}^{(1lm)}_\circ\to 8M^2\rho_{lm}e^{-i\Omega_m v}$. This yields
\begin{align}
R_{\rm E}= \frac{1}{2M^2}\left[1 + \sum_{lm}\mu^2\rho_{lm} Y_{lm}(\theta^A)e^{-i\Omega_m v}\right],
\end{align}
where, recall, $\mu^2:=(l+2)(l-1)$, and we again omit $O(\rho^2)$ terms.

Finally, equating the curvature $R_{\rm E}$ to the intrinsic curvature~\eqref{Ricci expansion} of the horizon, we obtain the radial profile through linear order:
\begin{equation}\label{rElm}
r_{\rm E} = 2M\Bigg[1 + 2\eps M^2\sum_{lm}\frac{{\cal R}^{(1)}_{lm}Y_{lm}(\theta^A)}{(l+2)(l-1)}e^{-i\Omega_m v}\Bigg].
\end{equation}
${\cal R}^{(1)}_{lm}$ is calculated from the coefficients $h^{(1lm)}_{\circ}$ and $h^{(1lm)}_{+}$ in the expansion~\eqref{two-timescale hab quasicircular}, evaluated at $r=2M$, using Eq.~\eqref{R1lm} with Eqs.~\eqref{gamma1 harmonic expansions} and \eqref{r1lm EH and AH}. Since the scalar curvature is invariant at first order, Eq.~\eqref{rElm} then provides an invariant representation of the horizon's radial profile.

There are several things to note about this construction. First, the metric~\eqref{metric of embedded surface} is manifestly identical to the intrinsic metric $\gamma_{AB}$ on ${\cal H}_v$ for a particular choice of gauge, as described in Sec.~\ref{sec:embedding locking}. In that section we imposed the gauge choice only on the slowly varying part of the intrinsic metric, but it can be imposed on the oscillatory part as well. The second thing to note is that because the two horizons are identical at linear order, this visualization represents the apparent horizon as well as the event horizon. Finally, we note that Eq.~\eqref{rElm} has a significant shortcoming as a representation of the horizon. While it provides a meaningful depiction of the horizon's {\em shape}, it does not contain any information about the horizon's {\em location}. In a binary system, both bodies orbit around the system's center of mass, and a faithful representation of the black hole horizon should encode this motion. 

The information about the horizon's location is encoded in the $l=1$ term in the metric perturbation. Since the construction leading to Eq.~\eqref{rElm} only requires that the embedded surface has the same intrinsic geometry as the horizon, and since ${\cal R}$ gets zero contribution from $l=1$ modes, we can in principle add any arbitrary $l=1$ term to $r_{\rm E}$. In previous analyses, the dipole term was simply set to zero, as we have implicitly done in Eq.~\eqref{rElm}. This is a sensible choice if we are only concerned with the horizon's shape, but not if we are concerned with its location. To construct a more faithful representation of the black hole horizon, we should instead replace the embedding~\eqref{Euclidean embedding v1} with
\begin{align}\label{Euclidean embedding v2}
x^i_{\rm E}(\theta^A;v) = x^i_{\rm BH} (v) + r_{\rm E}(\theta^A;v)\Omega^i(\theta^A).
\end{align}
In this representation, at each moment $v$, the surface $r_{\rm E}(\theta^A;v)$ is drawn around the black hole's ``center" $x^i_{\rm BH} (v)$. We must now find a meaningful way of measuring this center.

To see how $x^i_{\rm BH}$ relates to an $l=1$ mode, assume $x^i_{\rm BH}\sim \eps$. Then the radial location of a point on the surface $x^i_{\rm E}$ is
\beq
\sqrt{\delta_{ij}x^i_{\rm E}x^j_{\rm E}} = r_{\rm E} + \delta_{ij}x^i_{\rm BH}\Omega^j + O(\eps^2).
\eeq
The quantity $\delta_{ij}x^i_{\rm BH}\Omega^j$ is a pure dipole. If we define a new radial profile $r'_{\rm E}(\theta^A) =  r_{\rm E} + \delta_{ij}x^i_{\rm BH}\Omega^j$ and expand it as $r'_{\rm E} = \sum_{lm} r'_{lm}Y^{lm}$, then $x^i_{\rm BH}$ corresponds to the $l=1$ coefficients:
\beq
x^i_{\rm BH} = \sqrt{\frac{3}{4\pi}}\left[-\sqrt{2}\,{\rm Re}(r'_{11}),\sqrt{2}\,{\rm Im}(r'_{11}),r'_{10}\right].\label{BH center}
\eeq
In these expressions we absorb the $v$ dependence into the coefficients. Note that although we have this one-to-one relationship, the embedding~\eqref{Euclidean embedding v2} should be used rather than $x^i_{\rm E}(\theta^A;v)=r'_{\rm E}(\theta^A;v)\Omega^i(\theta^A)$; in Eq.~\eqref{Euclidean embedding v2}, the term $x^i_{\rm BH}$ represents a uniform translation, shifting the center of the black hole without altering its shape, while in $x^i_{\rm E}(\theta^A;v)=r'_{\rm E}(\theta^A;v)\Omega^i(\theta^A)$ it {\em does} visibly alter the shape. 

We now require a meaningful $r'_{1m}$ to use in Eq.~\eqref{BH center}. One option would be to simply use the $l=1$ mode of the coordinate profile~\eqref{r1lm EH and AH} in the Lorenz gauge. While the coordinate profile is a gauge-dependent quantity, there is good reason to think the Lorenz-gauge profile contains a meaningful dipole mode. As explained in Ref.~\cite{Detweiler:2003ci}, if the coordinate system is centered on the moving black hole, then the coordinates are noninertial, causing the metric at large $r$ to grow rather than decay. The transformation from such a gauge to the Lorenz gauge eliminates this feature, and in the Lorenz gauge one can identify the system's center of mass with the coordinate origin $r=0$ (of course keeping in mind that this is only an {\em effective} identification because the coordinate origin is deep inside the black hole and not a well-defined curve in the true spacetime manifold).

Unfortunately, the Lorenz-gauge dipole leads to counterintuitive behavior that does not scale correctly in the Newtonian limit. In Newtonian theory, if the black hole moves on a trajectory $Z^i$ and the particle on a trajectory $z^i$, then the system's center of mass is $\frac{M Z^i + m z^i}{M+m}$, and if the center of mass is at the coordinates' origin, then $Z^i = -\eps z^i$. The magnitude of the black hole's displacement should then be $r_{\rm BH} = \sqrt{\delta_{ij}Z^iZ^j} = \eps r_0$, growing linearly with the particle's orbital radius $r_0$. In fact, the Lorenz-gauge dipole does encode such linear growth~\cite{Ori:2003cm}, and we can even observe the growth in the metric perturbation on the horizon. However, the radial profile~\eqref{r1lm EH and AH} only depends on a specific piece of this perturbation, $h^{(1lm)}_{vv}$, and we find that this decays as $1/\sqrt{r_0}$ at large $r_0$.\footnote{The alternative formula~\eqref{first_rad_ah} does depend on other pieces of the dipole perturbation, but those pieces are found to precisely cancel.} We then arrive at a puzzling conclusion. If we follow an inspiral backward in time, such that the particle spirals outward toward greater radii, then far from the system, in the wave zone, the metric appears to behave as if the coordinates are centered on the center of mass, with the black hole also spiralling outward. But if we make a local measurement of the horizon's location, we find that while the particle spirals outward,  the black hole spirals {\em inward}, eventually settling precisely at the coordinate origin.

As an alternative to the Lorenz-gauge measure of the black hole's center, we can follow the idea that Eq.~\eqref{metric of embedded surface} is a gauge-fixed version of the horizon's intrinsic metric $\gamma_{AB}$. After eliminating $\gamma_{\<AB\>}$ via a gauge choice, we have $\gamma_{AB}=\gamma_\circ \Omega_{AB}:=(r'_{\rm E})^2\Omega_{AB}$. We can then put this in the form~\eqref{metric of embedded surface} by performing a uniform translation, by an amount $-x^i_{\rm BH}$, to eliminate the dipole mode from $\gamma_\circ$. This transformation would make the induced metric perfectly round up to $l>1$ contributions, as it should be if the coordinates are centered on the center of the black hole. We can then undo the translation, but now treating it exactly rather than perturbatively, to obtain the displaced surface~\eqref{Euclidean embedding v2}. In that expression, $x^i_{\rm BH}$ is given by Eq.~\eqref{BH center} with $r'_{1m} = \frac{\gamma^{(11m)}_\circ  e^{-i\Omega_m v}}{4M}$. Explicitly, appealing to Eq.~\eqref{gamma1 harmonic expansions} for $\gamma^{(11m)}_\circ$, we see that
\beq
r'_{1m} =  \left(\frac{h^{(11m)}_\circ}{4M} + r_{1m}\right)e^{-i\Omega_m v},\label{r'1m}
\eeq
where $r_{1m}$ is given by Eq.~\eqref{r1lm EH and AH}.

We use this definition of black hole's center in Figs.~\ref{fig:r07}--\ref{fig:r025}, which display the horizon surface~\eqref{Euclidean embedding v2} in the $x$--$y$ plane at several values of $v$ and for two values of $r_0$. In these plots, we see two effects: the horizon is tidally distorted by the orbiting companion, and it follows a counterclockwise orbit around the center of mass. (We have artificially multiplied $l>1$ modes of $r_{\rm E}$ by a factor of 200 to make the tidal distortion clearly visible. Other parameters are described in the captions.) 

Unfortunately, this definition of the black hole's position does not exhibit the expected Newtonian behavior. We find numerically that $x^i_{\rm BH}$ tends to $-\frac{1}{2}\eps x^i_p$ at large $r_0$ rather than to $x^i_{\rm BH} = -\eps x^i_p$. The linear scaling with $r_0$ is correct, and it is quite unlike the $\sim1/\sqrt{r_0}$ behavior of the Lorenz-gauge coordinate position. But there is a confounding factor-of-2 disagreement. This suggests that a different definition of $x^i_{\rm BH}$ is required. We leave investigation of this issue to future work.

For completeness, in Fig.~\ref{fig:r07 3D} we display complementary three-dimensional views of the horizon. Since this figure illustrates the shape (rather than motion) of the horizon, for it we use the traditional embedding~\eqref{Euclidean embedding v1}. 

\subsubsection{Horizon teleology}

Like the black hole, the companion follows a counterclockwise orbit, with a trajectory $x^i_p(t) = [r_0\cos(\Omega t),r_0\sin(\Omega t),0]$, such that at advanced time $v$ it is at a position 
\beq
x^i_p = \left\{r_0\cos[\Omega(v-r_0^*)],r_0\sin[\Omega(v-r_0^*)],0\right\},
\eeq
where $r_0^*$ is the tortoise coordinate at the particle's location. In each plot, the dotted line points toward this position. The dashed line points toward the location of greatest tidal distortion of the horizon---the tidal bulge. As we can see from the plots, the tidal bulge consistently leads the companion's orbit by a constant angle, seeming to point toward the companion's future location.\footnote{Note that the displacement of the black hole from the center of mass creates an $O(\eps r_0)$ correction to the angle between the two lines.}

This lead angle has been discussed many times in the past~\cite{Hartle:1974gy,Fang:2005qq,Vega_2011,Hughes1}. It markedly contrasts with the behavior of a fluid body. In Newtonian physics, the tidal bulge of an inviscid fluid body points toward the companion's instantaneous position. If the fluid is viscous, then the tidal bulge lags behind the companion's orbit. In relativistic physics, the tidal bulge will also lag due to retardation. In no case does the fluid body's bulge {\em lead} the companion's orbit. Previous literature has offered the explanation that the tidal lead is due to the teleological nature of the event horizon: the horizon  effectively anticipates the companion's future location. More mathematically, Eq.~\eqref{teleological} tells us that the horizon's deformation leads the metric perturbation by a time of order $1/\kappa_0$.

However, such an explanation cannot be wholly correct. At linear order, the event horizon is indistinguishable from the apparent horizon: their radial profiles~\eqref{teleological} (with $n=1$) and~\eqref{first_rad_ah} are identical. Therefore the event horizon's location on a slice of constant advanced time $v$ is completely determined by information on that slice. This rules out a truly teleological explanation, but it also suggests a nonteleological one. An apparent horizon is not pulled toward the orbiting companion; it is repelled by it. This is because the companion pulls light rays toward itself and hence away from the black hole, helping them to escape. Therefore the tidal bulge of the horizon cannot be interpreted in the same way as the tidal bulge of a fluid body. The bulge of a fluid body is created by the companion pulling the fluid toward itself, while the companion's pull on light rays has precisely the opposite effect on the black hole's horizon. Rather than thinking of the bulge leading the orbit, we can think of the depression lagging the orbit due to retardation. 

Of course, this only describes the behavior of the side of the horizon nearest the companion. But we can also explain the behaviour of the other sides. The bulge is created because on the bulging sides of the horizon, light rays pulled toward the companion fall into the hole, and the orientation of this effect again lags the orbit due to retardation. There is a less obvious reason why the far side of the horizon is depressed. Naively, one would expect that since the companion pulls light rays toward itself, it will pull them into the black hole if they are on the far side of the hole. This would suggest that the horizon should bulge outward on the far side, rather than being depressed. However, it is depressed there for the same reason that a fluid body bulges there: the companion's pull is weaker on the far side than at the black hole's ``center", meaning the companion pulls the black hole's center toward itself more than it pulls light rays into the black hole. 

Although these simplistic descriptions ignore the complexity of field propagation in the strong gravity of the black hole, it is clear that since the horizon is an apparent horizon, there must be a nonteleological explanation of its shape and orientation. This conclusion is reinforced by the fact that similar behavior has been observed in the apparent horizon in full numerical relativity~\cite{Stein:private}.

One might wonder (a) how this is reconciled with the teleological solution~\eqref{teleological}, and (b) whether the lead angle is fundamentally a teleological effect that only becomes effectively causal as a consequence of the temporal localization of the event horizon in a binary. We believe both questions are answered by the fact that the shape we have plotted is an apparent horizon regardless of whether or not it is the event horizon: the radial profile of the apparent horizon, \eqref{first_rad_ah}, encodes the lead angle. This tells us that the tidal deformation at time $v$ is determined solely by the behavior of null vectors at time $v$. In the case that the event horizon becomes temporally localized, reducing the teleological solution~\eqref{teleological} to the localized solution~\eqref{r1k EH}, then it too only depends on the behavior of null vectors at time $v$; the localized solution only encodes the teleological nature of~\eqref{teleological} in the weak sense of encoding the tangent vector to a curve at time $v$ (as opposed to only encoding the location of the curve at that time). The ``lead time" $1/\kappa_0$ is only teleological in that same weak sense. It is simply a statement about null tangent vectors, rather than a statement about the far future of null curves. 

Yet the teleological solution~\eqref{teleological} {\em does} impose boundary conditions in the future. What is their relevance? They only serve to rule out exponentially growing solutions, which would describe null curves that escape to infinity.\footnote{Note also that Eq.~\eqref{r1k EH} is the unique solution to Eq.~\eqref{first_rad_eh} if we assume a two-timescale ansatz.} The future boundary conditions therefore only ensure that the null curves are marginally trapped. In other words, they ensure that the event horizon remains close to the apparent horizon. Rather than inferring that the apparent horizon exhibits the event horizon's teleological behavior, we can infer that the event horizon, by hewing close to the apparent horizon, exhibits the apparent horizon's {\em non}teleological behavior.

We also stress that the identification of the two horizons does not depend on the localization: at linear perturbative order, the event horizon is an apparent horizon under fairly general conditions. In the context of a binary, the linear-order identification  only breaks down as the companion enters into its plunge. In those final moments, teleological effects {\em do} arise at linear order, as additional null rays join the event horizon in anticipation of the companion's entry into the black hole~\cite{Hamerly-Chen:10,Emparan-Martinez:16,Hussain-Booth:17,Emparan-Martinez-Zilhao:17}.

To more starkly illustrate that difference between the teleological and the nonteleological, we can consider a thought experiment in which the companion orbits the black hole while a pulse of null radiation approaches on a radial trajectory. Suppose the pulse is timed such that it strikes the companion and sends it into either a plunge or an escape trajectory. The pulse approaches on a surface $v=v_{\rm pulse}$. For all $v<v_{\rm pulse}$, the pulse is out of causal contact with the black hole, and the apparent horizon behaves just as we have illustrated it. But the event horizon anticipates the pulse, and additional generators begin to join it. At a time $v_{\rm teleo}\approx v_{\rm pulse}-\kappa_0^{-1}$, this teleological effect becomes significant. If the pulse ultimately knocks the companion into the black hole, then at $v_{\rm teleo}$, the event horizon must already be significantly reaching out to meet the companion's plunge. If instead the pulse sends the companion into an escape trajectory, then at $v_{\rm teleo}$ the event horizon must already be adjusting itself in a very different way. Yet at $v_{\rm teleo}$ the apparent horizon is identical in both cases. Unaffected by the future events, it continues to exhibit the same tidally deformed shape with the same lead angle.


\section{Conclusion}\label{sec:conclusion}

In this paper, motivated by the work that was necessary in Ref.~\cite{Pound_2020}, we have focused on a few features of perturbed horizons: their location and surface area, the quasilocal mass they contain, and their scalar curvature. We have primarily highlighted the degree to which the apparent and event horizon differ beginning at second perturbative order. In particular, we have established the relationships~\eqref{rAH vs rEH horizon locked}, \eqref{gamma AH vs gamma EH gauge fixed}, \eqref{R AH vs R EH gauge fixed}, and \eqref{A2hat AH} between the horizons' radii, intrinsic metrics, scalar curvatures, and surface areas, along with the equality~\eqref{MH AH = MH EH} of their Hawking masses. Although we have not discussed the Hawking-Hayward mass $M_{\rm HH}$~\cite{Hayward:1993ph}, an immediate corollary of our results is that the two horizons also have equal values of $M_{\rm HH}$ through order $\eps^2$.

Our results have illuminated several important, but subtle issues that arise in the study of perturbed black holes. First, we have highlighted the fact that the apparent and event horizons are identical at first perturbative order (except in certain cases such as shortly before merger in a binary). This was already known, as it is a simple consequence of the event horizon's expansion scalar vanishing at linear order, but to our knowledge it has not been emphasized in the literature. Because the event horizon is indistinguishable from an apparent horizon at linear order, one must be cautious in ascribing any teleological interpretation to its behavior at this order. In Sec.~\ref{visualizing embedding} we have explained how even seemingly teleological behavior such as the ``tidal lead" in a binary is actually nonteleological in origin.

Along the same lines, in Sec.~\ref{sec:localization} we have shown how, in a small-mass-ratio binary, the event horizon of the central black hole is effectively localized in time at all perturbative orders. This result establishes that in the small-mass-ratio context, teleological effects are negligible even beyond linear order.

Another important conclusion follows from our result that the apparent and event horizons have identical Hawking masses through second order. The Hawking mass of the apparent horizon grows monotonically, obeying a physically intuitive ``flux-balance" law~\cite{Ashtekar:2004cn}, which implies that the mass grows at the rate that energy crosses the horizon. Our result therefore shows that the Hawking mass of the event horizon obeys the same flux-balance law at least through second order. Besides its intrinsic interest, this may be useful in deriving and interpreting second-order balance laws for small-mass-ratio inspirals, which would relate the evolution of the companion's orbit to (i) the evolution of the central black hole, and (ii) the gravitational waves emitted to infinity.

All our results have been obtained in two formulations: in a form that does not specify a choice of gauge, and in a gauge-fixed form. In either case, the metric-perturbation inputs can be calculated in any gauge. If one uses the unfixed formulas, one's choice of gauge will influence one's results for the horizon area and scalar curvature, for example. That may or may not be advantageous depending on context; one might wish to use a gauge condition corresponding to a specific choice of horizon foliation, for example. On the other hand, the gauge-fixed formulation provides invariant results carrying a specific geometric interpretation.

From this starting point, there are many obvious follow-ons that we leave to future work. First, one might derive perturbative formulas for the black hole's spin and higher multipole moments. 

Second, our results have revealed a puzzling discrepancy between the location of the horizon and the expected Newtonian limit. To better characterize the black hole's orbit around the system's center of mass, one might derive a perturbative formula for an invariant linear momentum such as the Hawking momentum~\cite{Szabados:2009eka} or an analogous orbital angular momentum. 

Third, in a dynamically evolving spacetime with a two-timescale character, can we meaningfully identify an average horizon that only evolves on the slow timescale, and is it precisely a Kerr horizon with slowly evolving mass and spin, or does it contain higher moments that deviate from Kerr? Such a study might draw on the framework of slowly evolving horizons in Ref.~\cite{Booth:2003ji}.

To characterize this average horizon, a critical step would be obtaining flux-balance equations governing the evolution of the averaged mass, spin, and higher moments. This would involve higher-order extensions of standard relations for the black hole's mass and spin evolution in terms of the shear of the horizon generators~\cite{Poisson:04}, as well as equations for the higher moments that describe black hole memory effects~\cite{Hawking:2016msc,Rahman:2019bmk}.

Finally, an important goal will be to numerically compute these second-order effects on the horizon in realistic binary scenarios. That programme of research was initiated in the calculation of the apparent horizon's irreducible mass in Ref.~\cite{Pound_2020}, and more such calculations will soon be possible. Since the event horizon begins to differ from the apparent horizon at second order, these calculations may also allow us to identify genuine teleological effects during an inspiral.

\begin{acknowledgments}
	We are grateful to Niels Warburton and Barry Wardell, who kindly provided input data used within this paper, and to Eric Poisson for a sanity check and a helpful discussion. This work makes use of the Black Hole Perturbation Toolkit~\cite{BHPT}.
	RB thanks Stefano Ansoldi for introducing him to these topics. AP thanks Aaron Zimmerman for suggesting an analysis of the apparent horizon and for additional helpful discussions in the early stages of this work; Jonathan Thornburg for pointing out the exponential suppression of teleological effects on the event horizon; Leor Barack for helpful conversations and for independently checking some of our first-order results. ZS additionally thanks Tim Dietrich for his interest in this work and comments on an earlier version of the manuscript. RB additionally thanks the University of Trieste for the International Mobility Scholarship granted, and in turn the University of Southampton for the welcoming hospitality. AP also gratefully acknowledges the support of a Royal Society University Research Fellowship. AP's early work in this paper was supported by the European Research Council under the European Union’s Seventh Framework Programme (FP7/2007-2013)/ERC Grant No. 304978.
\end{acknowledgments}


\appendix


\section{Perturbative formulas for curvature tensors}\label{sec:curvature tensors}

In this appendix we review the derivation of perturbed curvature tensors through second order. Because we use these formulas for the curvature of the horizon, not solely for the Einstein equations, we allow the background metric to have nonvanishing Ricci curvature. Unlike in the body of the paper, here we use the inverse of the background metric to raise indices.

The derivation follows standard methods in Ref.~\cite{Wald}, for example. We consider a generic metric of the form $g_{\alpha\beta} = g^{(0)}_{\alpha\beta}+h_{\alpha\beta}$. The Christoffel symbols associated with $g_{\alpha\beta}$ and those associated with $g^{(0)}_{\alpha\beta}$ are related by
\beq
\Gamma^{\alpha}_{\beta\gamma} - {}^{(0)}\Gamma^{\alpha}_{\beta\gamma} = C^\alpha_{\beta\gamma},
\eeq
where
\beq\label{C tensor}
C^\alpha_{\beta\gamma}[h] := \tfrac{1}{2}g^{\alpha\mu}\left(h_{\mu\beta;\gamma}+h_{\mu\gamma;\beta}-h_{\beta\gamma;\mu}\right),
\eeq
and a semicolon denotes covariant differentiation compatible with $g^{(0)}_{\alpha\beta}$. 
In terms of this tensor, the Riemann tensors of the two metrics are related by
\beq
R^\alpha{}_{\beta\gamma\delta} = {}^{(0)}\!R^\alpha{}_{\beta\gamma\delta}+2C^\alpha_{\beta[\delta;\gamma]}+2C^\alpha_{\mu[\gamma}C^\mu_{\delta]\beta}.
\eeq
The Ricci tensors are therefore related  by
\beq
R_{\alpha\beta} = R^{(0)}_{\alpha\beta}+2C^\mu_{\alpha[\beta;\mu]}+2C^\nu_{\mu[\nu}C^\mu_{\beta]\alpha}.
\eeq

We now expand the Ricci tensor in powers of $h_{\alpha\beta}$ using
\beq
g^{\alpha\beta} = g^{\alpha\beta}_{(0)} - h^{\alpha\beta} + h^\alpha{}_\gamma h^{\gamma\beta}+O[(h)^3].
\eeq
The result is
\beq
R_{\alpha\beta} = R^{(0)}_{\alpha\beta} + \d R_{\alpha\beta}[h]+\d^2R_{\alpha\beta}[h] + O[(h)^3],
\eeq
where
\beq
\d R_{\alpha\beta} = -\tfrac{1}{2}( h_{\alpha\beta;\gamma}{}^{\gamma}+ h_{\gamma}{}^{\gamma}{}_{;\alpha\beta}) + h_{\mu(\alpha;\beta)}{}^\mu,
\eeq
and
\begin{align}
\d^2 R_{\alpha\beta} &= \tfrac{1}{4}h^{\mu\nu}{}_{;\alpha}h_{\mu\nu;\beta} + \tfrac{1}{2}h^{\mu}{}_{\beta}{}^{;\nu}\left(h_{\mu\alpha;\nu} - h_{\nu\alpha;\mu}\right) \nonumber\\
&\quad - \tfrac{1}{2} h^{\mu\nu}_*{}_{;\nu}\left(2h_{\mu(\alpha;\beta)}-h_{\alpha\beta;\mu}\right)\nonumber\\
&\quad -\tfrac{1}{2} h^{\mu\nu}\left(2h_{\mu(\alpha;\beta)\nu} - h_{\alpha\beta;\mu\nu} - h_{\mu\nu;\alpha\beta}\right),    
\end{align}
where a $*$ denotes trace reversal, as in $h^*_{\mu\nu}:=h_{\mu\nu}-\tfrac{1}{2}g_{\mu\nu}^{(0)}h_\gamma{}^\gamma$.

The perturbations of the Ricci scalar are straightforwardly deduced from those of the Ricci tensor. We have
\beq
R = R^{(0)} + \delta R[h] + \delta^2 R[h] + O[(h)^3] 
\eeq
with
\begin{align}
\delta R[h] &= g^{\alpha\beta}_{(0)}\delta R_{\alpha\beta}[h] - h^{\alpha\beta}R^{(0)}_{\alpha\beta},\\
\delta^2 R[h] &= g^{\alpha\beta}_{(0)}\delta^2 R_{\alpha\beta}[h] - h^{\alpha\beta}\delta R_{\alpha\beta}[h] \nonumber\\
&\quad + h^{\alpha}{}_\gamma h^{\gamma\beta}R^{(0)}_{\alpha\beta}.
\end{align}
Explicitly,
\begin{align}
\d R[h] = - h^{\a\b}R^{(0)}_{\a\b} - h^\a_{\ \a;\b}{}^\b + h^{\a\b}{}_{;\a\b},\label{dR}
\end{align}
and
\begin{align}
\d^2 R[h] &= \tfrac{3}{4}h^{\mu\nu;\beta}h_{\mu\nu;\beta} - \tfrac{1}{2}h^{\mu\beta;\nu}h_{\nu\beta;\mu}  -h^{\mu\nu}_*{}_{;\nu}h^*_{\mu\alpha}{}^{;\alpha}\nonumber\\
&\quad - h^{\mu\nu}\left(h_{\mu\alpha}{}^{;\alpha}{}_\nu + h_{\mu\alpha;\nu}{}^{;\alpha} - h_{\alpha}{}^{\alpha}{}_{;\mu\nu} - h_{\mu\nu}{}^{;\alpha}{}_{\alpha}\right)\nonumber\\
&\quad +h^{\alpha}{}_\gamma h^{\gamma\beta}R^{(0)}_{\alpha\beta}.\label{ddR} 
\end{align}


\section{Coupling of spherical harmonics}\label{sec:coupling}

In various places in the body of the paper, we must decompose products of angular functions into scalar spherical harmonics. That is, for a function $Z(\theta^B)=\chi_{A_1\cdots A_n}(\theta^B)\psi^{A_1\cdots A_n}(\theta^B)$, we require 
\beq
Z_{lm} = \int \bar Y^{lm}Z d\Omega.
\eeq
Each of the angular functions in the integrand is itself expressed as a sum of harmonics, leading to integrals with the schematic form
\beq
Z_{lm} = \sum_{\substack{l'm'\\l''m''}}\chi_{l'm'}\psi_{l''m''}\int \bar Y^{lm} {\cal Y}^{l'm'}_{A_1\cdots A_n}{\cal X}_{l''m''}^{A_1\cdots A_n} d\Omega,\label{coupling integral}
\eeq
where ${\cal Y}^{l'm'}_{A_1\cdots A_n}$ and ${\cal X}_{l''m''}^{A_1\cdots A_n}$ are constructed from $Y^{lm}$ through linear operations involving $\epsilon_{AB}$, $\Omega_{AB}$, $\Omega^{AB}$, and $D_A$.

To evaluate such derivatives, we express the integrand in terms of spin-weighted spherical harmonics, which are constructed from a complex basis on $S^2$. We define 
\begin{equation}
m^A:= \left(1,\frac{i}{\sin\theta}\right),
\end{equation}
in spherical coordinates $(\theta,\phi)$; the basis is then $\{m^A,\bar m^A\}$. These basis vectors have the useful properties
\begin{align}
\Omega_{AB}m^Am^B &=0, \label{m identity first}\\
\Omega_{AB}m^A \bar m^B &= 2,\\
m^B D_Bm^A &= m^A \cot\theta, \\
\bar m^{B} D_B m^A &= -m^A\cot\theta, \\
\epsilon_{AB}m^B &= i\,\Omega_{AB}m^B, \\
\Omega^{AB} &= \frac{1}{2}\left(m^A\bar m^{B}+\bar m^{A}m^B\right).\label{m identity last}
\end{align}
Our definition of $m^A$ differs by a factor of $\sqrt 2$ relative to that of Ref.~\cite{Newman-Penrose:66}, and it is normalized on the unit sphere rather than a sphere of radius $r$. In terms of $m^A$, we define derivative operators $\eth$ and $\bar\eth$ that act on a scalar of spin-weight $s$ as
\begin{align}
\eth v &= (m^AD_A-s\cot\theta)v,\label{eth}\\
\bar{\eth} v &= (\bar m^{A}D_A+s\cot\theta)v.\label{ethbar}
\end{align}
Our definitions here differ by an overall minus sign relative to those of Ref.~\cite{Newman-Penrose:66}. A quantity $v$ has spin-weight $s$ if it transforms as $v\to e^{is\varphi}v$ under the complex rotation $m^A\to e^{i\varphi}m^A$. $\eth v$ raises the spin weight by 1, while $\bar\eth$ lowers it by 1.

The spherical harmonics of spin-weight $s$ are
\begin{equation}
{}_sY^{l m} :=\frac{1}{\lambda_s}
	\begin{cases}
    (-1)^s\eth^sY^{l m}, & 0\leq s\leq l,\\
    \bar{\eth}^{|s|} Y^{l m}, & -l\leq s\leq0,
  \end{cases}
\end{equation}
where 
\beq
\lambda_s:=\sqrt{\frac{(l+|s|)!}{(l-|s|)!}}.\label{lambda_s}
\eeq
The integral of three spin-weighted harmonics is
\begin{equation}\label{Cdef}
C^{\ell m s}_{\ell'm's'\ell''m''s''}:=\oint {}_{s}\bar Y^{\ell m}{}_{s'}Y^{\ell' m'}{}_{s''}Y^{\ell'' m''}d\Omega. 
\end{equation}
In the case that $s=s'+s''$, the coupling constants are given by
\begin{align}\label{coupling}
\!\!C^{\ell m s}_{\ell'm's'\ell''m''s''} &= (-1)^{m+s}\sqrt{\frac{(2\ell+1)(2\ell'+1)(2\ell''+1)}{4\pi}}\nonumber\\
				&\quad\times\!	\begin{pmatrix}\ell & \ell' & \ell'' \\ s & -s' & -s''\end{pmatrix}\!\!	\begin{pmatrix}\ell & \ell' & \ell'' \\ -m & m' & m''\end{pmatrix}\!,
\end{align}
where the arrays are $3j$ symbols. These symbols inherit symmetries from the $3j$ symbols, specifically
\begin{align}
C^{ l m s}_{ l'm's' l''m''s''} &= (-1)^{l+l'+l''}C^{ l m -s}_{ l'm'-s' l''m''-s''},\\
C^{ l m s}_{ l'm's' l''m''s''} &= (-1)^{l+l'+l''}C^{ l -m s}_{ l'-m's' l''-m''s''},\\
C^{ l m s}_{ l'm's' l''m''s''} &= C^{ l m s}_{l''m''s''l'm's'}.
\end{align}
For $s=0$ and $s'=-s''$, they have the special value
\beq\label{coupling monopole}
C^{000}_{ l'm's' l''m''s''} = \frac{(-1)^{m' + s'}}{\sqrt{4\pi}} \delta_{l'l''}\delta_{m',-m''}.
\eeq

We evaluate all integrals of the form~\eqref{coupling integral} by expressing the integrand as a sum of terms of the form~\eqref{Cdef}. This is straightforwardly accomplished by applying the identities~\eqref{m identity first}--\eqref{m identity last} and the definitions~\eqref{YA}--\eqref{XAB}. For example, $D_AY^{\ell m}=\frac{1}{2}(m_Am^{B*}+m^*_Am^B)D_BY^{\ell m}$ together with Eqs.~\eqref{eth} and \eqref{ethbar} leads to
\begin{equation}\label{DYtosY}
Y^{lm}_A:= D_A Y^{\ell m} = \frac{1}{2}\lambda_1\left({}_{-1}Y^{\ell m}m_A-{}_1Y^{\ell m}m^*_A\right).
\end{equation}

The specific integrals involved in calculating $r^{(2)}_{lm}$ are found to be
\begingroup
\allowdisplaybreaks
\begin{align}
\int \bar Y^{lm} Y^{A}_{l'm'} Y_A^{l''m''}d\Omega  &= \int \bar Y^{lm} X^{A}_{l'm'} X_A^{l''m''} d\Omega\nonumber\\
&= -\frac{1}{2} \lambda_1' \lambda_1'' \sigma_+ C_{l' m' 1 l'' m'' -1}^{lm0}, \label{first_C_int} \\
\int \bar Y^{lm} Y^{A}_{l'm'} X_A^{l''m''}d\Omega &= \frac{i}{2} \lambda_1' \lambda_1'' \sigma_- C_{l' m' 1 l'' m'' -1}^{lm0}, \\
\int \bar Y^{lm} Y^{AB}_{l'm'} Y_{AB}^{l''m''}d\Omega  &= \int  \bar Y^{lm} X^{AB}_{l'm'} X_{AB}^{l''m''}d\Omega  \nonumber\\
&= \frac{1}{4} \lambda_2' \lambda_2'' \sigma_+ C_{l' m' 2 l'' m'' -2}^{lm0},  \\
\int \bar Y^{lm} Y^{AB}_{l'm'} X_{AB}^{l''m''}d\Omega  &= \frac{i}{4} \lambda_2' \lambda_2'' \sigma_- C_{l' m' 2 l'' m'' -2}^{lm0}, \label{last_C_int}
\end{align}
\endgroup
where $\lambda_s' :=\lambda_s(l')$,  $\lambda_s'' :=\lambda_s(l'')$,
\beq\label{sigma}
\sigma := (-1)^{l + l' + l''}\quad\text{and}\quad \sigma_\pm:=\sigma\pm1.
\eeq
In the calculation of ${\cal R}^{(2)}_{lm}$, third angular derivatives of $Y^{lm}$ arise. These are treated in the same way, with higher-order versions of Eq.~\eqref{DYtosY} containing spin weights up to $\pm 3$.


\section{Second-order expressions}\label{sec:second_order_expressions}

In this appendix we compile results for various second-order quantities. 

The functions of $v$ appearing in the mode decomposition~\eqref{R2lm} of the second-order scalar curvature ${\cal R}^{(2)}$
are
\begingroup
\allowdisplaybreaks
\begin{subequations}\label{R2lm coefficients}
\begin{align}
{}^{(3)}{\cal R}^{lm}_{l'm'l''m''}&= - \frac{\lambda_{3}' \lambda_{3}''}{2048M^6} \left(\breve{\c}'_{-}\c''_{\odot} + \breve{\c}'_{+}\c''_{\oplus}\right)
 ,\label{R coefficient 3}\\
{}^{(2)}{\cal R}^{lm}_{l'm'l''m''}&= \frac{\lambda_{2}' \lambda_{2}''}{256M^6} \left(\breve{\c}'_{-}\c''_{\odot} + \breve{\c}'_{+}\c''_{\oplus}\right)
 ,\label{R coefficient 2}\\
{}^{(1)}{\cal R}^{lm}_{l'm'l''m''} &=\frac{\lambda_{1}'\lambda_{1}''}{2048M^6} \Bigl\{(\mu')^2(\mu'')^2\left( \breve{\c}'_{-}\breve{\c}''_{\odot}  
+  \breve{\c}'_{+}\breve{\c}''_{\oplus}\right)\nonumber\\
&\quad-\breve{\c}'_{\circ} \left[4 \breve{\c}''_{\circ} \sigma_{+}{} - (\mu'')^2 \breve{\c}''_{\oplus}\right] \nonumber \\
&\quad - 12(\mu')^2\breve{\c}_{\circ}'' \breve{\c}'_{\ominus}\Bigr\} , \label{R coefficient 1}\\
{}^{(0)}{\cal R}^{lm}_{l'm'l''m''}&= \frac{\breve{\c}'_{\circ}}{64M^6}  \left\{2 \breve{\c}''_{\circ} \bigl[1 - \
\left(\lambda_{1}''\right)^2 \bigr] - \breve{\c}''_{+} \left(\lambda_{2}''\right)^2 \right\}, \label{R coefficient 0}
\end{align}
\end{subequations}
\endgroup
where $\lambda_s(l)$ is defined in Eq.~\eqref{lambda_s}, $\lambda_s' :=\lambda_s(l')$, $\lambda_s'' :=\lambda_s(l'')$, and $\sigma$ and $\sigma_\pm$ are defined in Eq.~\eqref{sigma}. In these expressions we have introduced the compact notation $\breve{\c}'_{\circ}:=\breve{\c}^{(1l'm')}_\circ(v)$, $\breve{\c}''_{\circ}:=\breve{\c}^{(1l''m'')}_\circ(v)$, and so on, along with the combinations $\c_{\otimes/\odot}:=\sigma_+ \c_{-} \pm i \sigma_- \c_{+}$,  and $\c_{\oplus/\ominus}:=\sigma_+ \c_{+} \pm i \sigma_- \c_{-}$  (suppressing $nlm$ labels).

The second-order term appearing in the expansion scalar~\eqref{theta expansion} is 
\begin{widetext}
\begin{align} 
	\vartheta^{(2)}_+ &= \frac{1}{32 M^4} \Big[ 8 M^2 (1-D^2)r^{(2)}- 16 M^3 h^{(2)}_{vv}+ 8 M^2\partial_v h^{(2)}_{\circ} - 8 M^2 D^Ah^{(2)}_{v A} + 4 M h^{(1)}_{vA}\Omega^{AB} h^{(1)}_{vB} + 4 M h^{(1)}_\circ h^{(1)}_{vv} \nonumber \\ 
	&\quad + 16 M^3 h^{(1)}_{vv} h^{(1)}_{vr}   -  2h^{(1)}_{\circ} r^{(1)} + 8 M^2 h^{(1)}_{vv} r^{(1)} -  2h^{(1)}_{vA} D^Ah^{(1)}_{\circ} - 8 M^2 h^{(1)}_{vr} r^{(1)}- 8 M (r^{(1)})^2  - \Omega^{AC}\Omega^{BD} h^{(1)}_{AB} \partial_v h^{(1)}_{CD} \nonumber \\
	&\quad - 8 M r^{(1)}\partial_v h^{(1)}_{\circ}   + 4 M D_Ar^{(1)}D^Ar^{(1)}  + 8 M^2 r^{(1)}\partial_r \partial_v h^{(1)}_{\circ}+ 8 M r^{(1)} D^Ah^{(1)}_{v  A} + 4 M^2 h^{(1)}_{vv} D^Ah^{(1)}_{r  A}  - 4 M^2 h^{(1)}_{vv}\partial_r h^{(1)}_{\circ} \nonumber \\ 
	&\quad + 2M r^{(1)}\partial_r h^{(1)}_{\circ} - 2 M r^{(1)}D^A h^{(1)}_{r  A} + 12 M h^{(1)}_{vA}D^A r^{(1)}  - 8 M^2 h^{(1)}_{vr}D^2 r^{(1)} + 8 M r^{(1)} D^2r^{(1)} + 8 M^2 \partial_v h^{(1)}_{rA}D^A r^{(1)} \nonumber \\ 
	&\quad  -  2D_Ar^{(1)}D^A h^{(1)}_\circ - 4 M^2 D_Ar^{(1)}D^A h^{(1)}_{vr} - 4 M h^{(1)}_{vA} D^Ar^{(1)}  - 8 M^2 \partial_r h^{(1)}_{vA}D^A r^{(1)} - 4 M^2 D_Ah^{(1)}_{vr}D^A r^{(1)} \nonumber \\ 
	&\quad + 2 \Omega^{AC}h^{(1) }_{vC} D^Bh^{(1)}_{AB}   + 2D^A r^{(1)} D^Bh^{(1)}_{AB} + 2 \Omega^{AC}\Omega^{BD}h^{(1)}_{CD} D_Bh^{(1)}_{vA}  + 2 h^{(1)}_{AB}D^AD^B r^{(1)} - 8 M^2 r^{(1)}  \partial_r D^Ah^{(1)}_{vA} \nonumber \\ 
	&\quad - 16 M^3 r^{(1)} \partial_r h^{(1)}_{vv} \Big].\label{theta2} 
\end{align}

The functions appearing in the mode decomposition~\eqref{theta2lm} of $\vartheta^{(2)}_+$ are given by
\begingroup
\allowdisplaybreaks
\begin{subequations}\label{theta2lm coefficients}
\begin{align}
{}^{(2)}\Theta^{lm}_{l'm'l''m''} & = \frac{ \lambda_2'  \lambda_2'' }{64 M^2\sqrt{\pi}} \left[\left(\partial_v h''_{-}-2 h''_{v-}\right) h'_\otimes+\left(\partial_v h''_{+}  - 2h''_{v+}-2r^{(1)}_{l'' m''}\right) h'_\oplus   \right], \label{Theta2} \\
{}^{(1)}\Theta^{lm}_{l'm'l''m''} & =-  \frac{\lambda_1' \lambda_1''}{32 M^2 \sqrt{\pi}} \biggl[-4Mh''_{v\ominus} h'_{v+} - 4Mh''_{v\odot}h'_{v-} + (\mu'')^2h'_{v-}h''_\odot + (\mu'')^2 \left(h'_{v+}+ r^{(1)}_{l' m'}\right) h''_\ominus - 8Mh'_{v\ominus} r^{(1)}_{l'' m''} \nonumber\\
&\quad + 4M^2 \sigma_+ \left(h'_{vr} r^{(1)}_{l'' m''}+  h''_{vr} r^{(1)}_{l' m'}\right)  - 4M \sigma_+ r^{(1)}_{l' m'} r^{(1)}_{l'' m''}  + 8M^2 r^{(1)}_{l'' m''}\left( \partial_r h'_{v\ominus}-\partial_v h'_{r\ominus}\right)  \biggr],\! \\
{}^{(0)}\Theta^{lm}_{l'm'l''m''} & = \frac{1}{8 M^2\sqrt{\pi}} \biggl\{-2 M h''_{vv} h'_{\circ} + (\lambda_1'')^2 h''_{v+} \left(h'_{\circ} + 4 M r^{(1)}_{l' m'}\right) -  M (\lambda_1'')^2 h''_{r+} r^{(1)}_{l'm'} + 4 M^2\left[1- (\lambda_1'')^2\right] h'_{vr} r^{(1)}_{l'' m''}  \nonumber \\
 &\quad + \left[1+(\lambda_1'')^2\right]\left(h'_{\circ}+4Mr^{(1)}_{l' m'}\right) r^{(1)}_{l'' m''}  + \left(h'_{\circ}+ 4 M r^{(1)}_{l' m'}\right) \partial_v h''_{\circ} - 4 M^2 r^{(1)}_{l' m'} \partial_r \partial_v h''_{\circ} -  M r^{(1)}_{l' m'} \partial_r h''_{\circ} \nonumber \\
 &\quad + 2 M^2 h'_{vv} \Bigl[-4 M h''_{vr} + (\lambda_1'')^2 h''_{r+} - 2 r^{(1)}_{l'' m''} + \partial_r h''_{\circ} \Bigr] + 8 M^3 r^{(1)}_{l' m'} \partial_r h''_{vv} - 4 M^2 (\lambda_1'')^2 r^{(1)}_{l' m'} \partial_r h''_{v+}\biggr\}, \label{Theta0}
\end{align}
\end{subequations}%
\endgroup
where $\mu'':=\mu(l'')=(l''+2)(l''-1)$, and we define $h'_{\circ}:=h^{(1l'm')}_\circ(v,2M)$, $h''_{\circ}:=h^{(1l''m'')}_\circ(v,2M)$, and so on. The combinations $h_{\otimes/\odot}$, $h_{\oplus/\ominus}$, $h_{a\otimes/\odot}$, and $h_{a\oplus/\ominus}$ are defined in terms of $h_\pm$ and $h_{a\pm}$ following the notation described below Eq.~\eqref{R2lm coefficients}. In the gauge-fixed form of Sec.~\ref{sec:horizon locking}, these coefficients become
\begin{subequations}\label{theta2lm coefficients gauge fixed}
\begin{align}
{}^{(2)}\hat\Theta^{lm}_{l'm'l''m''} & = \frac{ \lambda_2'\lambda_2'' }{64 M^4} \Bigl[\left(\partial_v\hat h''_{-}-2\hat h''_{v-}\right) \hat h'_\otimes+\left(\partial_v\hat h''_{+}  - 2 \hat h''_{v+}\right) \hat h'_\oplus \Bigr], \\
{}^{(1)}\hat\Theta^{lm}_{l'm'l''m''} & = \frac{\lambda_1^2\lambda_1'\lambda_1''}{64 M^4}\left(\hat h'_{v+}\hat h''_{\ominus}+\hat h'_{v-}\hat h''_\otimes\right).
\end{align}	
\end{subequations}

The $l=0$ term in the second-order correction to the apparent horizon's radius, ~\eqref{r2lm AH}, is
\begin{align}\label{r200 AH explicit}
	r^{(2)}_{00} &=2M h^{(200)}_{vv} - \partial_v h^{(200)}_\circ +\frac{1}{32 M^2 \sqrt{\pi}} \sum_{lm} \Big[
	4 \lambda_1^2 h_{v+}^{(1lm)} \bar h^{(1lm)}_\circ - 4 M \lambda_1^2 h_{r+}^{(1lm)} \bar r^{(1)}_{lm}  + 4 (\lambda_1^2+1)
	h^{(1lm)}_\circ\bar r^{(1)}_{lm} \nonumber \\
	&\quad + 8 M(2+\lambda_1^2) | r^{(1)}_{lm}|^2 - 8 M \lambda_1^2 \left(| h_{v-}^{(1lm)}|^2 + | h_{v+}^{(1lm)}|^2\right) + 16 M^2 \lambda_1^2 r^{(1)}_{lm} \partial_r \bar h_{v+}^{(1lm)} - 4 M r^{(1)}_{lm } \partial_r \bar h^{(1lm)}_\circ  \nonumber  \\
	&\quad - 8 M h^{(1lm)}_{vv} \Big(4 M^2 \bar h^{(1lm)}_{vr} -  M \lambda_1^2 \bar h_{r+}^{(1lm)} + \bar h^{(1lm)}_\circ + 2 M \bar r^{(1)}_{lm} -  M \partial_r \bar h^{(1lm)}_\circ \Big)  + 32 M^3 r^{(1)}_{lm} \partial_r\bar h^{(1lm)}_{vv} \nonumber \\
	&\quad   + 16 M^2 h^{(1lm)}_{vr} \bar r^{(1)}_{lm}  - 16 M^2 \lambda_1^2 r^{(1)}_{lm } \partial_v\bar h_{r+}^{(1lm)}  + \frac{1}{2}\lambda_2^2\, \partial_v\!\left(|h_{+}^{(1lm)}|^2+|h_{-}^{(1lm)}|^2\right)  + 2\partial_v| h^{(1lm)}_\circ|^2\nonumber \\
	&\quad + 16 M r^{(1)}_{lm} (1-M\partial_r) \partial_v\bar h^{(1lm)}_\circ \Big].
\end{align}
\end{widetext}

Finally, the quantities $H^{(200)}_{\circ}$  and $\tilde H^{(200)}_{vv}$ appearing in Eq.~\eqref{hhatcirc200} are given by 
\begingroup
\begin{align}
    \tilde H^{(200)}_{vv} &= h^{(200)}_{vv}+2\partial_{\tilde v}\zeta^r_{(100)} \nonumber\\
    &\quad + \frac{1}{\sqrt{4\pi}}\sum_{lm}\Big(\zeta^a_{(1lm)}\partial_a\bar p^{(lm)}_{vv}+\lambda^2_1\zeta^+_{(1lm)}\bar p^{(lm)}_{vv}\nonumber\\
    &\qquad\quad +2\partial_v\zeta^a_{(1lm)}\bar p^{(lm)}_{va}+2\lambda^2_1\partial_v\zeta^+_{(1lm)}\bar p^{(lm)}_{v+}\nonumber\\
    &\qquad\quad+2\lambda^2_1\partial_v\zeta^-_{(1lm)}\bar p^{(lm)}_{v-}\Big),\label{Hvv200}\\
    H^{(200)}_{\circ} &= h^{(200)}_{\circ} \nonumber\\
    &\quad+ \frac{1}{\sqrt{4\pi}}\sum_{lm}\Big[\zeta^a_{(1lm)}\partial_a\bar p^{(lm)}_\circ+\lambda^2_1\zeta^a_{(1lm)}\bar p^{(lm)}_{a+}\nonumber\\
    &\qquad\quad+\tfrac{1}{2}\lambda^2_2\left(\zeta^+_{(1lm)}\bar p^{(lm)}_++\zeta^-_{(1lm)}\bar p^{(lm)}_-\right)\!\Big] \label{Hcirc200}.
\end{align}
\endgroup
All quantities here are evaluated at $r=2M$. When acting on a nonconjugated quantity, $\partial_v = -i m\Omega$; on a conjugated quantity, $\partial_v = +im\Omega$. The harmonic coefficients of $\zeta^\alpha_{(1)}$ are given in Sec.~\ref{sec:horizon locking}, with the mode-number replacement $\bm{k}\to m$:
\begin{itemize}
    \item $\zeta^v_{(1lm)}$ is given by \eqref{zeta1v r=2M two-timescale} for $l>0$ and by $\zeta^v_{(100)}=0$ for $l=0$;
    \item $\zeta^r_{(1lm)}$ by Eq.~\eqref{zeta1r two-timescale};
    \item $\zeta^\pm_{(1lm)}$ by Eqs.~\eqref{eta+ r=2M} and \eqref{eta- r=2M} for $m\neq0$, by Eq.~\eqref{zeta1pm av} for $m=0$, $l>1$, by Eq.~\eqref{zeta1+ av dipole} for $m=0$ and $l=1$. $\zeta^-_{(110)}$ does not contribute because $\lambda_2=0$ for $l=1$.
\end{itemize}
The quantity $p_{\alpha\beta}:=h^{(1)}_{\alpha\beta}+\frac{1}{2}{\cal L}_{\zeta_{(1)}}g_{\alpha\beta}$ has tensor-harmonic coefficients $    p^{(lm)}_{vv} = \frac{1}{2}h^{(1lm)}_{vv}$, $    p^{(lm)}_{vr} = \frac{1}{2}h^{(1lm)}_{vr}$, $    p^{(lm)}_{r\pm} = \frac{1}{2}h^{(1lm)}_{r\pm}$, and
\begingroup
\allowdisplaybreaks
\begin{align}
    \partial_r p^{(lm)}_{vv} &= \partial_r h^{(1lm)}_{vv}+\frac{\kappa_0}{M}\zeta^r_{(1lm)} + (\partial_v-\kappa_0) \partial_r\zeta^r_{(1lm)}\nonumber\\
    &\quad -\frac{1}{2M}\partial_v\zeta^v_{(1lm)},\\
    p^{(lm)}_{v+} &= h^{(1lm)}_{v+}+2M^2\partial_v\zeta^+_{(1lm)}+\frac{1}{2}\zeta^r_{(1lm)},\\
    p^{(lm)}_{v-} &=  h^{(1)}_{v-}+2M^2\partial_v\zeta^-_{(1lm)},\\
    p^{(lm)}_\circ &= h^{(1lm)}_\circ +2M\zeta^r_{(1lm)}-2M^2\lambda^2_1\zeta^+_{(1lm)},\\
    \partial_r p^{(lm)}_\circ &= \partial_r h^{(1lm)}_\circ +(1+2M\partial_r)\zeta^r_{(1lm)}\nonumber\\
    &\quad -2M\lambda^2_1(1+M\partial_r)\zeta^+_{(1lm)},\\
    p^{(lm)}_\pm &= h^{(1lm)}_\pm + 4M^2\zeta^\pm_{(1lm)},
\end{align}
\endgroup
where, again, all quantities are evaluated at $r=2M$. $\partial_r\zeta^r_{(1lm)}$ and $\partial_r\zeta^+_{(1lm)}$ can be read off of Eqs.~\eqref{zeta1r r>2M} and \eqref{zeta+ r>2M}.

\medskip

\bibliography{references}

@article{Barack_2018,
    author = "Barack, Leor and Pound, Adam",
    title = "{Self-force and radiation reaction in general relativity}",
    eprint = "1805.10385",
    archivePrefix = "arXiv",
    primaryClass = "gr-qc",
    doi = "10.1088/1361-6633/aae552",
    journal = "Rept. Prog. Phys.",
    volume = "82",
    number = "1",
    pages = "016904",
    year = "2019"
}

@article{Pound_2015,
    author = "Pound, Adam",
    title = "{Gauge and motion in perturbation theory}",
    eprint = "1506.02894",
    archivePrefix = "arXiv",
    primaryClass = "gr-qc",
    doi = "10.1103/PhysRevD.92.044021",
    journal = "Phys. Rev. D",
    volume = "92",
    number = "4",
    pages = "044021",
    year = "2015"
}

@article{Martel_2005,
    author = "Martel, Karl and Poisson, Eric",
    title = "{Gravitational perturbations of the Schwarzschild spacetime: A Practical covariant and gauge-invariant formalism}",
    eprint = "gr-qc/0502028",
    archivePrefix = "arXiv",
    doi = "10.1103/PhysRevD.71.104003",
    journal = "Phys. Rev. D",
    volume = "71",
    pages = "104003",
    year = "2005"
}

@article{Vega_2011,
    author = "Vega, Ian and Poisson, Eric and Massey, Ryan",
    title = "{Intrinsic and extrinsic geometries of a tidally deformed black hole}",
    eprint = "1106.0510",
    archivePrefix = "arXiv",
    primaryClass = "gr-qc",
    doi = "10.1088/0264-9381/28/17/175006",
    journal = "Class. Quant. Grav.",
    volume = "28",
    pages = "175006",
    year = "2011"
}

@book{PoissonBook,
    author = {Eric Poisson},
    title = {A Relativist's Toolkit: The Mathematics of Black-Hole Mechanics},
    year = {2004},
    publisher = {Cambridge University Press}
}

@article{Penrose_1971,
       author = {Penrose, Roger and Floyd, R.M.},
        title = {Extraction of Rotational Energy from a Black Hole},
      journal = {Nature Physical Science},
         year = 1971,
        month = feb,
       volume = {229},
       number = {6},
        pages = {177-179},
          doi = {10.1038/physci229177a0},
       adsurl = {https://ui.adsabs.harvard.edu/abs/1971NPhS..229..177P}
}

@article{Ori_2000,
    author = "Ori, Amos and Thorne, Kip S.",
    title = "{The Transition from inspiral to plunge for a compact body in a circular equatorial orbit around a massive, spinning black hole}",
    eprint = "gr-qc/0003032",
    archivePrefix = "arXiv",
    doi = "10.1103/PhysRevD.62.124022",
    journal = "Phys. Rev. D",
    volume = "62",
    pages = "124022",
    year = "2000"
}

@article{Hinderer_2008,
    author = "Hinderer, Tanja and Flanagan, Eanna E.",
    title = "{Two timescale analysis of extreme mass ratio inspirals in Kerr. I. Orbital Motion}",
    eprint = "0805.3337",
    archivePrefix = "arXiv",
    primaryClass = "gr-qc",
    doi = "10.1103/PhysRevD.78.064028",
    journal = "Phys. Rev. D",
    volume = "78",
    pages = "064028",
    year = "2008"
}

@article{Pound_2020,
    author = "Pound, Adam and Wardell, Barry and Warburton, Niels and Miller, Jeremy",
    title = "{Second-Order Self-Force Calculation of Gravitational Binding Energy in Compact Binaries}",
    eprint = "1908.07419",
    archivePrefix = "arXiv",
    primaryClass = "gr-qc",
    doi = "10.1103/PhysRevLett.124.021101",
    journal = "Phys. Rev. Lett.",
    volume = "124",
    number = "2",
    pages = "021101",
    year = "2020"
}

@article{miller2020twotimescale,
    author = "Miller, Jeremy and Pound, Adam",
    title = "{Two-timescale evolution of extreme-mass-ratio inspirals: waveform generation scheme for quasicircular orbits in Schwarzschild spacetime}",
    eprint = "2006.11263",
    archivePrefix = "arXiv",
    primaryClass = "gr-qc",
    doi = "10.1103/PhysRevD.103.064048",
    journal = "Phys. Rev. D",
    volume = "103",
    number = "6",
    pages = "064048",
    year = "2021"
}

@article{Campanelli_1999,
    author = "Campanelli, Manuela and Lousto, Carlos O.",
    title = "{Second order gauge invariant gravitational perturbations of a Kerr black hole}",
    eprint = "gr-qc/9811019",
    archivePrefix = "arXiv",
    reportNumber = "AEI-096",
    doi = "10.1103/PhysRevD.59.124022",
    journal = "Phys. Rev. D",
    volume = "59",
    pages = "124022",
    year = "1999"
}

@article{Green_2020,
   title={Teukolsky formalism for nonlinear Kerr perturbations},
   volume={37},
   ISSN={1361-6382},
   url={http://dx.doi.org/10.1088/1361-6382/ab7075},
   DOI={10.1088/1361-6382/ab7075},
   number={7},
   journal={Classical and Quantum Gravity},
   publisher={IOP Publishing},
   author={Green, Stephen R and Hollands, Stefan and Zimmerman, Peter},
   year={2020},
   month={Feb},
   pages={075001}
}

@article{Christodoulou:70,
	title = {Reversible and Irreversible Transformations in Black-Hole Physics},
	author = {Demetrios Christodoulou},
	journal = {Phys. Rev. Lett.},
	volume = {25}, 
	pages = {1596},
	year = {1970}
}

@article{Jaramillo:2010ay,
    author = "Jaramillo, J. L. and Gourgoulhon, E.",
    editor = "Blanchet, Luc and Spallicci, Alessandro and Whiting, Bernard",
    title = "{Mass and Angular Momentum in General Relativity}",
    eprint = "1001.5429",
    archivePrefix = "arXiv",
    primaryClass = "gr-qc",
    doi = "10.1007/978-90-481-3015-3_4",
    journal = "Fundam. Theor. Phys.",
    volume = "162",
    pages = "87--124",
    year = "2011"
}

@article{Ashtekar:2004cn,
    author = "Ashtekar, Abhay and Krishnan, Badri",
    title = "{Isolated and dynamical horizons and their applications}",
    eprint = "gr-qc/0407042",
    archivePrefix = "arXiv",
    doi = "10.12942/lrr-2004-10",
    journal = "Living Rev. Rel.",
    volume = "7",
    pages = "10",
    year = "2004"
}

@article{Newman-Penrose:66,
	title		= {Note on the {B}ondi-{M}etzner-{S}achs Group},
	author	= {E. T. Newman and R. Penrose},
	journal	= {J. Math. Phys.},
	volume	= {7}, 
	pages	= {863},
	year		= {1966}
}

@book{Wald,
		author 			= "Wald, Robert M.",
		booktitle		= "General Relativity",
		publisher		= "The University of Chicago Press",
		address			= "Chicago and London",
		year				= "1984"	
}

@unpublished{Pound-Wardell:2021,
    author = "Pound, Adam and Wardell, Barry",
    title = "{Black hole perturbation theory and gravitational self-force}",
    eprint = "2101.04592",
    archivePrefix = "arXiv",
    primaryClass = "gr-qc",
    month = "1",
    year = "2021"
}

@article{Brizuela:09,
    author = "Brizuela, David and Martin-Garcia, Jose M. and Tiglio, Manuel",
    title = "{A Complete gauge-invariant formalism for arbitrary second-order perturbations of a Schwarzschild black hole}",
    eprint = "0903.1134",
    archivePrefix = "arXiv",
    primaryClass = "gr-qc",
    doi = "10.1103/PhysRevD.80.024021",
    journal = "Phys. Rev. D",
    volume = "80",
    pages = "024021",
    year = "2009"
}

@article{Hughes1,
    author = "O'Sullivan, Stephen and Hughes, Scott A.",
    title = "{Strong-field tidal distortions of rotating black holes: Formalism and results for circular, equatorial orbits}",
    eprint = "1407.6983",
    archivePrefix = "arXiv",
    primaryClass = "gr-qc",
    doi = "10.1103/PhysRevD.91.109901",
    journal = "Phys. Rev. D",
    volume = "90",
    number = "12",
    pages = "124039",
    year = "2014",
    note = "[Erratum: Phys.Rev.D 91, 109901 (2015)]"
}

@article{Hughes2,
    author = "O'Sullivan, Stephen and Hughes, Scott A.",
    title = "{Strong-field tidal distortions of rotating black holes: II. Horizon dynamics from eccentric and inclined orbits}",
    eprint = "1505.03809",
    archivePrefix = "arXiv",
    primaryClass = "gr-qc",
    doi = "10.1103/PhysRevD.94.044057",
    journal = "Phys. Rev. D",
    volume = "94",
    number = "4",
    pages = "044057",
    year = "2016"
}

@article{Hughes3,
    author = "Penna, Robert F. and Hughes, Scott A. and O'Sullivan, Stephen",
    title = "{Strong-field tidal distortions of rotating black holes. III. Embeddings in hyperbolic three-space}",
    eprint = "1704.05471",
    archivePrefix = "arXiv",
    primaryClass = "gr-qc",
    doi = "10.1103/PhysRevD.96.064030",
    journal = "Phys. Rev. D",
    volume = "96",
    number = "6",
    pages = "064030",
    year = "2017"
}

@article{Poisson:04,
      author         = "Poisson, Eric",
      title          = "{Absorption of mass and angular momentum by a black hole:
                        Time-domain formalisms for gravitational perturbations,
                        and the small-hole / slow-motion approximation}",
      journal        = "Phys. Rev. D",
      volume         = "70",
      pages          = "084044",
      doi            = "10.1103/PhysRevD.70.084044",
      year           = "2004",
      eprint         = "gr-qc/0407050",
      archivePrefix  = "arXiv"
}

@article{Price-Thorne:86,
    author = "Price, R. H. and Thorne, K. S.",
    title = "{Membrane Viewpoint on Black Holes: Properties and Evolution of the Stretched Horizon}",
    doi = "10.1103/PhysRevD.33.915",
    journal = "Phys. Rev. D",
    volume = "33",
    pages = "915--941",
    year = "1986"
}

@article{Hawking-Hartle:72,
    author = "Hawking, S. W. and Hartle, J. B.",
    title = "{Energy and angular momentum flow into a black hole}",
    doi = "10.1007/BF01645515",
    journal = "Commun. Math. Phys.",
    volume = "27",
    pages = "283--290",
    year = "1972"
}

@article{Teukolsky-Press:74,
    author = "Teukolsky, S. A. and Press, W. H.",
    title = "{Perturbations of a rotating black hole. III - Interaction of the hole with gravitational and electromagnet ic radiation}",
    doi = "10.1086/153180",
    journal = "Astrophys. J.",
    volume = "193",
    pages = "443--461",
    year = "1974"
}

@article{Hamerly-Chen:10,
    author = "Hamerly, Ryan and Chen, Yanbei",
    title = "{Event Horizon Deformations in Extreme Mass-Ratio Black Hole Mergers}",
    eprint = "1007.5387",
    archivePrefix = "arXiv",
    primaryClass = "gr-qc",
    doi = "10.1103/PhysRevD.84.124015",
    journal = "Phys. Rev. D",
    volume = "84",
    pages = "124015",
    year = "2011"
}

@article{Cohen-Pfeiffer-Scheel:08,
    author = "Cohen, Michael I. and Pfeiffer, Harald P. and Scheel, Mark A.",
    title = "{Revisiting Event Horizon Finders}",
    eprint = "0809.2628",
    archivePrefix = "arXiv",
    primaryClass = "gr-qc",
    doi = "10.1088/0264-9381/26/3/035005",
    journal = "Class. Quant. Grav.",
    volume = "26",
    pages = "035005",
    year = "2009"
}

@article{Bohn-Kidder-Teukolsky:16,
    author = "Bohn, Andy and Kidder, Lawrence E. and Teukolsky, Saul A.",
    title = "{Parallel adaptive event horizon finder for numerical relativity}",
    eprint = "1606.00437",
    archivePrefix = "arXiv",
    primaryClass = "gr-qc",
    doi = "10.1103/PhysRevD.94.064008",
    journal = "Phys. Rev. D",
    volume = "94",
    number = "6",
    pages = "064008",
    year = "2016"
}

@article{Thornburg:03,
    author = "Thornburg, Jonathan",
    title = "{A Fast apparent horizon finder for three-dimensional Cartesian grids in numerical relativity}",
    eprint = "gr-qc/0306056",
    archivePrefix = "arXiv",
    reportNumber = "AEI-2003-049",
    doi = "10.1088/0264-9381/21/2/026",
    journal = "Class. Quant. Grav.",
    volume = "21",
    pages = "743--766",
    year = "2004"
}

@article{Emparan-Martinez-Zilhao:17,
    author = "Emparan, Roberto and Martinez, Marina and Zilhao, Miguel",
    title = "{Black hole fusion in the extreme mass ratio limit}",
    eprint = "1708.08868",
    archivePrefix = "arXiv",
    primaryClass = "gr-qc",
    doi = "10.1103/PhysRevD.97.044004",
    journal = "Phys. Rev. D",
    volume = "97",
    number = "4",
    pages = "044004",
    year = "2018"
}

@article{Emparan-Martinez:16,
    author = "Emparan, Roberto and Martinez, Marina",
    title = "{Exact Event Horizon of a Black Hole Merger}",
    eprint = "1603.00712",
    archivePrefix = "arXiv",
    primaryClass = "gr-qc",
    doi = "10.1088/0264-9381/33/15/155003",
    journal = "Class. Quant. Grav.",
    volume = "33",
    number = "15",
    pages = "155003",
    year = "2016"
}

@article{Hussain-Booth:17,
    author = "Hussain, Uzair and Booth, Ivan",
    title = "{Deformation of horizons during a merger}",
    eprint = "1705.01510",
    archivePrefix = "arXiv",
    primaryClass = "gr-qc",
    doi = "10.1088/1361-6382/aa9959",
    journal = "Class. Quant. Grav.",
    volume = "35",
    number = "1",
    pages = "015013",
    year = "2018"
}

@article{Bruni-etal:96,
      author         = "Bruni, Marco and Matarrese, Sabino and Mollerach, Silvia
                        and Sonego, Sebastiano",
      title          = "{Perturbations of space-time: Gauge transformations and
                        gauge invariance at second order and beyond}",
      journal        = "Class.Quant.Grav.",
      volume         = "14",
      pages          = "2585-2606",
      doi            = "10.1088/0264-9381/14/9/014",
      year           = "1997",
      eprint         = "gr-qc/9609040",
      archivePrefix  = "arXiv",

      reportNumber   = "IC-96-174, SISSA-136-96-A"
}

@book{Kevorkian-Cole:96,
	author = "Kevorkian, J. and Cole, Julian D.",
	title = "Multiple scale and singular perturbation methods",
	publisher = "Springer",
	address = "New York",
	year = "1996"
}

@article{Pound:10,
    author = "Pound, Adam",
    title = "{Singular perturbation techniques in the gravitational self-force problem}",
    eprint = "1003.3954",
    archivePrefix = "arXiv",
    primaryClass = "gr-qc",
    doi = "10.1103/PhysRevD.81.124009",
    journal = "Phys. Rev. D",
    volume = "81",
    pages = "124009",
    year = "2010"
}

@article{vandeMeent:14,
    author = "van de Meent, Maarten",
    title = "{Resonantly enhanced kicks from equatorial small mass-ratio inspirals}",
    eprint = "1406.2594",
    archivePrefix = "arXiv",
    primaryClass = "gr-qc",
    doi = "10.1103/PhysRevD.90.044027",
    journal = "Phys. Rev. D",
    volume = "90",
    number = "4",
    pages = "044027",
    year = "2014"
}

@article{Schnetter:2006yt,
    author = "Schnetter, Erik and Krishnan, Badri and Beyer, Florian",
    title = "{Introduction to dynamical horizons in numerical relativity}",
    eprint = "gr-qc/0604015",
    archivePrefix = "arXiv",
    reportNumber = "AEI-2006-018, LSU-REL-033006",
    doi = "10.1103/PhysRevD.74.024028",
    journal = "Phys. Rev. D",
    volume = "74",
    pages = "024028",
    year = "2006"
}

@article{Prasad:2020xgr,
    author = "Prasad, Vaishak and Gupta, Anshu and Bose, Sukanta and Krishnan, Badri and Schnetter, Erik",
    title = "{News from horizons in binary black hole mergers}",
    eprint = "2003.06215",
    archivePrefix = "arXiv",
    primaryClass = "gr-qc",
    reportNumber = "LIGO Preprint number LIGO-P2000098",
    doi = "10.1103/PhysRevLett.125.121101",
    journal = "Phys. Rev. Lett.",
    volume = "125",
    number = "12",
    pages = "121101",
    year = "2020"
}

@unpublished{Pook-Kolb:2020zhm,
    author = "Pook-Kolb, Daniel and Birnholtz, Ofek and Jaramillo, Jos\'e Luis and Krishnan, Badri and Schnetter, Erik",
    title = "{Horizons in a binary black hole merger I: Geometry and area increase}",
    eprint = "2006.03939",
    archivePrefix = "arXiv",
    primaryClass = "gr-qc",
    month = "6",
    year = "2020"
}

@unpublished{Pook-Kolb:2020jlr,
    author = "Pook-Kolb, Daniel and Birnholtz, Ofek and Jaramillo, Jos\'e Luis and Krishnan, Badri and Schnetter, Erik",
    title = "{Horizons in a binary black hole merger II: Fluxes, multipole moments and stability}",
    eprint = "2006.03940",
    archivePrefix = "arXiv",
    primaryClass = "gr-qc",
    month = "6",
    year = "2020"
}

@article{Booth:2003ji,
    author = "Booth, Ivan and Fairhurst, Stephen",
    title = "{The First law for slowly evolving horizons}",
    eprint = "gr-qc/0307087",
    archivePrefix = "arXiv",
    doi = "10.1103/PhysRevLett.92.011102",
    journal = "Phys. Rev. Lett.",
    volume = "92",
    pages = "011102",
    year = "2004"
}

@article{Hartle:1974gy,
    author = "Hartle, James B.",
    title = "{Tidal shapes and shifts on rotating black holes}",
    doi = "10.1103/PhysRevD.9.2749",
    journal = "Phys. Rev. D",
    volume = "9",
    pages = "2749--2759",
    year = "1974"
}

@article{Hartle:1973zz,
    author = "Hartle, James B.",
    title = "{Tidal Friction in Slowly Rotating Black Holes}",
    doi = "10.1103/PhysRevD.8.1010",
    journal = "Phys. Rev. D",
    volume = "8",
    pages = "1010--1024",
    year = "1973"
}

@article{Rahman:2019bmk,
    author = "Rahman, Adel A. and Wald, Robert M.",
    title = "{Black Hole Memory}",
    eprint = "1912.12806",
    archivePrefix = "arXiv",
    primaryClass = "gr-qc",
    doi = "10.1103/PhysRevD.101.124010",
    journal = "Phys. Rev. D",
    volume = "101",
    number = "12",
    pages = "124010",
    year = "2020"
}

@article{Wald:1991zz,
    author = "Wald, Robert M. and Iyer, Vivek",
    title = "{Trapped surfaces in the Schwarzschild geometry and cosmic censorship}",
    doi = "10.1103/PhysRevD.44.R3719",
    journal = "Phys. Rev. D",
    volume = "44",
    pages = "3719--3722",
    year = "1991"
}

@article{Szabados:2009eka,
    author = "Szabados, L\'aszl\'o B.",
    title = "{Quasi-Local Energy-Momentum and Angular Momentum in General Relativity}",
    doi = "10.12942/lrr-2009-4",
    journal = "Living Rev. Rel.",
    volume = "12",
    pages = "4",
    year = "2009"
}

@article{Gralla:2020srx,
    author = "Gralla, Samuel E. and Lupsasca, Alexandru and Marrone, Daniel P.",
    title = "{The shape of the black hole photon ring: A precise test of strong-field general relativity}",
    eprint = "2008.03879",
    archivePrefix = "arXiv",
    primaryClass = "gr-qc",
    doi = "10.1103/PhysRevD.102.124004",
    journal = "Phys. Rev. D",
    volume = "102",
    number = "12",
    pages = "124004",
    year = "2020"
}

@article{EHT,
    author = "Akiyama, Kazunori and others",
    collaboration = "Event Horizon Telescope",
    title = "{First M87 Event Horizon Telescope Results. I. The Shadow of the Supermassive Black Hole}",
    eprint = "1906.11238",
    archivePrefix = "arXiv",
    primaryClass = "astro-ph.GA",
    doi = "10.3847/2041-8213/ab0ec7",
    journal = "Astrophys. J.",
    volume = "875",
    number = "1",
    pages = "L1",
    year = "2019"
}

@unpublished{Volkel:2020xlc,
    author = {V\"olkel, Sebastian H. and Barausse, Enrico and Franchini, Nicola and Broderick, Avery E.},
    title = "{EHT tests of the strong-field regime of General Relativity}",
    eprint = "2011.06812",
    archivePrefix = "arXiv",
    primaryClass = "gr-qc",
    month = "11",
    year = "2020"
}

@article{Psaltis:2020lvx,
    author = "Psaltis, Dimitrios and others",
    collaboration = "Event Horizon Telescope",
    title = "{Gravitational Test Beyond the First Post-Newtonian Order with the Shadow of the M87 Black Hole}",
    eprint = "2010.01055",
    archivePrefix = "arXiv",
    primaryClass = "gr-qc",
    doi = "10.1103/PhysRevLett.125.141104",
    journal = "Phys. Rev. Lett.",
    volume = "125",
    number = "14",
    pages = "141104",
    year = "2020"
}

@article{Hayward:1993ph,
    author = "Hayward, Sean A.",
    title = "{Quasilocal gravitational energy}",
    eprint = "gr-qc/9303030",
    archivePrefix = "arXiv",
    reportNumber = "PRINT-93-0302 (GARCHING)",
    doi = "10.1103/PhysRevD.49.831",
    journal = "Phys. Rev. D",
    volume = "49",
    pages = "831--839",
    year = "1994"
}

@article{Hayward:1993mw,
    author = "Hayward, Sean A.",
    title = {General laws of black hole dynamics}, journal = {Phys. Rev. D},
    volume = {49},
    pages = {6467--6474},
    eprint = "gr-qc/9303006",
    archivePrefix = "arXiv",
    year = "1994"
}

@unpublished{Hayward:2008ti,
    author = "Hayward, Sean A.",
    title = "{Dynamics of black holes}",
    eprint = "0810.0923",
    archivePrefix = "arXiv",
    primaryClass = "gr-qc",
    month = "10",
    year = "2008"
}

@article{Booth:2005qc,
    author = "Booth, Ivan",
    title = "{Black hole boundaries}",
    eprint = "gr-qc/0508107",
    archivePrefix = "arXiv",
    doi = "10.1139/p05-063",
    journal = "Can. J. Phys.",
    volume = "83",
    pages = "1073--1099",
    year = "2005"
}

@article{Krishnan:2007va,
    author = "Krishnan, Badri",
    editor = "Scott, Susan M. and McClelland, David E.",
    title = "{Fundamental properties and applications of quasi-local black hole horizons}",
    eprint = "0712.1575",
    archivePrefix = "arXiv",
    primaryClass = "gr-qc",
    reportNumber = "AEI-2007-170",
    doi = "10.1088/0264-9381/25/11/114005",
    journal = "Class. Quant. Grav.",
    volume = "25",
    pages = "114005",
    year = "2008"
}

@article{Gourgoulhon:2008pu,
    author = "Gourgoulhon, Eric and Jaramillo, Jose Luis",
    title = "{New theoretical approaches to black holes}",
    eprint = "0803.2944",
    archivePrefix = "arXiv",
    primaryClass = "astro-ph",
    doi = "10.1016/j.newar.2008.03.026",
    journal = "New Astron. Rev.",
    volume = "51",
    pages = "791--798",
    year = "2008"
}

@article{Hawking:2016msc,
    author = "Hawking, Stephen W. and Perry, Malcolm J. and Strominger, Andrew",
    title = "{Soft Hair on Black Holes}",
    eprint = "1601.00921",
    archivePrefix = "arXiv",
    primaryClass = "hep-th",
    doi = "10.1103/PhysRevLett.116.231301",
    journal = "Phys. Rev. Lett.",
    volume = "116",
    number = "23",
    pages = "231301",
    year = "2016"
}

@article{Cardoso:2019rvt,
    author = "Cardoso, Vitor and Pani, Paolo",
    title = "{Testing the nature of dark compact objects: a status report}",
    eprint = "1904.05363",
    archivePrefix = "arXiv",
    primaryClass = "gr-qc",
    doi = "10.1007/s41114-019-0020-4",
    journal = "Living Rev. Rel.",
    volume = "22",
    number = "1",
    pages = "4",
    year = "2019"
}

@article{Abbott:2020niy,
    author = "Abbott, R. and others",
    collaboration = "LIGO Scientific, Virgo",
    title = "{GWTC-2: Compact Binary Coalescences Observed by LIGO and Virgo During the First Half of the Third Observing Run}",
    eprint = "2010.14527",
    archivePrefix = "arXiv",
    primaryClass = "gr-qc",
    reportNumber = "P2000061",
    doi = "10.1103/PhysRevX.11.021053",
    journal = "Phys. Rev. X",
    volume = "11",
    pages = "021053",
    year = "2021"
}

@article{Barausse:2020rsu,
    author = "Barausse, Enrico and others",
    title = "{Prospects for Fundamental Physics with LISA}",
    eprint = "2001.09793",
    archivePrefix = "arXiv",
    primaryClass = "gr-qc",
    doi = "10.1007/s10714-020-02691-1",
    journal = "Gen. Rel. Grav.",
    volume = "52",
    number = "8",
    pages = "81",
    year = "2020"
}

@article{Gair:2012nm,
    author = "Gair, Jonathan R. and Vallisneri, Michele and Larson, Shane L. and Baker, John G.",
    title = "{Testing General Relativity with Low-Frequency, Space-Based Gravitational-Wave Detectors}",
    eprint = "1212.5575",
    archivePrefix = "arXiv",
    primaryClass = "gr-qc",
    doi = "10.12942/lrr-2013-7",
    journal = "Living Rev. Rel.",
    volume = "16",
    pages = "7",
    year = "2013"
}

@article{Berti:2016lat,
    author = "Berti, Emanuele and Sesana, Alberto and Barausse, Enrico and Cardoso, Vitor and Belczynski, Krzysztof",
    title = "{Spectroscopy of Kerr black holes with Earth- and space-based interferometers}",
    eprint = "1605.09286",
    archivePrefix = "arXiv",
    primaryClass = "gr-qc",
    doi = "10.1103/PhysRevLett.117.101102",
    journal = "Phys. Rev. Lett.",
    volume = "117",
    number = "10",
    pages = "101102",
    year = "2016"
}

@article{Maggiore:2019uih,
    author = "Maggiore, Michele and others",
    title = "{Science Case for the Einstein Telescope}",
    eprint = "1912.02622",
    archivePrefix = "arXiv",
    primaryClass = "astro-ph.CO",
    doi = "10.1088/1475-7516/2020/03/050",
    journal = "JCAP",
    volume = "03",
    pages = "050",
    year = "2020"
}

@article{Loutrel:2020wbw,
    author = "Loutrel, Nicholas and Ripley, Justin L. and Giorgi, Elena and Pretorius, Frans",
    title = "{Second Order Perturbations of Kerr Black Holes: Reconstruction of the Metric}",
    eprint = "2008.11770",
    archivePrefix = "arXiv",
    primaryClass = "gr-qc",
    doi = "10.1103/PhysRevD.103.104017",
    journal = "Phys. Rev. D",
    volume = "103",
    number = "10",
    pages = "104017",
    year = "2021"
}

@article{Ripley:2020xby,
    author = "Ripley, Justin L. and Loutrel, Nicholas and Giorgi, Elena and Pretorius, Frans",
    title = "{Numerical computation of second order vacuum perturbations of Kerr black holes}",
    eprint = "2010.00162",
    archivePrefix = "arXiv",
    primaryClass = "gr-qc",
    doi = "10.1103/PhysRevD.103.104018",
    journal = "Phys. Rev. D",
    volume = "103",
    pages = "104018",
    year = "2021"
}

@article{Ashtekar:2005ez,
    author = "Ashtekar, Abhay and Galloway, Gregory J.",
    title = "{Some uniqueness results for dynamical horizons}",
    eprint = "gr-qc/0503109",
    archivePrefix = "arXiv",
    doi = "10.4310/ATMP.2005.v9.n1.a1",
    journal = "Adv. Theor. Math. Phys.",
    volume = "9",
    number = "1",
    pages = "1--30",
    year = "2005"
}

@article{Detweiler:2003ci,
    author = "Detweiler, Steven L. and Poisson, Eric",
    title = "{Low multipole contributions to the gravitational selfforce}",
    eprint = "gr-qc/0312010",
    archivePrefix = "arXiv",
    doi = "10.1103/PhysRevD.69.084019",
    journal = "Phys. Rev. D",
    volume = "69",
    pages = "084019",
    year = "2004"
}

@article{Thornburg:2006zb,
    author = "Thornburg, Jonathan",
    title = "{Event and apparent horizon finders for 3+1 numerical relativity}",
    eprint = "gr-qc/0512169",
    archivePrefix = "arXiv",
    reportNumber = "AEI-2005-184",
    journal = "Living Rev. Rel.",
    volume = "10",
    pages = "3",
    year = "2007"
}

@book{Hawking:1973uf,
    author = "Hawking, S. W. and Ellis, G. F. R.",
    title = "{The Large Scale Structure of Space-Time}",
    doi = "10.1017/CBO9780511524646",
    isbn = "978-0-521-20016-5, 978-0-521-09906-6, 978-0-511-82630-6, 978-0-521-09906-6",
    publisher = "Cambridge University Press",
    series = "Cambridge Monographs on Mathematical Physics",
    month = "2",
    year = "2011"
}

@incollection{Hawking:1973,
    author = {Hawking, S. W.}, 
    title = {The Event Horizon},
    booktitle = {Black Holes}, 
    pages = {1--55}, 
    editor = {De{W}itt, C. and De{W}itt, B. S.},
    publisher = {Gordon and Breach}, 
    address = {New York}, 
    year = {1973}
}

@article{Abbott:1982jh,
    author = "Abbott, L. F. and Deser, Stanley",
    title = "{Charge Definition in Nonabelian Gauge Theories}",
    reportNumber = "Print-82-0511 (BRANDEIS)",
    doi = "10.1016/0370-2693(82)90338-0",
    journal = "Phys. Lett. B",
    volume = "116",
    pages = "259--263",
    year = "1982"
}

@article{Dolan:2012jg,
    author = "Dolan, Sam R. and Barack, Leor",
    title = "{Self-force via $m$-mode regularization and 2+1D evolution: III. Gravitational field on Schwarzschild spacetime}",
    eprint = "1211.4586",
    archivePrefix = "arXiv",
    primaryClass = "gr-qc",
    doi = "10.1103/PhysRevD.87.084066",
    journal = "Phys. Rev. D",
    volume = "87",
    pages = "084066",
    year = "2013"
}

@article{Amaro-Seoane:2014ela,
    author = "Amaro-Seoane, Pau and Gair, Jonathan R. and Pound, Adam and Hughes, Scott A. and Sopuerta, Carlos F.",
    editor = "Ciani, Giacomo and Conklin, John W. and Mueller, Guido",
    title = "{Research Update on Extreme-Mass-Ratio Inspirals}",
    eprint = "1410.0958",
    archivePrefix = "arXiv",
    primaryClass = "astro-ph.CO",
    doi = "10.1088/1742-6596/610/1/012002",
    journal = "J. Phys. Conf. Ser.",
    volume = "610",
    number = "1",
    pages = "012002",
    year = "2015"
}

@unpublished{Amaro-Seoane:2020zbo,
    author = "Amaro-Seoane, Pau",
    title = "{The gravitational capture of compact objects by massive black holes}",
    eprint = "2011.03059",
    archivePrefix = "arXiv",
    primaryClass = "gr-qc",
    month = "11",
    year = "2020"
}

@misc{BHPT,
note = {The Black Hole Perturbation Toolkit is available at \url{https://bhptoolkit.org/toolkit.html}}
}

@article{Stein:2019buj,
    author = "Stein, Leo C. and Warburton, Niels",
    title = "{Location of the last stable orbit in Kerr spacetime}",
    eprint = "1912.07609",
    archivePrefix = "arXiv",
    primaryClass = "gr-qc",
    doi = "10.1103/PhysRevD.101.064007",
    journal = "Phys. Rev. D",
    volume = "101",
    number = "6",
    pages = "064007",
    year = "2020"
}

@article{Taracchini:2014zpa,
    author = "Taracchini, Andrea and Buonanno, Alessandra and Khanna, Gaurav and Hughes, Scott A.",
    title = "{Small mass plunging into a Kerr black hole: Anatomy of the inspiral-merger-ringdown waveforms}",
    eprint = "1404.1819",
    archivePrefix = "arXiv",
    primaryClass = "gr-qc",
    doi = "10.1103/PhysRevD.90.084025",
    journal = "Phys. Rev. D",
    volume = "90",
    number = "8",
    pages = "084025",
    year = "2014"
}

@article{Barack:1999ya,
    author = "Barack, Leor and Ori, Amos",
    title = "{Late time decay of gravitational and electromagnetic perturbations along the event horizon}",
    eprint = "gr-qc/9907085",
    archivePrefix = "arXiv",
    doi = "10.1103/PhysRevD.60.124005",
    journal = "Phys. Rev. D",
    volume = "60",
    pages = "124005",
    year = "1999"
}

@article{Berti:2009kk,
    author = "Berti, Emanuele and Cardoso, Vitor and Starinets, Andrei O.",
    title = "{Quasinormal modes of black holes and black branes}",
    eprint = "0905.2975",
    archivePrefix = "arXiv",
    primaryClass = "gr-qc",
    doi = "10.1088/0264-9381/26/16/163001",
    journal = "Class. Quant. Grav.",
    volume = "26",
    pages = "163001",
    year = "2009"
}

@article{Price:1971fb,
    author = "Price, Richard H.",
    title = "{Nonspherical perturbations of relativistic gravitational collapse. 1. Scalar and gravitational perturbations}",
    doi = "10.1103/PhysRevD.5.2419",
    journal = "Phys. Rev. D",
    volume = "5",
    pages = "2419--2438",
    year = "1972"
}

@article{Preston:2006ze,
    author = "Preston, Brent and Poisson, Eric",
    title = "{A light-cone gauge for black-hole perturbation theory}",
    eprint = "gr-qc/0606094",
    archivePrefix = "arXiv",
    doi = "10.1103/PhysRevD.74.064010",
    journal = "Phys. Rev. D",
    volume = "74",
    pages = "064010",
    year = "2006"
}

@article{Akcay:2013wfa,
    author = "Akcay, Sarp and Warburton, Niels and Barack, Leor",
    title = "{Frequency-domain algorithm for the Lorenz-gauge gravitational self-force}",
    eprint = "1308.5223",
    archivePrefix = "arXiv",
    primaryClass = "gr-qc",
    doi = "10.1103/PhysRevD.88.104009",
    journal = "Phys. Rev. D",
    volume = "88",
    number = "10",
    pages = "104009",
    year = "2013"
}

@article{Fujita:2009us,
    author = "Fujita, Ryuichi and Hikida, Wataru and Tagoshi, Hideyuki",
    title = "{An Efficient Numerical Method for Computing Gravitational Waves Induced by a Particle Moving on Eccentric Inclined Orbits around a Kerr Black Hole}",
    eprint = "0904.3810",
    archivePrefix = "arXiv",
    primaryClass = "gr-qc",
    doi = "10.1143/PTP.121.843",
    journal = "Prog. Theor. Phys.",
    volume = "121",
    pages = "843--874",
    year = "2009"
}

@article{Ori:2003cm,
    author = "Ori, Amos",
    title = "{Harmonic gauge dipole metric perturbations for weak field circular orbits in Schwarzschild space-time}",
    eprint = "gr-qc/0312013",
    archivePrefix = "arXiv",
    doi = "10.1103/PhysRevD.70.124027",
    journal = "Phys. Rev. D",
    volume = "70",
    pages = "124027",
    year = "2004"
}

@article{Fang:2005qq,
    author = "Fang, Hua and Lovelace, Geoffrey",
    title = "{Tidal coupling of a Schwarzschild black hole and circularly orbiting moon}",
    eprint = "gr-qc/0505156",
    archivePrefix = "arXiv",
    doi = "10.1103/PhysRevD.72.124016",
    journal = "Phys. Rev. D",
    volume = "72",
    pages = "124016",
    year = "2005"
}

@miscellaneous{Stein:private,
    author = {Leo Stein},
    note = {private communication}
}

@unpublished{Wardell-etal:2021,
    author = "Wardell, Barry and Pound, Adam and Warburton, Niels and Miller, Jeremy and Durkan, Leanne and Le Tiec, Alexandre",
    title = "{Gravitational waveforms for compact binaries from second-order self-force theory}",
    eprint = "2112.12265",
    archivePrefix = "arXiv",
    primaryClass = "gr-qc",
    month = "12",
    year = "2021"
}

\end{document}